\documentclass[prd,showpacs,superscriptaddress,nofootinbib,floatfix,11pt]{revtex4-1} 
\usepackage[utf8]{inputenc}
\usepackage{amsmath,amssymb}
\usepackage{slashed}
\usepackage{graphicx}
\usepackage{subfigure}
\usepackage[colorlinks=true,
linkcolor=red,
breaklinks=true,
urlcolor=magenta,
citecolor=blue]{hyperref}
\usepackage[usenames,dvipsnames]{color}

\usepackage{float}
\usepackage{mathrsfs}
\usepackage{color}
\usepackage{geometry}
\usepackage{bbm}               
\usepackage{xspace}				
\usepackage{epstopdf}
\usepackage{pgffor}					
\usepackage{multirow}
\usepackage[dvipsnames]{xcolor}

\allowdisplaybreaks
\usepackage{newtxtext, newtxmath}
\usepackage{braket,bm}
\usepackage{multirow}
\usepackage{enumitem}
\usepackage{epstopdf}

\newcommand{\itp}{\affiliation{CAS Key Laboratory of Theoretical Physics, Institute of Theoretical Physics,\\ Chinese Academy of Sciences, Beijing 100190, China}}

\newcommand{\ucas}{\affiliation{School of Physical Sciences, University of Chinese Academy of Sciences, Beijing 100049, China}}

\newcommand{\ific}{\affiliation{Instituto de F\'isica Corpuscular (centro mixto CSIC-UV), \\
Institutos de Investigaci\'on de Paterna, Apartado 22085, 46071, Valencia, Spain}}

\newcommand{\qfnu}{\affiliation{College of Physics and Engineering, Qufu Normal University, Qufu 273165, China}}

\begin{document}

\title{Production of hidden-heavy and double-heavy hadronic molecules at the \texorpdfstring{$Z$}{Z} factory of CEPC} 

\author{Zhao-Sai Jia}\qfnu \itp 
\author{Gang Li}\email{gli@qfnu.edu.cn} \qfnu
\author{Pan-Pan Shi}\email{panpan@ific.uv.es}\itp \ucas \ific 
\author{Zhen-Hua Zhang} \email{zhangzhenhua@itp.ac.cn} \itp \ucas  

\date{\today}

\begin{abstract} 
  \rule{0ex}{3ex}
With a clean environment and high collision energy, the Circular Electron Positron Collider (CEPC) would be an excellent facility for heavy flavor physics.  
Using the Monte Carlo event generator Pythia, we simulate the production of the charmed (bottom) hadron pairs in the electron-positron collisions at the $Z$ factory of CEPC, and the inclusive production rates for typical candidates of the hidden/double-charm and hidden/double-bottom $S$-wave hadronic molecules are estimated at an order-of-magnitude level with the final state interactions after the hadron pair production. The predicted cross sections for the hidden-charm meson-meson molecules $X(3872)$ and $Z_c(3900)$ are at $\rm{pb}$ level, which are about two to three orders of magnitude larger than the production cross sections for the double-charm meson-meson molecules $T_{cc}$ and $T_{cc}^{*}$, as the double-charmed ones require the production of two pairs of $c\bar{c}$ from the $Z$ boson decay. The production cross sections for the hidden-charm pentaquark states $P_{c}$ and $P_{cs}$ as
meson-baryon molecules are a few to tens of fb, which are about one magnitude larger than those of the possible hidden-charm baryon-antibaryon and double-charm meson-baryon molecules.
In the bottom sector, the production cross sections for the $Z_b$ states as 
$B^{(*)}\bar{B}^{*}$ molecules are about tens to hundreds of fb, 
indicating $10^6$ -- $10^7$ events from a two-year operation of CEPC, and the expected events from the double-bottom molecules are about 2 -- 5 orders of magnitude smaller than the $Z_b$ states.
Our results shows great prospects of probing heavy exotic hadrons at CEPC.

\end{abstract}

\maketitle

\section{Introduction}
Since 2003, many of so-called $XYZ$ states have been observed in the hidden-charm sector, which exhibit different properties from the traditional hadrons in the quark model and thus are candidates of exotic hadrons. One of the most famous $XYZ$ states is the $X(3872)$~\cite{Belle:2003nnu},
also called $\chi_{c1}(3872)$~\cite{ParticleDataGroup:2022pth}, observed in the $J/\psi\pi\pi$ invariant mass distribution by the Belle Collaboration and confirmed by many other experiments~\cite{CDF:2003cab, D0:2004zmu, LHCb:2013kgk, CMS:2013fpt, BESIII:2013fnz, LHCb:2014jvf, LHCb:2011zzp, BaBar:2010wfc, BaBar:2008flx, BaBar:2008qzi,BaBar:2007cmo}. Another important state is the isovector $Z_c(3900)$ discovered in the $J/\psi\pi$ invariant mass distribution by the BESIII and Belle Collaborations~\cite{BESIII:2013ris, Belle:2013yex,BESIII:2015cld}, which must have nontrivial structures beyond the traditional charmonium. In the past few years, the LHCb Collaboration has reported the hidden-charm pentaquark candidates $P_c(4380)^+$, $P_c(4440)^+$, $P_c(4450)^+$, and $P_c(4312)^+$ in the $J/\psi p$ invariant mass distribution~\cite{LHCb:2015yax,LHCb:2019kea}; the strange pentaquark candidates $P_{cs}(4459)$ and $P_{cs}(4338)$ states in the $J/\psi\Lambda$ spectrum~\cite{LHCb:2020jpq,LHCb:2022ogu}; and the double-charm tetraquark candidate $T_{cc}(3875)^+$ in the $D^0D^0\pi^+$
distribution~\cite{LHCb:2021vvq}. The masses of these states are close to the thresholds of at least a pair of heavy hadrons, making them candidates of hadronic molecules. There are also other interpretations. For instance, the compact multiquark picture has also been utilized to understand the structures of exotic states (for recent reviews, see~\cite{
Chen:2016qju,Hosaka:2016pey,Esposito:2016noz,Lebed:2016hpi,Ali:2017jda,Olsen:2017bmm,Guo:2017jvc,Albuquerque:2018jkn,Liu:2019zoy,Guo:2019twa,Brambilla:2019esw,Chen:2022asf}). According to heavy quark flavor symmetry (HQFS),
the bottom counterparts of these states like the $X_b$, $Z_b$, $P_b$, and $T_{bb}$ should also exist and have been predicted in various models~\cite{Ebert:2005nc,Hou:2006it,Belle:2011aa,Yang:2018oqd,Sharma:2024ern,Ren:2021dsi,Deng:2021gnb}, but so far only two bottomonium-like $Z_b$ states, $Z_b^{\pm}(10610)$ and $Z_b^{\pm}(10650)$, have been observed by the Belle experiment in the $\Upsilon(5S)$ decay processes~\cite{Belle:2011aa}. 
One of the decisive reasons for the deficiency of the signals from exotic states in the bottom sector is the limitation of the collision energy and detection efficiency of the present high-energy colliders.
The $Z$ factory of future Circular Electron Positron Collider (CEPC)~\cite{Gao:2022lew} with high center-of-mass (c.m.) energy (close to the mass of the $Z$ boson), clean background, high resolution and detection ability can provide great opportunity to study the heavy exotic states.

\begin{table}[htbp]
\caption{\label{tab_CEPC_parameters} The energy configurations, instantaneous luminosity ($L$), integrated luminosity ($\int{L}$), and event yields of the $Z$ factory at CEPC~\cite{Gao:2022lew, Sun:2023lut}.}
\renewcommand{\arraystretch}{1.2}
\begin{tabular*}{\columnwidth}{@{\extracolsep\fill}ccccccccccc}
\hline\hline 
$\sqrt{s}$ & Run time & Instantaneous luminosity & Integrated luminosity & \multirow{2}{*}{Event yields} \\[3pt]
$(\rm{GeV})$ &(year) &($10^{34}~\rm{cm^{-2} s^{-1}}$, per IP) & ($\text{ab}^{-1}$, 2 IPs) \\[3pt]
\hline
   $91.2$ &$2$ &$191.7$ &$100$ &$3 \times 10^{12}$ \\ [3pt]
\hline\hline
\end{tabular*}
\end{table}
The c.m. energy, luminosity and event yields of the $Z$ factory are listed in Table~\ref{tab_CEPC_parameters}~\cite{Gao:2022lew, Sun:2023lut}. At parton level, the hidden-charm, hidden-bottom, double-charm, and double-bottom exotic states are produced in the processes $Z\to c\bar{c}$, $Z\to b\bar{b}$, $Z\to c\bar{c}c\bar{c}$, and $Z\to b\bar{b}b\bar{b}$, respectively.
The branching radios measured at LEP are $\mathrm{Br}[Z\to c\bar{c}]=12.03\pm 0.21\%$, $\mathrm{Br}[Z\to b\bar{b}]=15.12\pm 0.05\%$, and $\mathrm{Br}[Z \to b\bar{b}b\bar{b}]=(3.6\pm 1.3)\times 10 ^{-4}$~\cite{ParticleDataGroup:2022pth}. The production of the doubly heavy baryons $\Xi_{cc}$, $\Xi_{bc}$, and $\Xi_{bb}$ at the $Z$ factory with the underlying processes $Z\to c\bar{c}c\bar{c}$, $Z\to b\bar{b}c\bar{c}$, and $Z\to b\bar{b}b\bar{b}$ has been studied in Ref.~\cite{Niu:2023ojf} in the NRQCD framework, and the total production cross sections for $\Xi_{cc}$, $\Xi_{bc}$, and $\Xi_{bb}$ are 848.03~fb, 2260.51~fb, and 41.16~fb, respectively. The production of the $T_{bb}$ state in the compact tetraquark configuration at the $Z$ factory has also been studied in Refs.~\cite{Ali:2018ifm,Ali:2019tli} by employing the Monte Carlo (MC) event generators MadGraph5\_aMC@NLO~\cite{Alwall:2014hca} and Pythia6~\cite{Sjostrand:2006za}, and the estimated production cross section is about 36~fb, corresponding to $3.6\times 10^6$ events produced at the $Z$ factory during a two-year operation.
The MC generator Pythia has also been widely used to simulate 
the production of the multiquark states in the $e^+e^-$~\cite{Artoisenet:2010uu,Qin:2020zlg}, $ep$~\cite{Yang:2021jof,Shi:2022ipx}, $pp$~\cite{Guo:2013ufa,Guo:2014ppa,Guo:2014sca,Albaladejo:2017blx,Ling:2021sld,Shi:2021hzm,Jin:2021cxj,Hua:2023zpa}, and $p\bar{p}$~\cite{Guo:2013ufa,Bignamini:2009sk,Bignamini:2009sk,Guo:2014sca,Albaladejo:2017blx} collisions. 

In this work, we will employ the Pythia8~\cite{Bierlich:2022pfr} 
to estimate the production cross sections of the charmonium-like states $X(3872)$ and $Z_c(3900)$,
the double-charm tetraquark state $T_{cc}(3875)^+$, the hidden-charm pentaquark states $P_c(4380)^+$, $P_c(4440)^+$, $P_c(4450)^+$, $P_c(4312)^+$, and $P_{cs}(4459)$,
and the bottomonium-like states $Z_b(10610)$ and $Z_b(10650)$ in the $e^+e^-$ collisions at the $Z$ factory, assuming they are $S$-wave hadronic molecules. The production cross sections of some typical hidden-charm and double-charm hadronic molecules predicted in Refs.~\cite{Dong:2021bvy,Dong:2021juy} as well as the $T_{bb}$ state as a $\bar B \bar B^*$ molecule will also be calculated.

The remaining parts of this paper is organized as follows. In Sec.~\ref{Sec_Hadroproduction}, we introduce the inclusive production mechanism of the hadronic molecules. The numerical results of the cross sections are presented in Sec.~\ref{Sec_Numerical Results}. A brief summary is given in Sec.~\ref{Sec_Summary}.

\section{Leptoproduction}\label{Sec_Hadroproduction}

\begin{figure}[htbp]
    \centering   
    \includegraphics[scale=0.45]{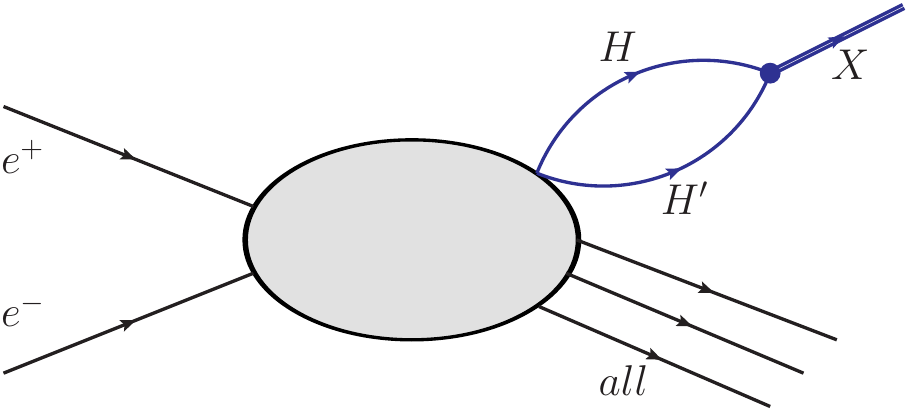}
    \caption{The inclusive production of the $X$ as a $HH^{\prime}$ hadronic molecule in $e^+ e^-$ collisions, where “all” denotes the other particles produced in this process.}
    \label{fig_prouction}
\end{figure}
In this section, we introduce the inclusive production mechanism of the hadronic molecule $X$ in $e^+e^-$ collisions. As shown in Fig.~\ref{fig_prouction}, a heavy-hadron pair $HH^{\prime}$ is inclusively generated in the $e^+ e^-$ collision, which is of short-distance nature, and then they are bound through the final state interaction (FSI) to form a hadronic molecule. At the $Z$ factory, the typical short-distance pair-production process is mediated by a virtual photon and $Z$ boson, while the contribution from the intermediate photon is negligible~\cite{Niu:2023ojf}. To give an order-of-magnitude estimation for the production cross sections of the hadronic molecules in $e^{+}e^{-}$ collision, we utilize the Pythia8 to simulate the short-distance inclusive production of the $HH^{\prime}$ pair and other particles, and the long-distance FSI is derived in the nonrelativistic effective field theory (NREFT) framework. In general, the amplitude for the production of the hadronic molecule state $X$ can be factorized as~\cite{Braaten:2005jj,Artoisenet:2009wk,Guo:2014sca}
\begin{align}
  \mathcal{M}[X+\text{all}]=\mathcal{M}[HH^{\prime}+\text{all}] \times G \times T_{X},
  \label{Eq_MX}
\end{align}
where $\mathcal{M}[HH^{\prime}+\text{all}]$ is the short-distance amplitude for the inclusive production process $e^+ e^- \to HH^{\prime}+\text{all}$, $G$ is the Green function of the intermediate heavy hadron $HH^{\prime}$ pair, and $T_{X}$ is the long-distance amplitude for $HH^{\prime} \to X$. Here the Green's function $G$ is UV divergent and is regularized by the Gaussian regulator~\cite{Nieves:2012tt}
\begin{align}
  G(E, \Lambda)=-\frac{\mu}{\pi^{2}} \left\{ \sqrt{2\pi}\frac{\Lambda}{4} + \frac{\gamma \pi}{2} e^{-2 \gamma^2 / \Lambda^2} \left[ i-\text{erfi}\left( \frac{\sqrt{2} \gamma}{\Lambda} \right) \right] \right\},
\end{align}
where $\gamma=\sqrt{2\mu(E-m_H-m_{H^\prime})}$ is the binding momentum of the $HH'$ pair with $\mu$ the reduced mass of $H$ and $H'$,  $\text{erfi}(z)=2/\sqrt{\pi} \int_0^z e^{t^2} dt$ is the imaginary error function, and the cutoff $\Lambda$ is in the range of 0.5--1.0~GeV, following Refs.~\cite{Guo:2013sya,Guo:2014sca, Shi:2022ipx}.

For the $S$-wave shallow hadronic molecule, the amplitude $T_{X}$ can be approximated by the effective coupling constant $g_{X}$, which can be extracted from the residues of the low-energy $HH'\to HH'$ scattering amplitude $T(E)$ as
\begin{align}
g_{X}^2 = \lim_{E \to E_0} (E^2 - E_0^2) T(E),
\label{eq_gX}
\end{align}
where $E_0$ is the pole position in the complex $E$-plane, satisfying $\text{det}[1-VG(E_0, \Lambda)] = 0$. One has $E_0 = M_{X}$ for a bound state on the physical Riemann sheet (RS) or virtual state on the unphysical RS, and $E_0 = M_{X} - i\Gamma/2$ for a resonance on the unphysical RS with mass $M_{X}$ and width $\Gamma$.
For the near-threshold hadronic molecules, one can use a constant separateable potential $V$ for the $HH'\to HH'$ scattering, and the scattering amplitude $T(E)$ can be solved from the Lippmann-Schwinger equation as
\begin{align}
T(E) = \frac{V}{1-VG(E, \Lambda)}.
\label{eq_T(E)}
\end{align}

The production cross section of the hadronic molecule $X$ can also be factorized into short-distance and long-distance parts. The short-distance part is given by the differential MC cross section of the inclusive $HH^{\prime}$ production,
\begin{align}
d \sigma[HH^{\prime}(k)+\text{all}]_{\text{MC}}=&\, K_{HH^{\prime}} \frac{1}{\text{flux}} \sum_{\text{all}} \int d \phi_{HH^{\prime}+\text{all}} \vert \mathcal{M}[HH^{\prime}(k)+\text{all}] \vert^2 \frac{d^3 k}{(2 \pi)^3 2 \mu},
\label{Eq_dsigma_HHMC}
\end{align}
where $k$ is the three-momentum in the c.m. frame of the $HH^{\prime}$ system. The overall factor $K_{HH^{\prime}} \sim \mathcal{O}(1)$ represents the difference between the MC simulation and the experimental data, and can be roughly taken as $K_{HH^{\prime}} \simeq 1$ for an order-of-magnitude estimate~\cite{Guo:2014sca, Shi:2022ipx}. The short-distance production amplitude $\mathcal{M}[HH^{\prime}+\text{all}]$, which is insensitive to the final-state relative momentum $k$
~\cite{Guo:2014sca, Shi:2022ipx}, can be approximated as a constant and taken outside from the integration of the final-state momentum $k$.
 Consequently, the differential cross section for $HH^{\prime}$ production in the MC event generator is proportional to $k^2$,
\begin{align}
\left( \frac{d \sigma[HH^{\prime}(k)+\text{all}]}{d k} \right)_{\text{MC}} \propto k^2.
\label{Eq_dsigma_HHMCdk}
\end{align}
The total cross section for the hadronic molecule $X$ production is given by
\begin{align}
\sigma[X+\text{all}] = \frac{1}{\text{flux}} \sum_{\text{all}} \int d\phi_{X+\text{all}} \vert \mathcal{M}[X+\text{all}] \vert^2,
\label{Eq_sigma_X}
\end{align}
where the phase-space integration is the same as that in Eq.~\eqref{Eq_dsigma_HHMC}~\cite{Guo:2014sca}. With the use of Eqs.~\eqref{Eq_MX} and ~\eqref{Eq_dsigma_HHMC}, the production cross section of $X$ can be derived as
\begin{align}
\sigma[X+\text{all}] = \frac{1}{4 m_{H} m_{H^{\prime}}} \vert G g_{X} \vert^2 \left( \frac{d \sigma[HH^{\prime}(k)+\text{all}]}{d k} \right)_{\text{MC}} \frac{4 \pi^2 \mu}{k^2},
\label{Eq_sigma_X(HHMC)}
\end{align}
where $m_{H}$ and $m_{H^{\prime}}$ are the masses of the heavy hadrons $H$ and $H^{\prime}$, respectively.

\section{Numerical Results}\label{Sec_Numerical Results}
In this section, the production cross sections of the typical hidden-heavy and double-heavy hadronic molecules at the $Z$ factory are estimated at an order-of-magnitude level using Eq.~\eqref{Eq_sigma_X(HHMC)}.
The differential cross sections $(d\sigma[HH^\prime+\text{all}]/dk)_{\text{MC}}$ of the $HH'$ pair production in the $e^+e^-$ collisions are obtained using the MC event generator Pythia~\cite{Sjostrand:2006za}. Some typical differential cross sections for the production of the charm-anticharm, double-charm, bottom-antibottom, and double-bottom hadron pairs are shown in Fig.~\ref{Fig:diffxsec_HHprime}, and the differential cross sections for other heavy hadron pairs can be found in Appendix~\ref{appen:differendial_pair_productions}. The formation of a hadronic molecule requires the constituent hadrons move collinearly with a small relative momentum. The choice of the cut of momentum $k$ has a small effect to the cross section and does not change our order-of-magnitude estimate. Therefore we follow the works in Refs. \cite{Yang:2021jof,Shi:2022ipx} and choose a small relative momentum range $|k|<350\,\rm{MeV}$ where
\begin{align}
    \left(\frac{d\sigma[HH^\prime+\text{all}]}{dk}\right)_{\text{MC}}=a k^2,
\end{align}
and the coefficient $a$ is obtained by fitting the differential pair-production cross sections simulated by the MC event generator.
The final expression of the production cross section of $X$ can be written as 
\begin{align}
\sigma[X+\text{all}] = \frac{\pi^2}{ m_{H}+m_{H^{\prime}}} \vert G g_{X} \vert^2 a.
\label{Eq_sigma_X_final}
\end{align}
The predicted cross sections for the hidden-charm, double-charm, and hidden-bottom hadronic molecules with $\Lambda=0.5$~GeV (out of the parentheses) and $\Lambda=1.0$~GeV (in the parentheses) are listed in Tables~\ref{tab_hidden-charm X cross sections}, \ref{tab_double-charm X cross sections}, and \ref{tab_hidden-bottom X cross sections}, respectively,
where the binding energy is defined as $E_B=m_H+m_{H^{\prime}}-m_X$\footnote{Here we use the isospin averaged mass for the heavy hadron $H^{(\prime)}$ in the isospin multiplet.} with $m_X$ the mass of the produced hadronic molecule. The binding energies at the outside and inside of the square brackets in these tables correspond to $\Lambda=0.5$ and $1.0$~GeV, respectively. 
\begin{figure}[tb]
\subfigure[]{\includegraphics[width=0.45\textwidth]{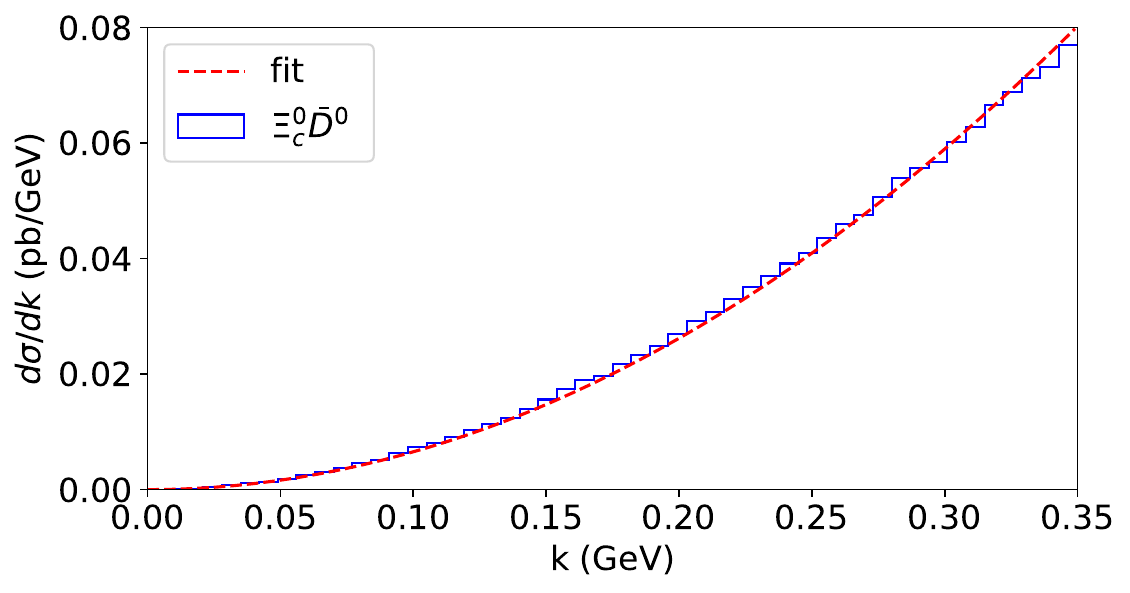}}
\subfigure[]{\includegraphics[width=0.45\textwidth]{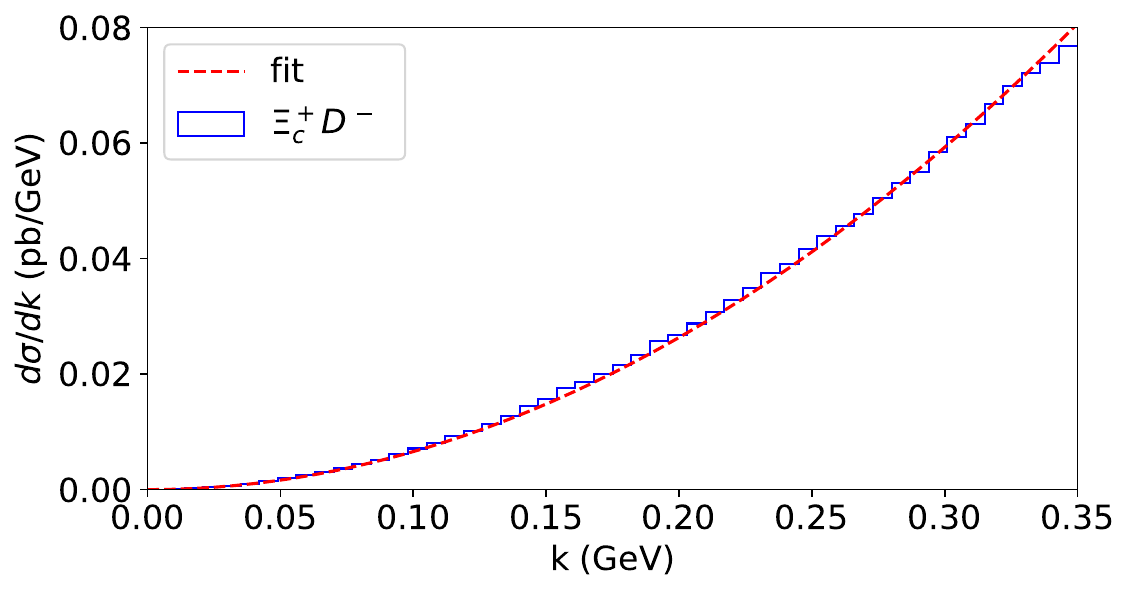}}
\subfigure[]{\includegraphics[width=0.45\textwidth]{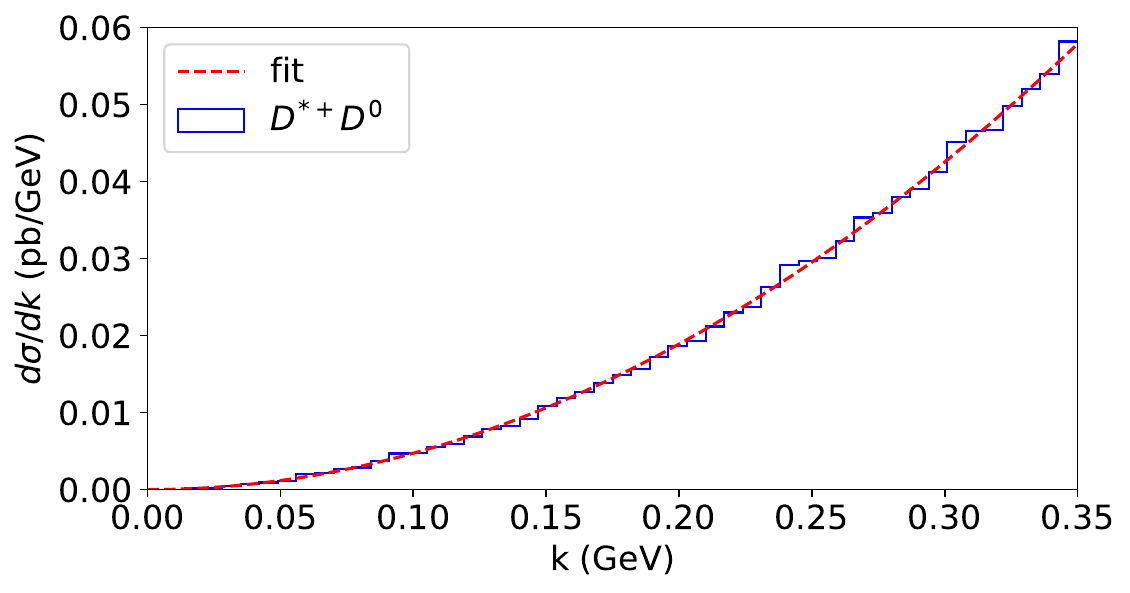}}
\subfigure[]{\includegraphics[width=0.45\textwidth]{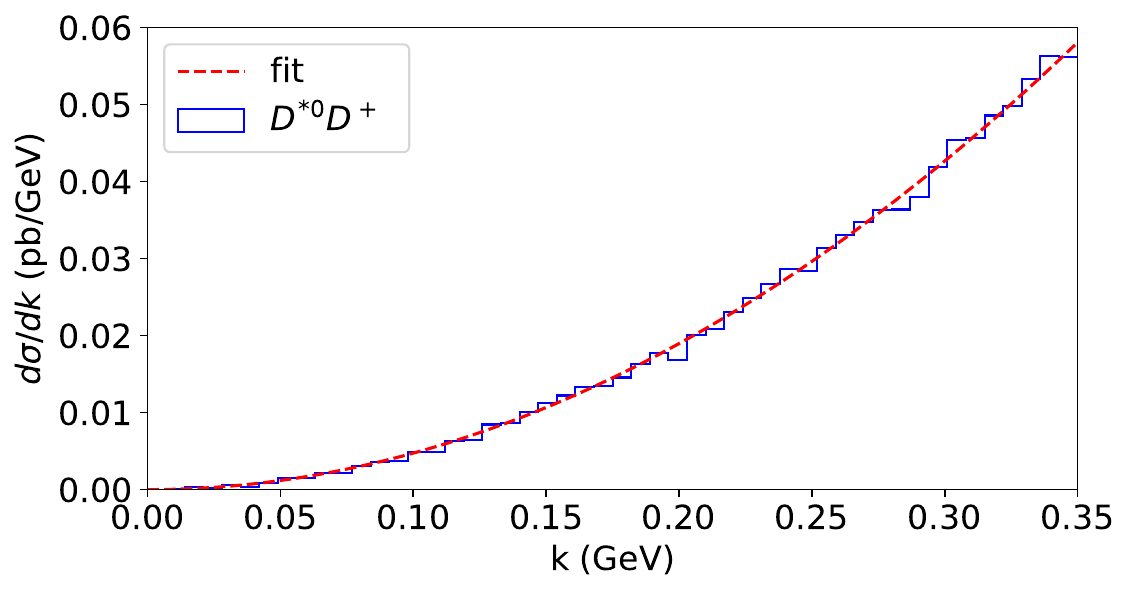}}
\subfigure[]{\includegraphics[width=0.45\textwidth]{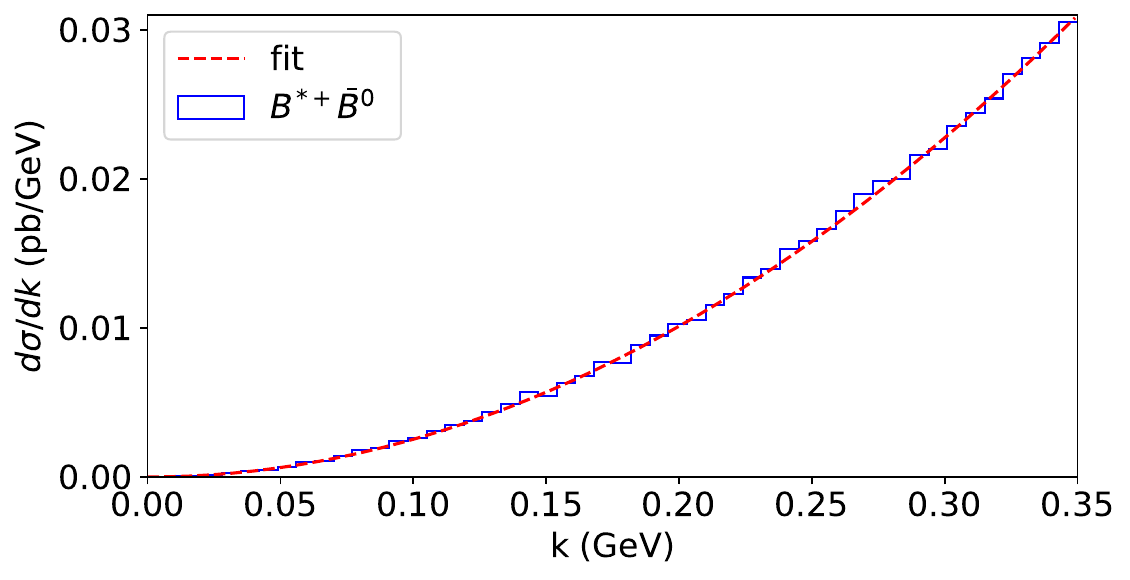}}
\subfigure[]{\includegraphics[width=0.45\textwidth]{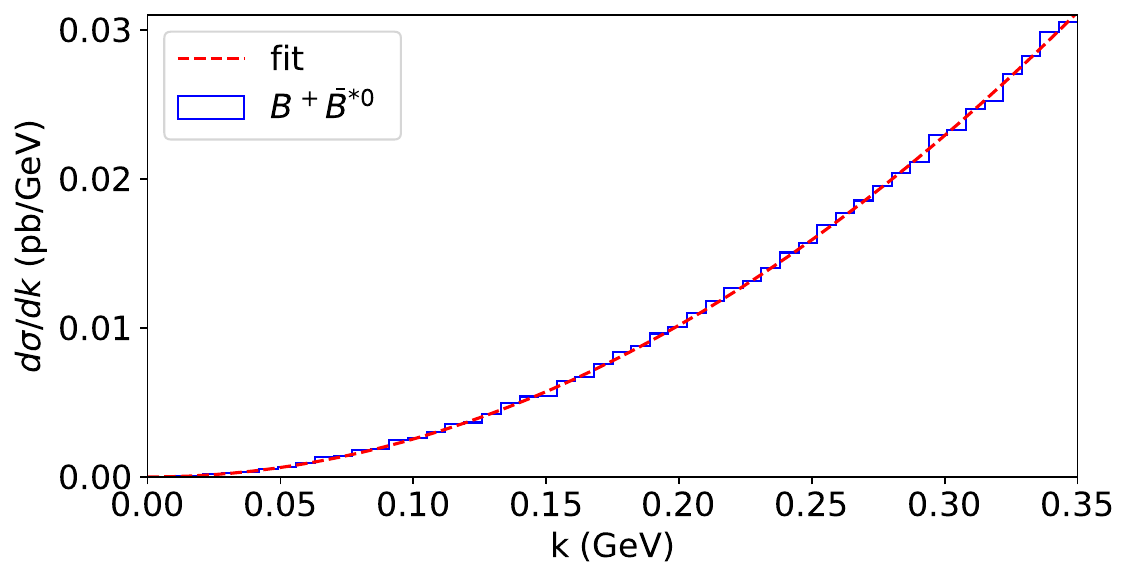}}
\subfigure[]{\includegraphics[width=0.46\textwidth]{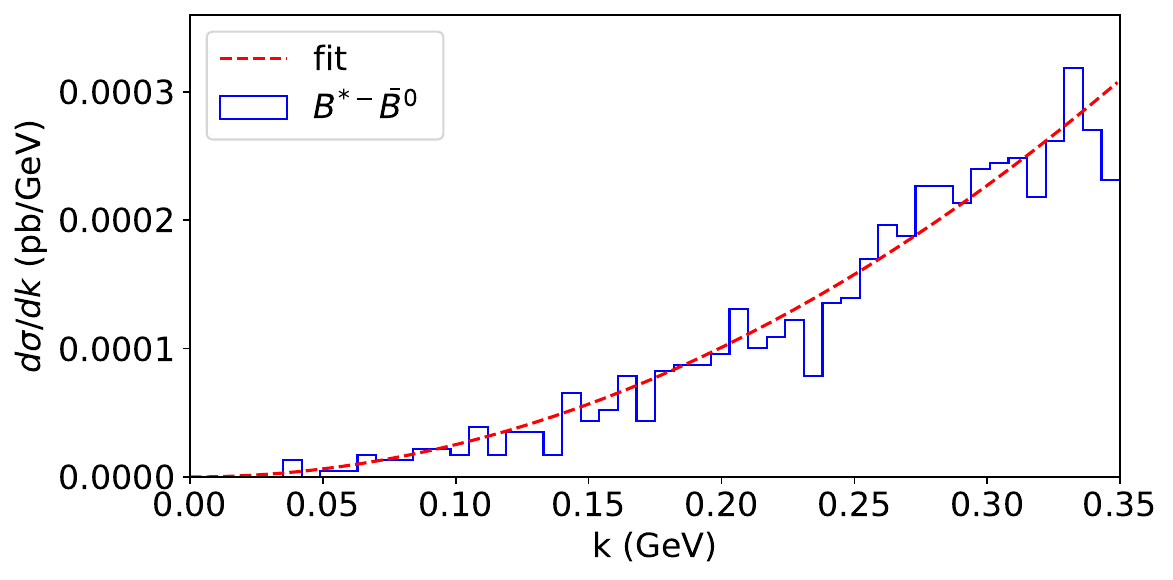}}
\subfigure[]{\includegraphics[width=0.46\textwidth]{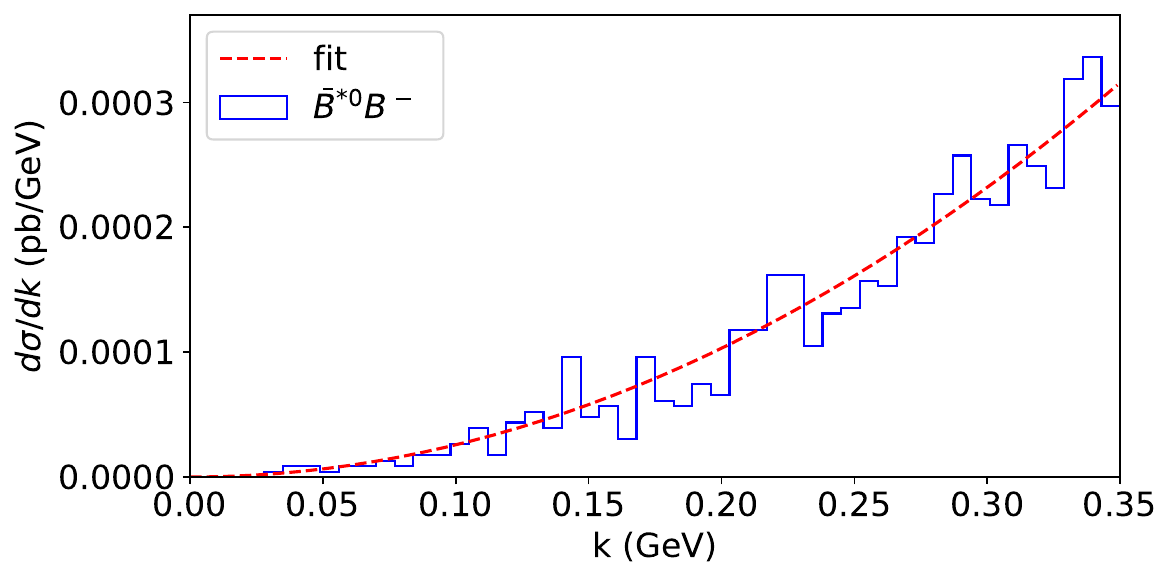}}
 \caption{Differential cross sections d$\sigma$/d$k$ (in units of $\rm{pb/GeV}$) for the process $e^+e^- \to Z^0 \to HH^{\prime}$. The range of the relative momentum between the two produced hadrons is chosen as $\vert k \vert < 350~\rm{MeV} $. The histograms denote the differential cross sections simulated by Pythia8, and the red dashed curves are obtained from the fits using d$\sigma$/d$k$ $\propto k^2$. The subfigures demonstrate the differential production cross sections for (a) $\Xi_c^0\bar{D}^0$; (b) $\Xi_c^+D^-$; (c) $D^{\ast +}D^0$;
                            (d) $D^{\ast 0}D^+$; (e) $B^{\ast +}\bar{B}^0$; (f) $B^+\bar{B}^{\ast 0}$; (g) $B^{\ast-}\bar{B}^{0}$; and (h) $\bar{B}^{\ast 0}B^-$.}
\label{Fig:diffxsec_HHprime}
\end{figure}

\begin{table}[htbp]
\caption{\label{tab_hidden-charm X cross sections} Order-of-magnitude estimations of the inclusive production cross sections (in units of $\rm{pb}$) for the hidden-charm hadronic molecules at the $Z$ factory. The values of the binding energies at the outside (inside) of the square brackets and the cross sections at the outside (inside) of the parentheses 
correspond to the cutoff $\Lambda=0.5~\rm{GeV}~(1.0~\rm{GeV})$.}
\renewcommand{\arraystretch}{1.2}
\begin{tabular*}{\columnwidth}{@{\extracolsep\fill}cccccc}
\hline\hline                                  
        & Constituents
        & $I(J^{PC})$ 
        & Binding energy (MeV)
        & $\sigma[X]~(\rm{pb})$
        \\[3pt]        
\hline
  $X(3872)$ &$D\bar{D}^*$ &$0(1^{++})$ &$4.15$~\cite{ParticleDataGroup:2022pth}&$0.3(1.6)$ \\[3pt]
  ${Z_c(3900)^0}$ &$D\bar{D}^*$ &$1(1^{+-})$ &$-11.30$~\cite{ParticleDataGroup:2022pth} &$4.0(8.1)$ \\[3pt]
  ${Z_c(3900)^+}$ &$D\bar{D}^*$ &$1(1^+)$ &$-12.00$~\cite{ParticleDataGroup:2022pth} &$3.8(7.4)$ \\[3pt]
  $P_{cs}$ &$\Xi_c\bar{D}$ &$0(\frac{1}{2}^-)$ &$2.14[7.53]$~\cite{Dong:2021juy} &$4.4 \times 10^{-3}(0.03)$ \\[3pt]
  $P_{cs}(4459)$ &$\Xi_c\bar{D}^*$ &$0(\frac{1}{2}^-)$ &$18.83$~\cite{ParticleDataGroup:2022pth} &$1.7 \times 10^{-3}(0.01)$ \\[3pt]
  $P_{cs}(4459)$ &$\Xi_c\bar{D}^*$ &$0(\frac{3}{2}^-)$ &$18.83$~\cite{ParticleDataGroup:2022pth} &$3.3 \times 10^{-3}(0.02)$ \\[3pt]
  $P_c(4312)^+$ &$\Sigma_c\bar{D}$ &$\frac{1}{2}(\frac{1}{2}^-)$ &$6.71$~\cite{ParticleDataGroup:2022pth} &$2.1 \times 10^{-3}(0.01)$ \\[3pt]
  $P_c(4380)^+$ &$\Sigma_c^*\bar{D}$ &$\frac{1}{2}(\frac{3}{2}^-)$ &$7.35$~\cite{ParticleDataGroup:2022pth} &$4.1 \times 10^{-3}(0.02)$ \\[3pt]
  $P_c(4440)^+$ &$\Sigma_c\bar{D}^*$ &$\frac{1}{2}(\frac{3}{2}^-)$ &$21.01$~\cite{ParticleDataGroup:2022pth} &$1.8 \times 10^{-3}(0.01)$ \\[3pt]
  $P_c(4457)^+$ &$\Sigma_c\bar{D}^*$ &$\frac{1}{2}(\frac{1}{2}^-)$ &$3.01$~\cite{ParticleDataGroup:2022pth} &$1.4 \times 10^{-3}(6.4 \times 10^{-3})$ \\[3pt]
  &$\Lambda_c\bar{\Lambda}_c$ &$0(0^{-+})$ &$1.98[33.8]$~\cite{Dong:2021juy} &$4.9 \times 10^{-3}(0.06)$ \\[3pt]
  &$\Sigma_c\bar{\Sigma}_c$ &$0(0^{-+})$ &$11.1[60.8]$~\cite{Dong:2021juy} &$7.2 \times 10^{-5}(5.7 \times 10^{-4})$ \\[3pt]
  &$\Sigma_c\bar{\Sigma}_c$ &$1(0^{-+})$ &$8.28[53.3]$~\cite{Dong:2021juy} &$1.1 \times 10^{-4}(8.7 \times 10^{-4})$ \\[3pt]
  &$\Xi_c\bar{\Xi}_c$ &$0(0^{-+})$ &$4.72[42.2]$~\cite{Dong:2021juy} &$9.3 \times 10^{-4}(7.5 \times 10^{-3})$ \\[3pt]
  &$\Xi_c\bar{\Xi}_c$ &$1(0^{-+})$ &$18.2[0.39]$~\cite{Dong:2021juy} &$9.5 \times 10^{-4}(1.4 \times 10^{-3})$ \\[3pt]
  &$\Lambda_c\bar{\Sigma}_c$ &$1(0^{-})$ &$2.19[33.9]$~\cite{Dong:2021juy} &$1.5 \times 10^{-4}(1.6 \times 10^{-3})$ \\[3pt]
  &$\Lambda_c\bar{\Xi}_c$ &$\frac{1}{2}(0^{-})$ &$1.29[8.42]$~\cite{Dong:2021juy} &$4.7 \times 10^{-4}(2.9 \times 10^{-3})$ \\[3pt]
\hline\hline
\end{tabular*}
\end{table}

The results in Tables~\ref{tab_hidden-charm X cross sections}, \ref{tab_double-charm X cross sections}, and \ref{tab_hidden-bottom X cross sections} reveal that:
\begin{itemize}
    \item The production cross sections of the hidden-charm hadronic molecules $X(3872)$ and $Z_c(3900)$ at the $Z$ factory are at pb level,
    and the production cross section of the $Z_c(3900)$ state is about $5$ -- $10$ times larger than that of the $X(3872)$, which is comparable with the prediction in the semi-inclusive leptoproduction process~\cite{Yang:2021jof}. Considering the integrated luminosity $\int L = 100~\mathrm{ab}^{-1}$ as listed in Table~\ref{tab_CEPC_parameters}, there will be 
    approximately $3 \times 10^7$ -- $1.6 \times 10^8$ and $4 \times 10^8$ -- $8 \times 10^8$ events of the $X(3872)$ and $Z_c(3900)$ produced in the two-year operation of the $Z$ factory, respectively. 
    
    \item The production cross sections of the hidden-charm pentaquark candidates $P_c$ and $P_{cs}$ states at the $Z$ factory are at the same level (about a few to tens of fb), two to three orders of magnitude smaller than those of the $X(3872)$ and $Z_c(3900)$. Such cross sections give $10^5$ -- $10^6$ production events at the $Z$ factory during the two-year operation.
    
\begin{table}[htbp]
\caption{\label{tab_double-charm X cross sections} Order-of-magnitude estimations of the inclusive production cross sections (in units of $\rm{pb}$) for the double-charm hadronic molecules at the $Z$ factory. The values of the binding energies at the outside (inside) of the square brackets and the cross sections at the outside (inside) of the parentheses 
correspond to the cutoff $\Lambda=0.5~\rm{GeV}~(1.0~\rm{GeV})$.}
\renewcommand{\arraystretch}{1.2}
\begin{tabular*}{\columnwidth}{@{\extracolsep\fill}cccccc}
\hline\hline                                  
        & Constituents
        & $I(J^{PC})$ 
        & Binding energy (MeV)
        & $\sigma[X]~(\rm{pb})$
        \\[3pt]     
\hline
  $T_{cc}^{+}$ &$DD^*$ &$0(1^{+})$ &$0.273$\cite{LHCb:2021auc,LHCb:2021vvq} &$2.3 \times 10^{-3}(9.7 \times 10^{-3})$ \\[3pt]
  $T_{cc}^{*+}$ &$D^*D^*$ &$0(1^{+})$ &$0.503$\cite{Du:2021zzh} &$1.3 \times 10^{-3}(5.5 \times 10^{-3})$ \\[3pt]
  &$\Lambda_cD$ &$\frac{1}{2}(\frac{1}{2}^-)$ &$3.44[5.62]$\cite{Dong:2021bvy} &$2.4 \times 10^{-4}(1.8 \times 10^{-3})$ \\[3pt]
  &$\Lambda_cD^*$ &$\frac{1}{2}(\frac{1}{2}^-)$ &$2.53[6.73]$\cite{Dong:2021bvy} &$5.6 \times 10^{-5}(5.3 \times 10^{-4})$ \\[3pt]
  &$\Lambda_cD^*$ &$\frac{1}{2}(\frac{3}{2}^-)$ &$2.53[6.73]$\cite{Dong:2021bvy} &$1.1 \times 10^{-4}(1.1 \times 10^{-3})$ \\[3pt]
\hline\hline
\end{tabular*}
\end{table}

     \item The production cross sections of the double-charm tetraquark candidates $T_{cc}^+$ and its heavy-quark-spin symmetry (HQSS) partner $T_{cc}^{\ast +}$~\cite{Du:2021zzh} are comparable with the cross sections of the $P_c$ and $P_{cs}$ states, two to three orders of magnitude smaller than those of the hidden-charm tetraquarks. Such a large gap between the production cross sections of the $T_{cc}$ and the hidden-charm tetraquarks can be attributed to the parton level where the production of double-charm molecules requires two pairs of $c\bar{c}$ produced from the $Z$ boson decay. The branching ratio of $Z\to c\bar{c}c\bar{c}$ is much smaller than that of $Z\to c\bar{c}$, which is the underlying process for the production of the hidden-charm hadronic molecules.
    There will be about $2.3 \times 10^5$ -- $9.7 \times 10^5$ and $1.3 \times 10^5$ -- $5.5 \times 10^5$ events for the production of $T_{cc}$ and $T_{cc}^*$ at the $Z$ factory, respectively. The $T_{cc}$ events produced at the $Z$ factory is roughly three times larger than those in the proposed electron-ion colliders in US in the two-year operation~\cite{Shi:2022ipx}. Furthermore, assuming Br$[T_{cc}^+\to D^0\bar D^0 \pi^+]\simeq 59.6\%$ in terms of the leading-order estimation of the XEFT \cite{Dai:2023mxm} and Br$[D^0\to K^-\pi^+]=3.9\%$ \cite{ParticleDataGroup:2022pth}, 
    the number of $T_{cc}$ events reconstructed in the $D^0D^0\pi^+$ invariant mass distribution can reach ${\cal O}(10^3)$ at the $Z$ factory. 
    The event number for $T_{cc}^+ \to D^0(\to K^- \pi^+) \bar{D}^0(\to K^- \pi^+)\pi^+$ observed by the LHCb Collaboration is $117 \pm 16$ with an integrated luminosity of $9~\rm{fb}^{-1}$~\cite{LHCb:2021vvq}. Therefore, the $Z$ factory could be a much better platform to study the $T_{cc}$ in detail and to search for its spin partner $T_{cc}^*$. 
 \item The production cross sections of the hidden-charm baryon-antibaryon hadronic molecules predicted in Ref.~\cite{Dong:2021juy} and the double-charm meson-baryon hadronic molecules predicted in Ref.~\cite{Dong:2021bvy} are at the same order of magnitude, about $0.1$ -- $1.0$~fb, one order of magnitude smaller than those of the $P_{c}$, $P_{cs}$, and $T_{cc}^{(*)}$. An exception is the $\Lambda_c\bar{\Lambda}_c$ molecule, whose production cross section is one magnitude larger than other hidden-charm baryon-antibaryon hadronic molecules, and the resulting events for the $\Lambda_c \bar{\Lambda}_c$ molecule production at the $Z$ factory is about $4.9 \times 10^5$ -- $6 \times 10^6$. Therefore it is purposeful to search for the $\Lambda_c \bar{\Lambda}_c$ molecule at the $Z$ factory of CEPC.
 \begin{table}[htbp]
\caption{\label{tab_hidden-bottom X cross sections} Order-of-magnitude estimations of the inclusive production cross sections (in units of $\rm{pb}$) for the hidden-bottom hadronic molecules at the $Z$ factory. The values of the cross sections at the outside (inside) of the parentheses correspond to the cutoff $\Lambda=0.5~\rm{GeV}~(1.0~\rm{GeV})$.}
\renewcommand{\arraystretch}{1.2}
\begin{tabular*}{\columnwidth}{@{\extracolsep\fill}cccccc}
\hline\hline                                  
        & Constituents
        & $I(J^{PC})$ 
        & Binding energy (MeV)
        & $\sigma[X]~(\rm{pb})$
        \\[3pt]     
\hline
  $Z_{b}(10610)^{\pm}$ &$B\bar{B}^*$ &$1(1^{+})$ &$-2.99$~\cite{ParticleDataGroup:2022pth} &$6.5 \times 10^{-2}(8.7 \times 10^{-2})$ \\[3pt]
  $Z_{b}(10610)^0$ &$B\bar{B}^*$ &$1(1^{+-})$ &$-4.79$~\cite{ParticleDataGroup:2022pth} &$7.2 \times 10^{-2}(1.0 \times 10^{-1})$ \\[3pt]
  $Z_{b}(10650)^{\pm}$ &$B^*\bar{B}^*$ &$1(1^{+})$ &$-2.78$~\cite{ParticleDataGroup:2022pth} &$6.6 \times 10^{-2}(1.2 \times 10^{-1})$ \\[3pt]
  $Z_{b}(10650)^0$ &$B^*\bar{B}^*$ &$1(1^{+-})$ &$-2.78$~\cite{ParticleDataGroup:2022pth} &$7.8 \times 10^{-2}(1.4 \times 10^{-1})$ \\[3pt]
\hline\hline
\end{tabular*}
\end{table}
\item In the bottom sector,
the production cross sections of the hidden-bottom hadronic molecules $Z_b(10610)$ and $Z_b(10650)$ can reach tens to hundreds of fb, about one to two order(s) of magnitude smaller than those of the  $X(3872)$ and $Z_c(3900)$, and one order of magnitude larger than the production cross section of the double-charm states $T_{cc}$ and $T_{cc}^{*}$.  
The expected number of $Z_b$ events at CEPC over a two-year period is around $10^7$, indicating promising prospects for its discovery and detailed study. 
Considering that the branching ratios of the $Z_b(10610)$ decays to the $\Upsilon(1S)\pi$, $\Upsilon(2S)\pi$, and $\Upsilon(3S)\pi$ final 
states are $5.4_{-1.5}^{+1.9} \times 10^{-3}$, $3.6_{-0.8}^{+1.1}\%$, and $2.1_{-0.6}^{+0.8}\%$
, respectively,
and 
$Z_b(10650)$ decays to the $\Upsilon(1S)\pi$, $\Upsilon(2S)\pi$, and $\Upsilon(3S)\pi$ final states are $1.7_{-0.6}^{+0.8} \times 10^{-3}$, $1.4_{-0.4}^{+0.6}\%$, and $1.6_{-0.5}^{+0.7}\%$,
respectively~\cite{ParticleDataGroup:2022pth},
the event yields of $e^+e^-\to Z\to Z_b(10650)/Z_b(10610)+\mathrm{all} \to \Upsilon(nS)\pi+\mathrm{all},\;n=1,2,3$ are about $10^4$--$10^5$. Although the production cross sections of 
$e^+e^-\to Z\to Z_b(10650)/Z_b(10610)+\mathrm{all} \to \Upsilon(nS)\pi+\mathrm{all},\;n=1,2,3$ at the $Z$ factory is about two orders of magnitude smaller than the cross sections at Belle\footnote{The exclusive production of $Z_b$ can occur through the process $e^+e^-\to\Upsilon(5S)\to Z_b \pi$ via the intermediate states $B_1^{\prime}\bar B$ and $B_0^{*}\bar B^*$. The enhancement of this process, as discussed in Ref.~\cite{Wu:2018xaa}, can lead to a large exclusive cross section.}~\cite{Belle:2011aa,Belle:2014vzn}, the integrated luminosity at the $Z$ factory, roughly three orders of magnitude higher than Belle's ($121.4~\mathrm{fb}^{-1}$), results in approximately one order of magnitude larger $Z_b$ event yields compared to Belle.
\end{itemize}
Despite the absence of the experimental signal from the double-bottom tetraquark ($bb\bar{u}\bar{d}$) $T_{bb}$ state at present, the existence of the $T_{bb}$ with quantum numbers $I(J^P)=0(1^+)$ has been approved by the lattice QCD (LQCD) calculation~\cite{Aoki:2023nzp} in the HAL QCD method, where the $T_{bb}$ is predicted to be a deeply bound state with a binding energy $E_{T_{bb}}^{\mathrm{single}}=155\pm 17$~MeV only considering the $\bar{B}\bar{B}^{\ast}$ single channel and a binding energy $E_{T_{bb}}^{\mathrm{coupled}}=83\pm 10$~MeV considering the $\bar{B}\bar{B}^{\ast}-\bar{B}^{\ast}\bar{B}^{\ast}$ coupled channel treatment relative to the $\bar{B}\bar{B}^{\ast}$ threshold. In addition, the $\bar{b}\bar{b}ud$ and $\bar{b}\bar{b}us$ tetraquarks have also been predicted by the LQCD calculation in Ref.~\cite{Alexandrou:2024iwi}, with binding energies $100\pm 10 ^{+43}_{-36}$~MeV and $30\pm 3^{+31}_{-11}$~MeV
relative to the $B B^{\ast}$ and $B B_s^{\ast}$ thresholds, respectively. The large binding energies for both the double bottom and anti-bottom tetraquarks indicate the $T_{bb}$ could not be simply regarded as a pure $\bar{B}\bar{B}^{\ast}$ hadronic molecule. To give an order-of-magnitude estimate, we assume that the $T_{bb}$ and $T_{bb}^{*}$ have the same binding energy (83~MeV), and still calculate their production cross sections in the hadronic molecule picture. The results are about 
$\sigma[T_{bb}^{(\ast)}] \approx 10^{-3}(10^{-1})$~fb for $\Lambda=0.5(1.0)$~GeV, 2 -- 5 orders of magnitude smaller than those of the $Z_b$ states and the $T_{bb}$ state in the compact tetraquark configuration predicted in Refs.~\cite{Ali:2018ifm,Ali:2019tli}.

\section{Summary}\label{Sec_Summary}

In summary, we have investigated the inclusive differential production cross sections of the $e^+e^- \to HH^{\prime}+\text{all}$ processes using the Monte Carlo event generator Pythia8 and estimated the production cross sections of typical
hidden/double-charm, and hidden/double-bottom hadronic molecules at the $Z$ factory by considering the FSI between the hadron pairs $HH'$. The predicted production cross sections of the hidden-charm molecules $X(3872)$ and $Z_c(3900)$ are at the $\rm{pb}$ level, and the expected event yields of these molecules are about $10^7$--$10^8$.
The production cross sections of the hidden-charm pentaquark candidates $P_c$ and $P_{cs}$, and the double-charm tetraquark candidates $T_{cc}$ and $T_{cc}^{\ast}$ as $S$-wave hadronic molecules are at the same order of magnitude, about two to three orders of magnitude smaller than the cross sections of $X(3872)$ and $Z_c(3900)$. The production cross sections of some possible hidden-charm baryon-antibaryon and double-charm meson-baryon hadronic molecules predicted in Refs.~\cite{Dong:2021bvy,Dong:2021juy} are further smaller than those of the $P_c$, $P_{cs}$, and $T_{cc}^{(*)}$ by about one order of magnitude, except the $\Lambda_c\bar{\Lambda}_c$ molecule whose production cross section is one order of magnitude larger than those of other hidden-charm baryon-antibaryon molecules. 
As the $Z$ boson can decay to one and two pairs of $b\bar{b}$ in the parton level with sizeable branching ratios, the $Z$ factory is a ideal platform for the study of hidden/double bottom exotic states. The estimated production cross sections of the $Z_b$ states can reach tens to hundreds of pb, giving $10^4$--$10^5$ event yields of $e^+e^-\to Z\to Z_b(10650)/Z_b(10610)+\mathrm{all} \to \Upsilon(nS)\pi+\mathrm{all},\;n=1,2,3$ with the two-year integrated luminosity $\int L = 100~\mathrm{ab}^{-1}$, which is about one order of magnitude larger than the event yields in the Belle experiment. 
The production cross sections of the double-bottom tetraquark candidate $T_{bb}$ as a deeply bound $S$-wave $\bar{B}\bar{B}^{\ast}$ molecule is also estimated using the binding energy from LQCD calculation as an input, and the result is about $\sigma[T_{bb}^{(\ast)}] \approx 10^{-3}(10^{-1})$~fb for $\Lambda=0.5(1.0)$~GeV, 2--5 orders of magnitude smaller than those of the $Z_b$ states and the $T_{bb}$ state in the compact tetraquark configuration predicted in Refs.~\cite{Ali:2018ifm,Ali:2019tli}. Our order-of-magnitude estimates indicate appreciable production event yields of these hidden/double-charm and hidden/double-bottom hadronic molecules at the $Z$ factory.

\section{Acknowledgments}\label{sec: ACKNOWLEDGMENTS}
We are grateful to Feng-Kun Guo for suggestions, Shu-Ming Wu for useful discussions, and Shi-Dong Liu for a careful reading of this manuscript. The numerical calculations were done at the HPC Cluster of ITP-CAS. This work is supported in part by the Chinese Academy of Sciences under Grant No. XDB34030000; by the National Natural Science Foundation of China (NSFC) under Grants No. 12125507, No. 11835015, No. 12047503, and No. 12075133; and by the NSFC and the Deutsche Forschungsgemeinschaft
(DFG) through the funds provided to the TRR110 “Symmetries and the Emergence of Structure in QCD” (NSFC Grant No. 12070131001, DFG Project-ID No. 196253076). This work is also supported by the Natural Science Foundation of Shandong province under the Grant No. ZR2022ZD26, Taishan Scholar Project of Shandong Province under Grant No. tsqn202103062 and the Higher Educational Youth Innovation Science and Technology
Program Shandong Province  under Grant No. 2020KJJ004.
P.-P.S. also acknowledges the Generalitat valenciana (GVA) for the project with ref. CIDEGENT/2019/015.

\appendix
\section{Differential cross sections of the hadron pairs}\label{appen:differendial_pair_productions}
In this section, we show all the differential cross sections for the constituent hadron pairs of the hadronic molecules considered in the main text. In the charm sector, Fig.~\ref{Fig:diffxsec_X3872Zc3900} shows the 
differential cross sections of the constituents of $X(3872)$ and $Z_c(3900)$.
Fig.~\ref{Fig:diffxsec_Pcs} shows the 
differential cross sections of the constituents of $P_{cs}$ states.
Fig.~\ref{Fig:diffxsec_Pc} shows the 
differential cross sections of the constituents of $P_{c}(4312)$, $P_{c}(4380)$, $P_{c}(4440)$, and $P_{c}(4457)$.
Fig.~\ref{Fig:diffxsec_baryonbaryon} shows the 
differential cross sections of the $\Lambda_c \bar{\Lambda}_c$, $\Sigma_c \bar{\Sigma}_c$, $\Xi_c \bar{\Xi}_c$, $\Lambda_c \bar{\Sigma}_c$, and $\Lambda_c \bar{\Xi}_c$ pairs as constituents of the hidden-charm baryon-antibaryon hadronic molecules predicted in Ref.~\cite{Dong:2021juy}.
Fig.~\ref{Fig:diffxsec_TccTccstarLD} shows the 
differential cross sections of the constituents of $T_{cc}$, $T_{cc}^*$, and the $\Lambda_c D$, $\Lambda_c D^*$ pairs as constituents of double-charm meson-baryon hadronic molecules predicted in Ref.~\cite{Dong:2021bvy}.

In the bottom sector, Fig.~\ref{Fig:diffxsec_Zbpm} shows the 
differential cross sections of the constituents of $Z_b(10610)^\pm$ and $Z_b(10650)^\pm$.
Fig.~\ref{Fig:diffxsec_Zb0} shows the 
differential cross sections of the constituents of $Z_b(10610)^0$ and $Z_b(10650)^0$.
Fig.~\ref{Fig:diffxsec_BB} shows the 
differential cross sections of the constituents of $T_{bb}$ and $T_{bb}^*$.
\begin{figure}[htbp]
   \subfigure[]{\includegraphics[width=0.45\textwidth]{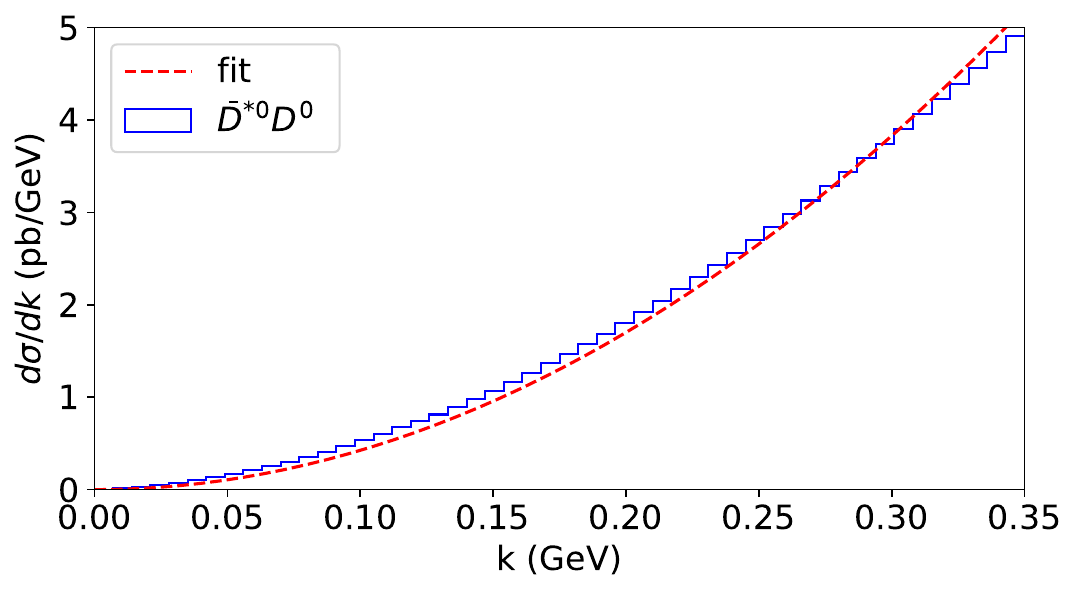}}
   \subfigure[]{\includegraphics[width=0.45\textwidth]{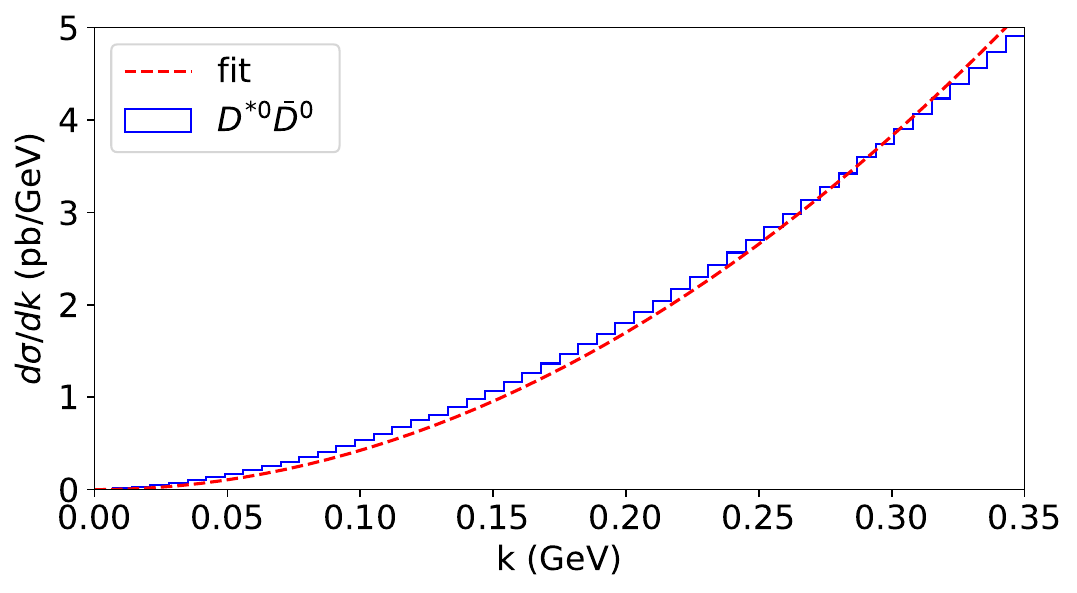}}
   \subfigure[]{\includegraphics[width=0.45\textwidth]{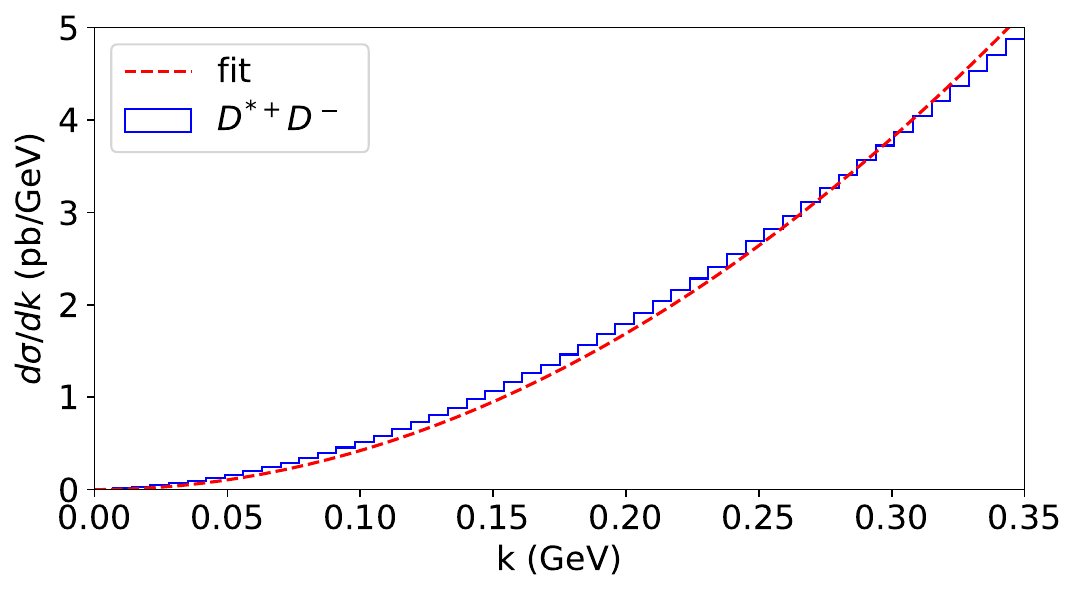}}
   \subfigure[]{\includegraphics[width=0.45\textwidth]{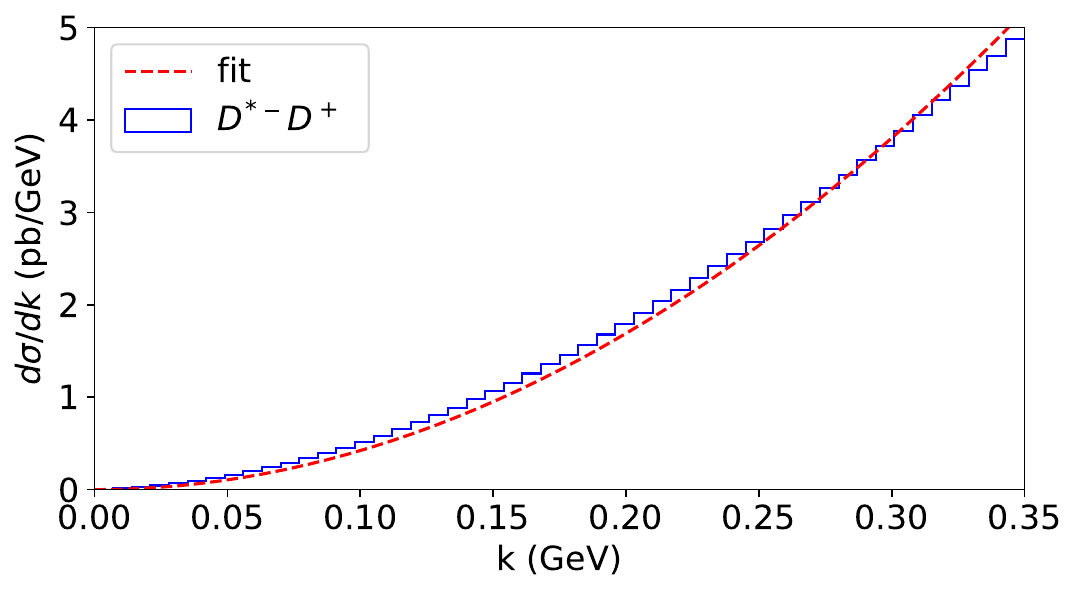}}
   \subfigure[]{\includegraphics[width=0.45\textwidth]{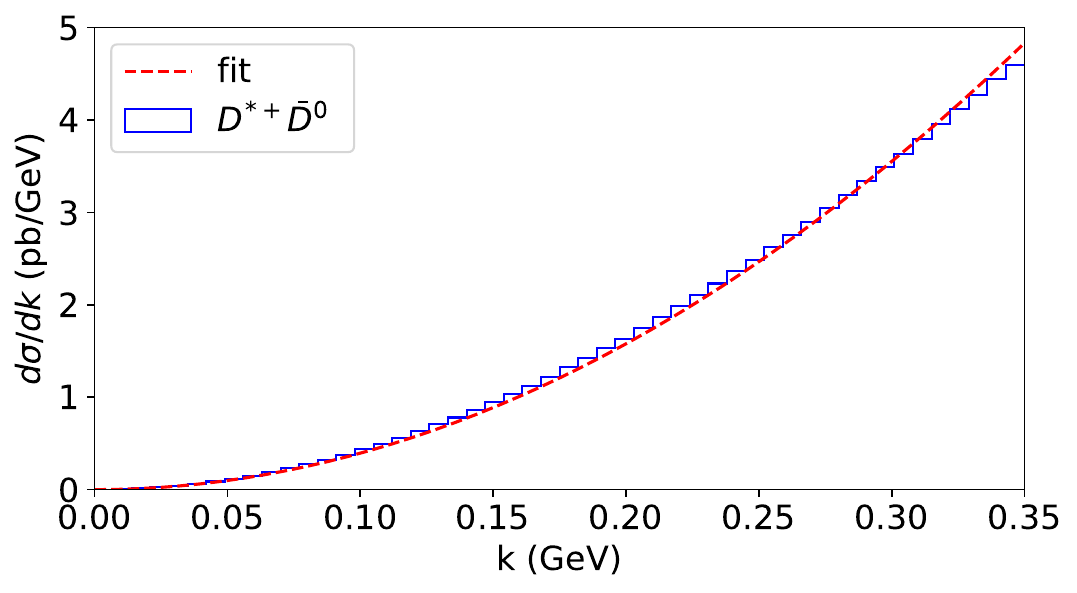}}
   \subfigure[]{\includegraphics[width=0.45\textwidth]{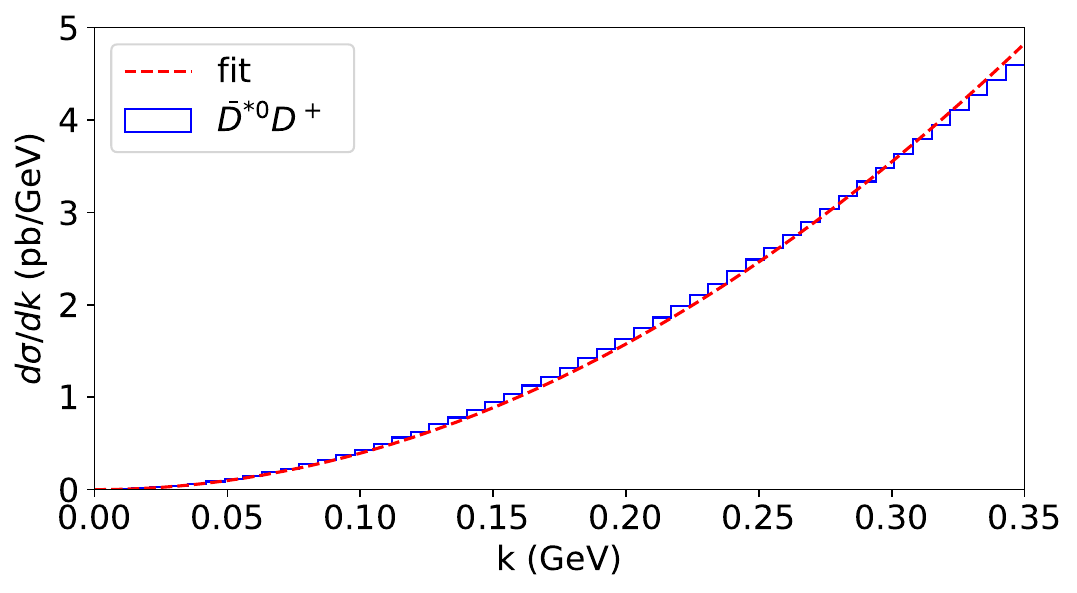}}
\caption{Differential cross sections d$\sigma$/d$k$ (in units of $\rm{pb/GeV}$) for the process $e^+e^- \to Z^0 \to D^{(*)}\bar{D}^{(*)}$. The subfigures demonstrate the differential production cross sections for (a) $\bar{D}^{\ast 0}D^0$; (b) $D^{\ast 0}\bar{D}^0$; (c) $D^{\ast +}D^-$;
                            (d) $D^{\ast -}D^+$; (e) $D^{\ast +}\bar{D}^{0}$; and (f) $\bar{D}^{\ast 0}D^+$.}
   \label{Fig:diffxsec_X3872Zc3900}
\end{figure}

\begin{figure}[htbp]
   \subfigure[]{\includegraphics[width=0.45\textwidth]{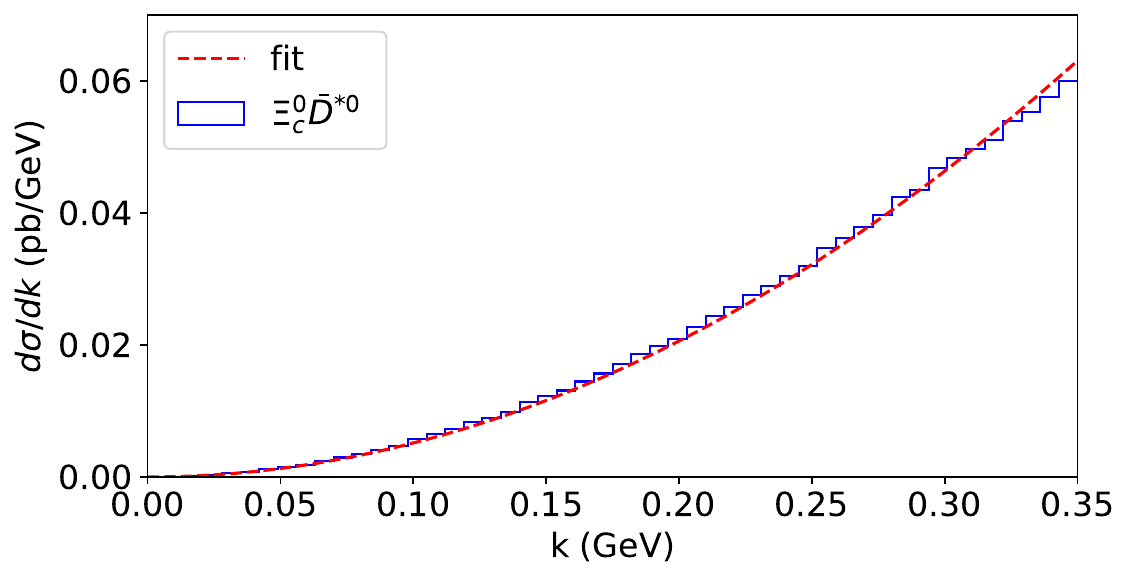}}
   \subfigure[]{\includegraphics[width=0.45\textwidth]{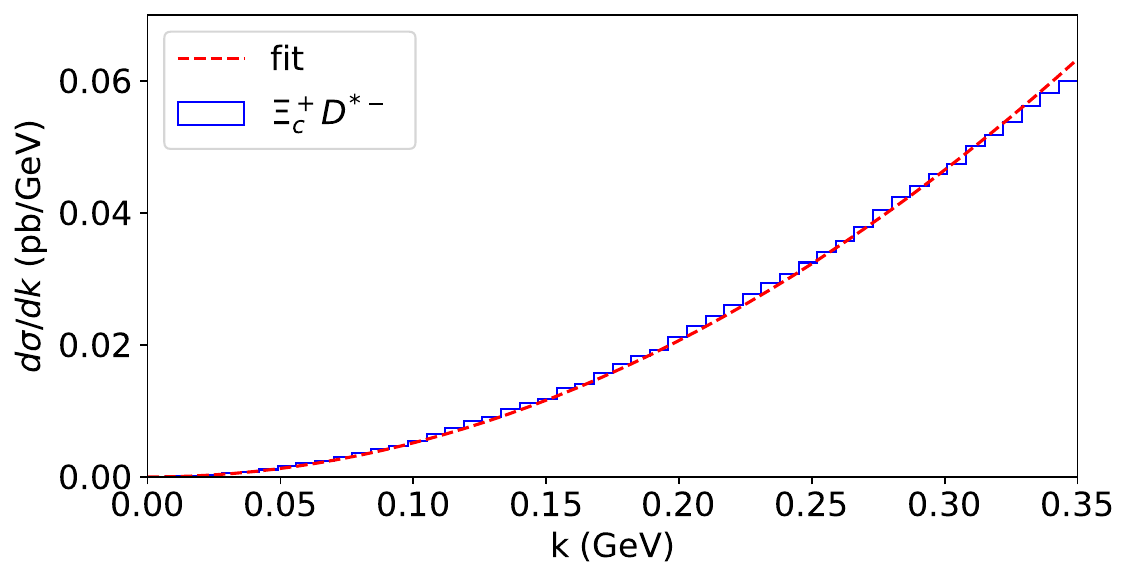}}
   \subfigure[]{\includegraphics[width=0.45\textwidth]{Fig_Xic0D0bar.pdf}}
   \subfigure[]{\includegraphics[width=0.45\textwidth]{Fig_XicpDm.pdf}}
  \caption{Differential cross sections d$\sigma$/d$k$ (in units of $\rm{pb/GeV}$) for the process $e^+e^- \to Z^0 \to \Xi_c \bar{D}^{(*)}$. The subfigures demonstrate the differential production cross sections for (a) $\Xi_c^0\bar{D}^{\ast 0}$; (b) $\Xi_c^+D^{\ast -}$; (c) $\Xi_c^0\bar{D}^0$;
                          and (d) $\Xi_c^+D^-$.}
   \label{Fig:diffxsec_Pcs}
\end{figure}

\begin{figure}
   \subfigure[]{\includegraphics[width=0.45\textwidth]{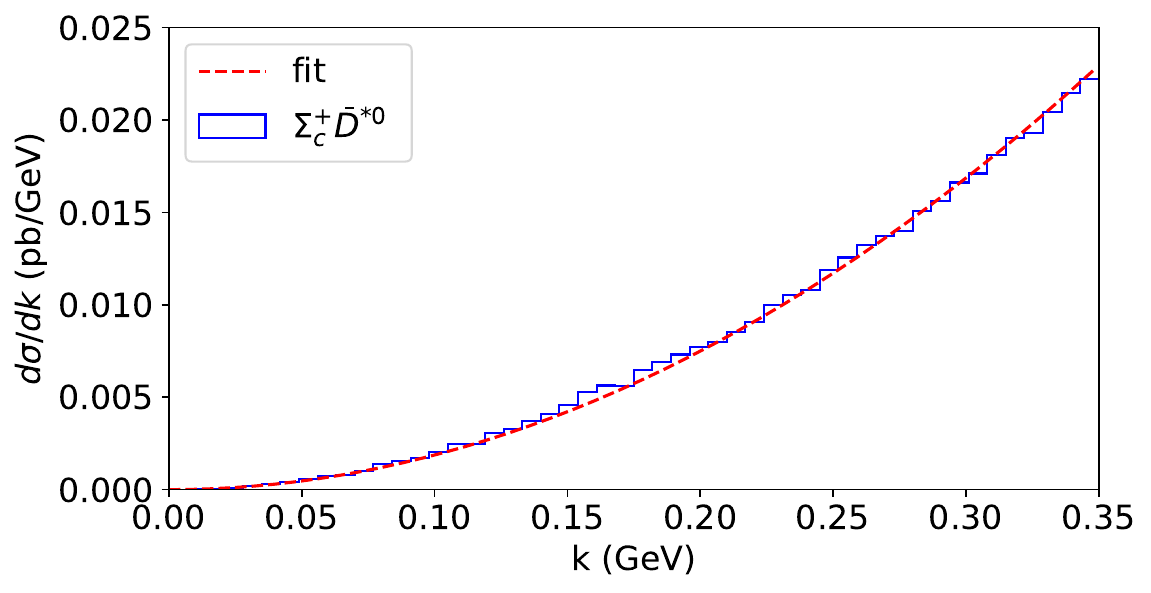}}
   \subfigure[]{\includegraphics[width=0.45\textwidth]{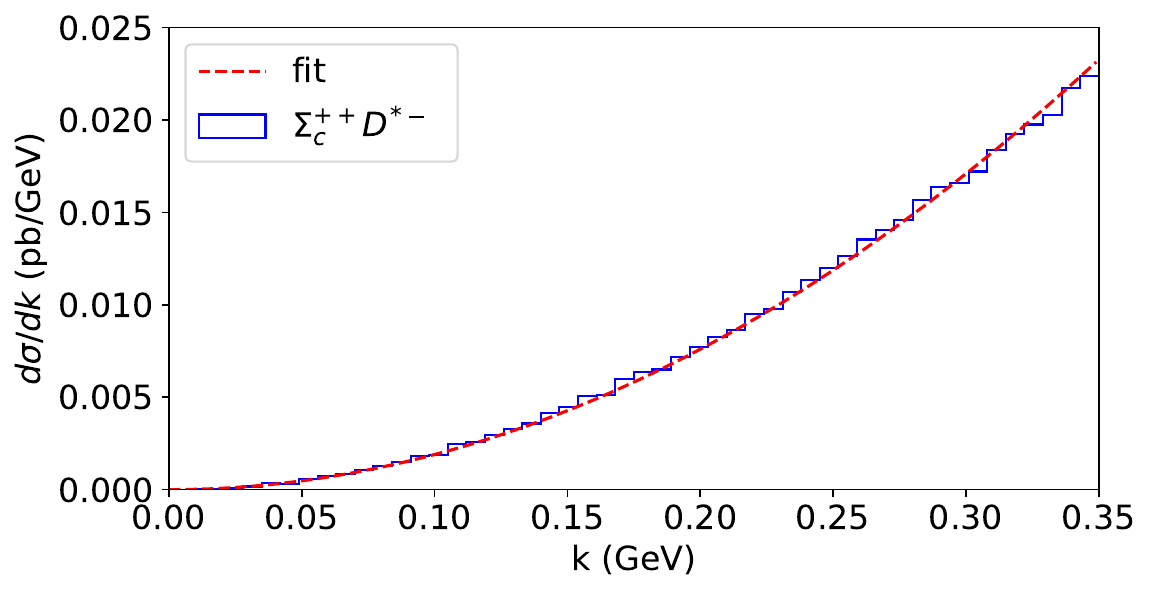}}
   \subfigure[]{\includegraphics[width=0.45\textwidth]{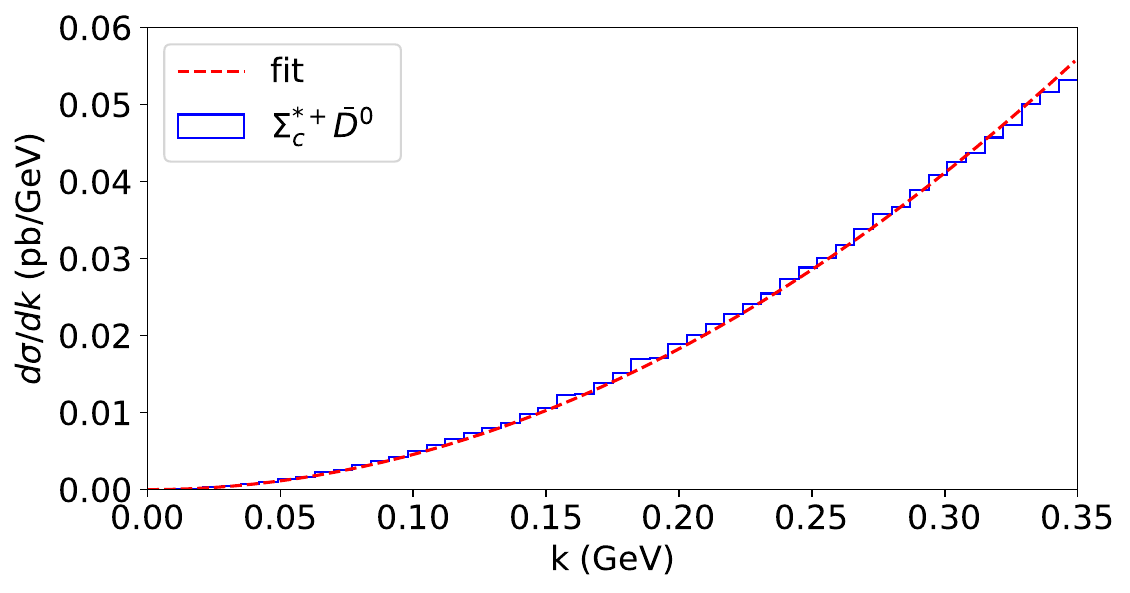}}
   \subfigure[]{\includegraphics[width=0.45\textwidth]{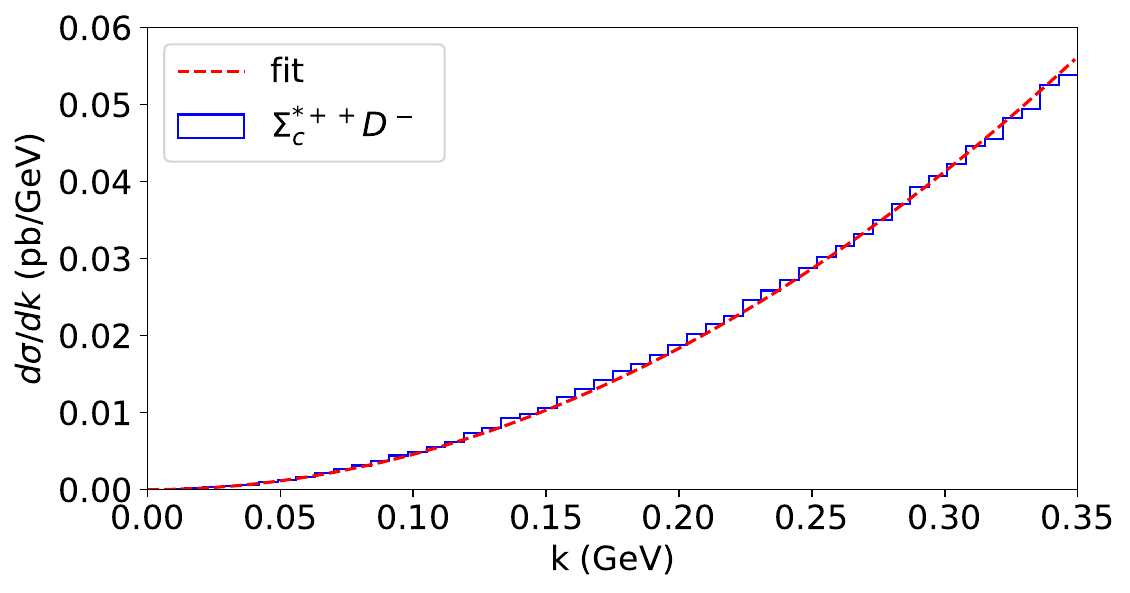}}
   \subfigure[]{\includegraphics[width=0.45\textwidth]{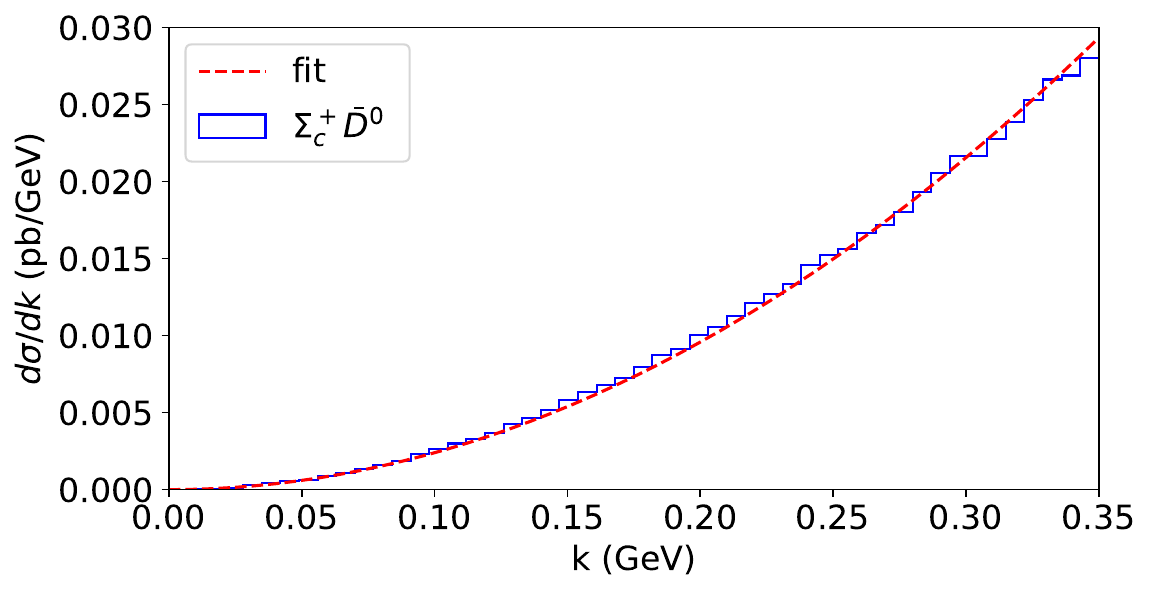}}
   \subfigure[]{\includegraphics[width=0.45\textwidth]{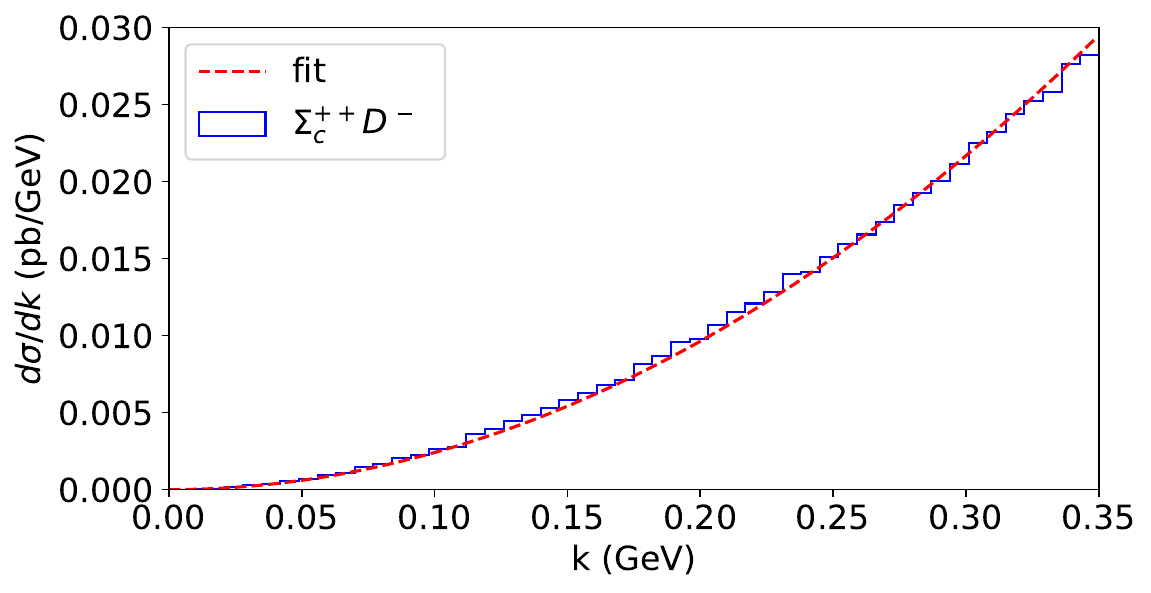}}
  \caption{Differential cross sections d$\sigma$/d$k$ (in units of $\rm{pb/GeV}$) for the process $e^+e^- \to Z^0 \to \Sigma_c^{(*)} \bar{D}^{(*)}$. The subfigures demonstrate the differential production cross sections for (a) $\Sigma_c^+\bar{D}^{\ast 0}$; (b) $\Sigma_c^{++}D^{\ast -}$; (c) $\Sigma_c^{\ast +} \bar{D}^0$; 
                            (d) $\Sigma_{c}^{\ast ++} D^-$; (e) $\Sigma_c^+\bar{D}^{0}$; and (f) $\Sigma_{c}^{++} D^-$. }
   \label{Fig:diffxsec_Pc}
\end{figure}

\begin{figure} 
 \subfigure[]{\includegraphics[width=0.45\textwidth]{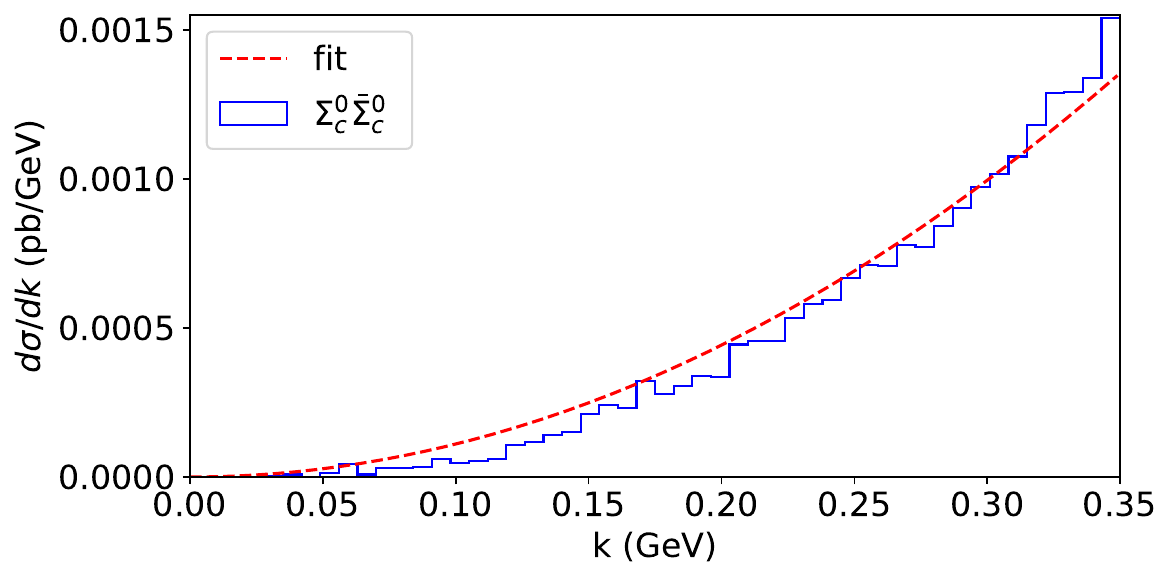}}
 \subfigure[]{\includegraphics[width=0.45\textwidth]{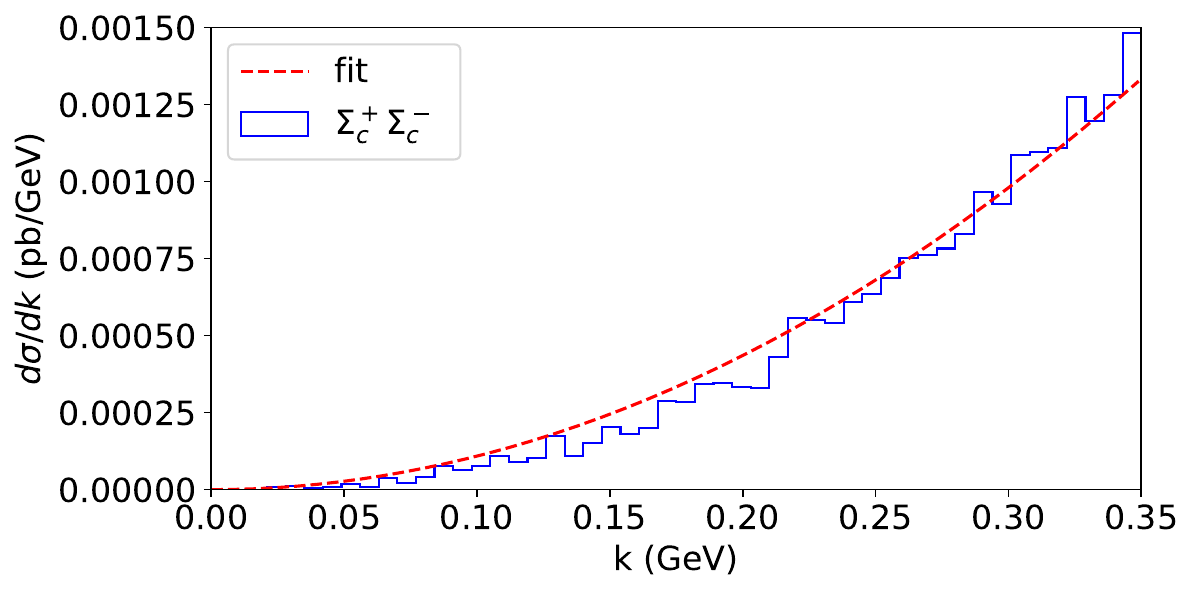}}
 \subfigure[]{\includegraphics[width=0.45\textwidth]{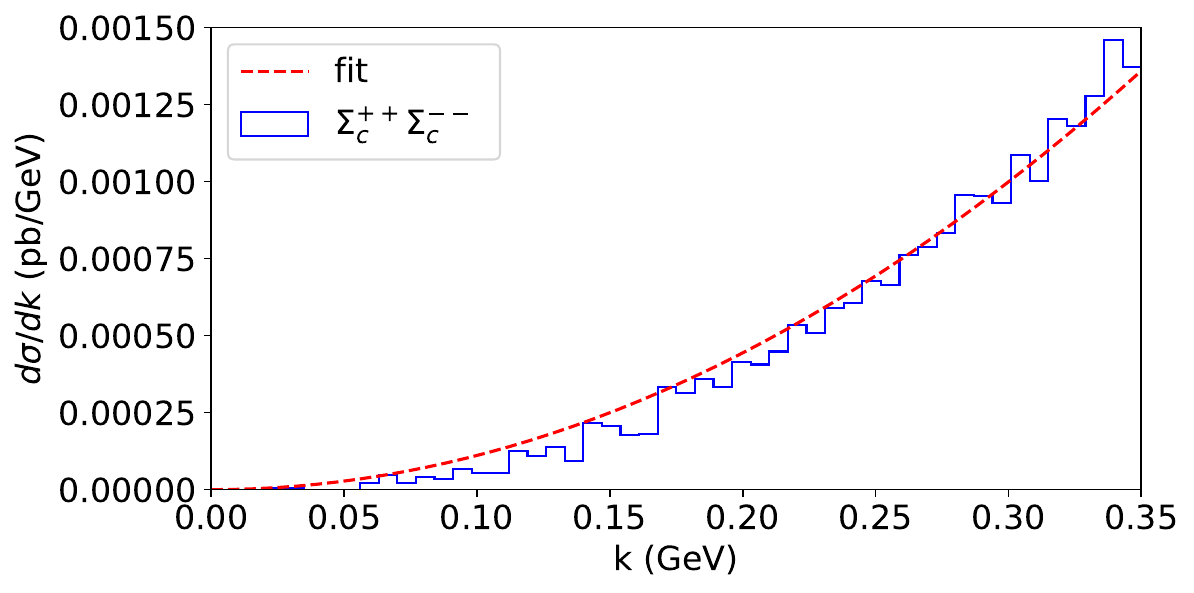}}
 \subfigure[]{\includegraphics[width=0.45\textwidth]{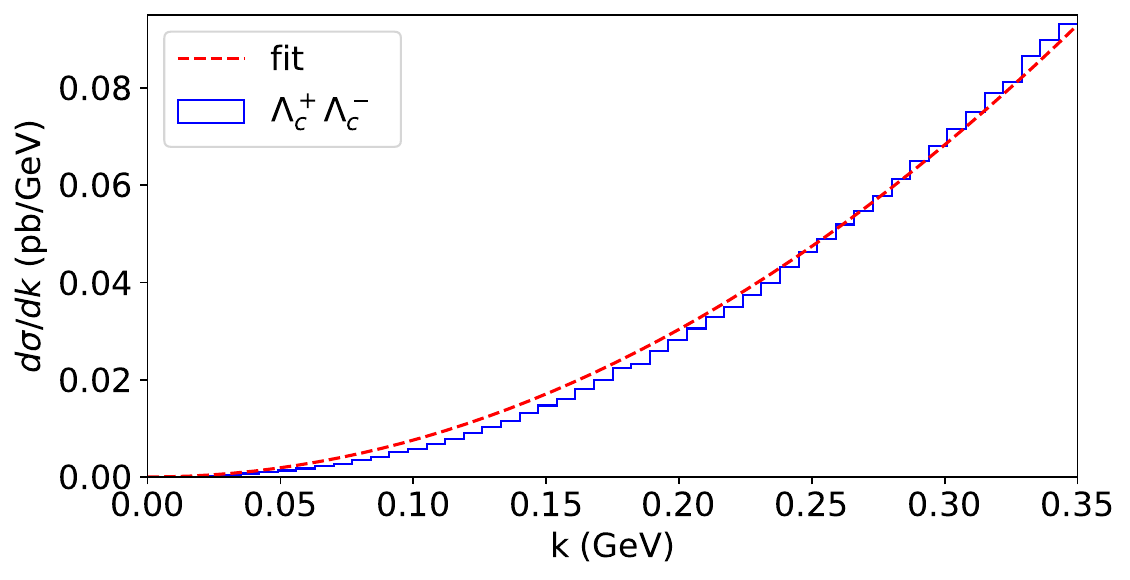}}
 \subfigure[]{\includegraphics[width=0.45\textwidth]{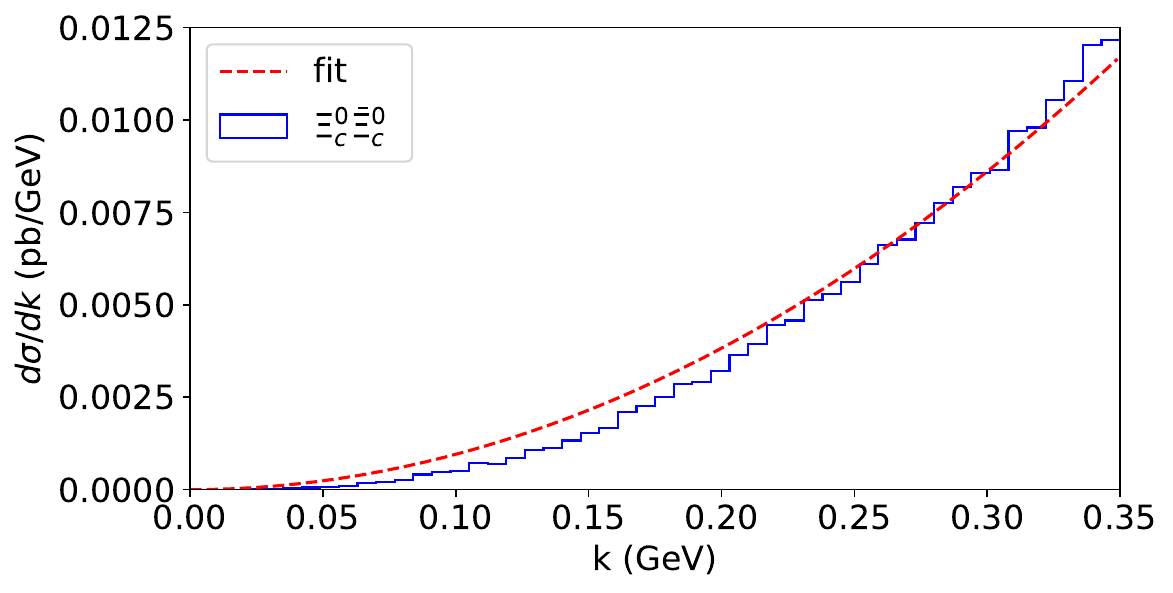}}
 \subfigure[]{\includegraphics[width=0.45\textwidth]{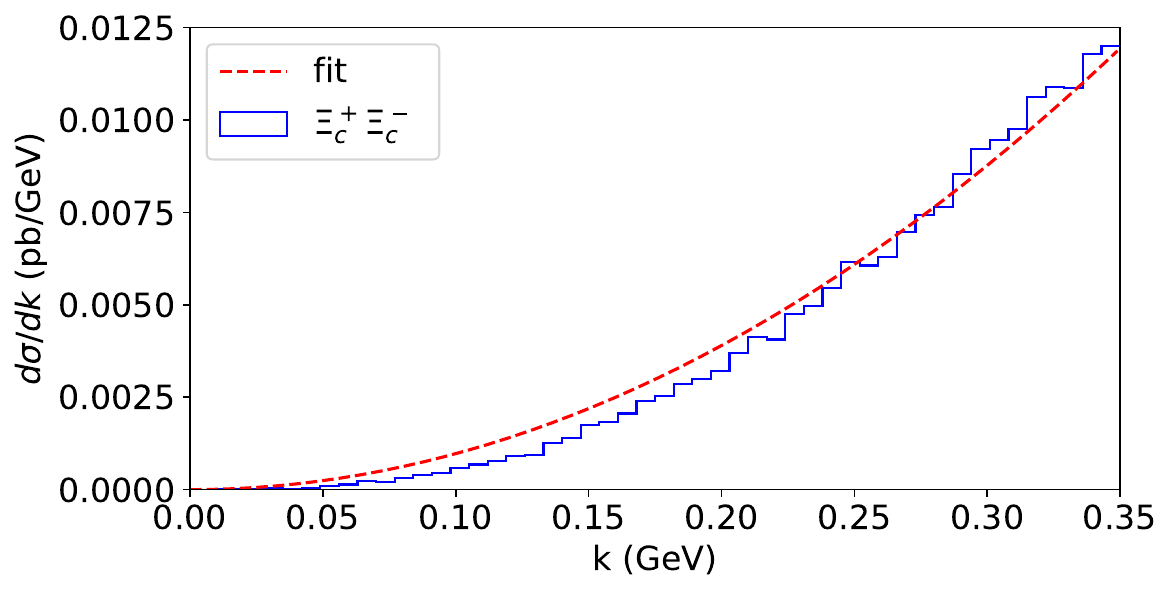}}
 \subfigure[]{\includegraphics[width=0.45\textwidth]{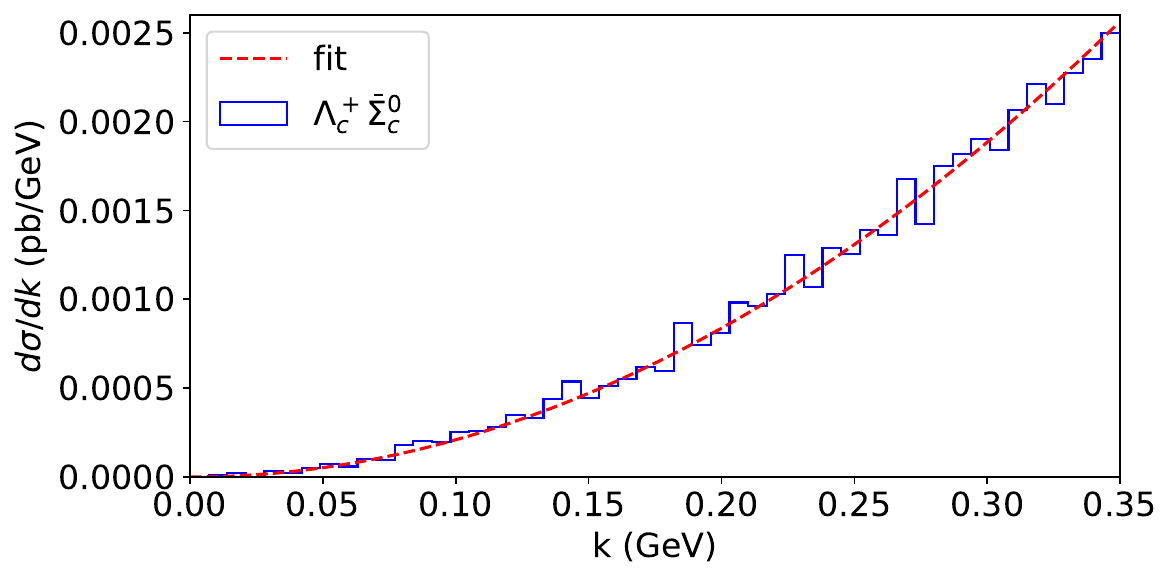}}
 \subfigure[]{\includegraphics[width=0.45\textwidth]{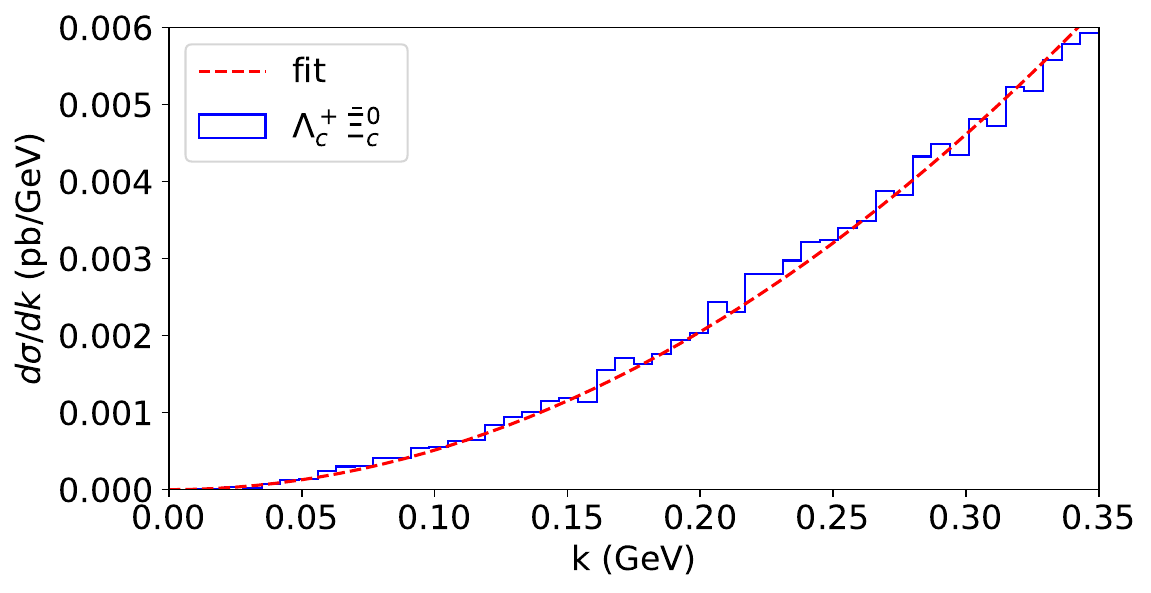}}
   \caption{Differential cross sections d$\sigma$/d$k$ (in units of $\rm{pb/GeV}$) for the charmed baryon-antibaryon pairs as constituents of hidden-charm hadronic molecules predicted in Ref.~\cite{Dong:2021juy}.  The subfigures demonstrate the differential production cross sections for (a) $\Sigma_{c}^0\bar{\Sigma}_{c}^{0}$; (b) $\Sigma_{c}^+\Sigma_{c}^-$; (c) $\Sigma_{c}^{++}\Sigma_{c}^{--}$;
                            (d) $\Lambda_{c}^+\Lambda_{c}^-$; (e) $\Xi_{c}^0\bar{\Xi}_c^0$; (f) $\Xi_c^+\Xi_c^-$; (g) $\Lambda_c^+\bar{\Sigma}_c^0$; and (h) $\Lambda_{c}^+\bar{\Xi}_{c}^0$.}
   \label{Fig:diffxsec_baryonbaryon}
\end{figure}

\begin{figure}[htbp]
 \subfigure[]{\includegraphics[width=0.45\textwidth]{Fig_DstarpD0.pdf}}
 \subfigure[]{\includegraphics[width=0.45\textwidth]{Fig_Dstar0Dp.pdf}}
 \subfigure[]{\includegraphics[width=0.45\textwidth]{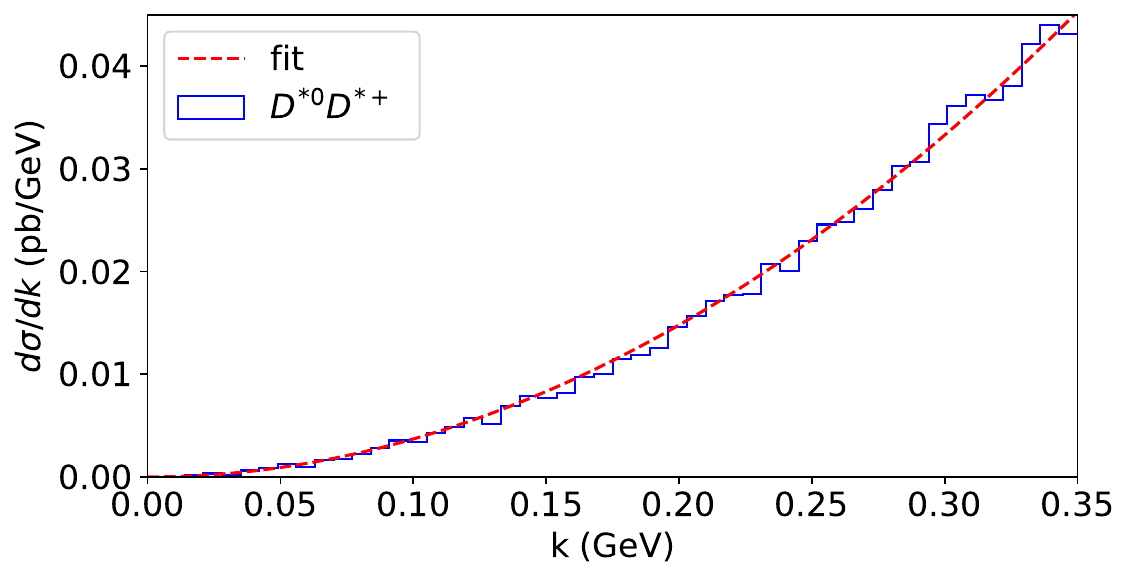}}
 \subfigure[]{\includegraphics[width=0.45\textwidth]{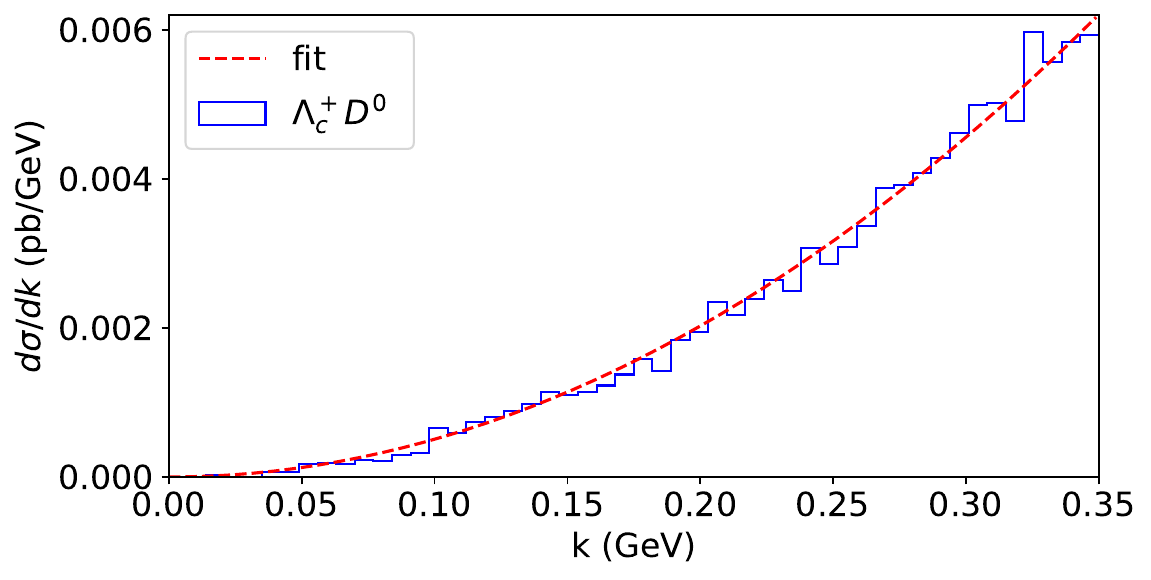}}
 \subfigure[]{\includegraphics[width=0.45\textwidth]{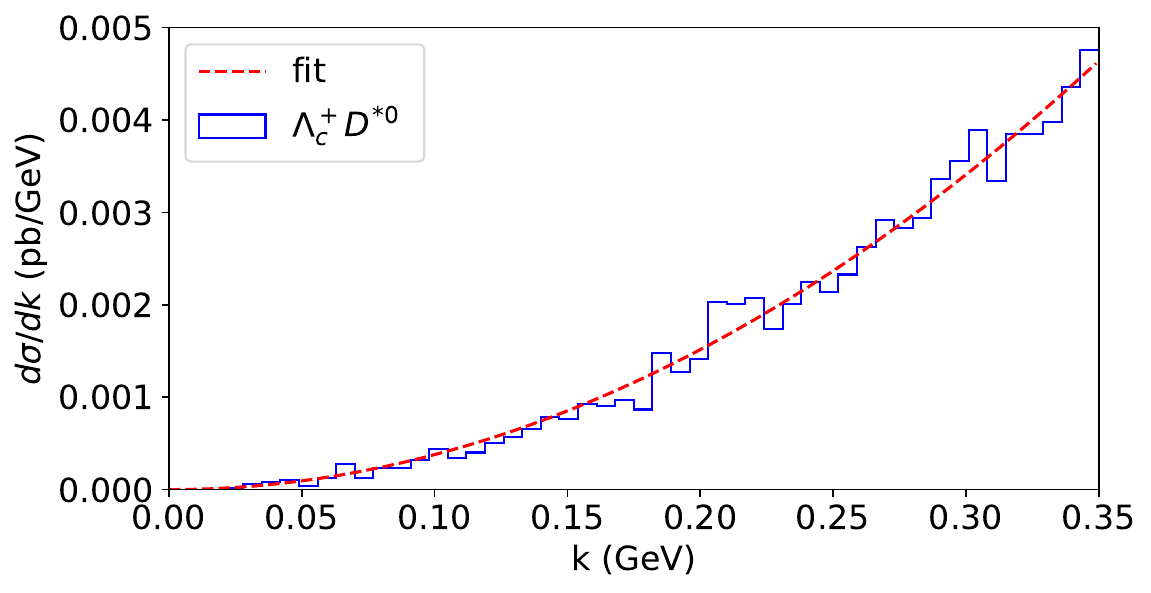}}
   \caption{Differential cross sections d$\sigma$/d$k$ (in units of $\rm{pb/GeV}$) for the processes $e^+e^- \to Z^0 \to D^{(*)0}D^{(*)+}$ and $e^+e^- \to Z^0 \to \Lambda_c^+D^{(*)0}$. The subfigures demonstrate the differential production cross sections for (a) $D^{\ast +} D^0$; (b) $D^{\ast 0} D^+$; (c) $D^{\ast 0}D^{\ast +}$; (d) $\Lambda_c^+D^0$; and (e) $\Lambda_c^+D^{\ast 0}$.}
   \label{Fig:diffxsec_TccTccstarLD}
\end{figure}

\begin{figure}[htbp]
 \subfigure[]{\includegraphics[width=0.45\textwidth]{Fig_BstarpB0bar.pdf}}
 \subfigure[]{\includegraphics[width=0.45\textwidth]{Fig_BpBstar0bar.pdf}}
 \subfigure[]{\includegraphics[width=0.45\textwidth]{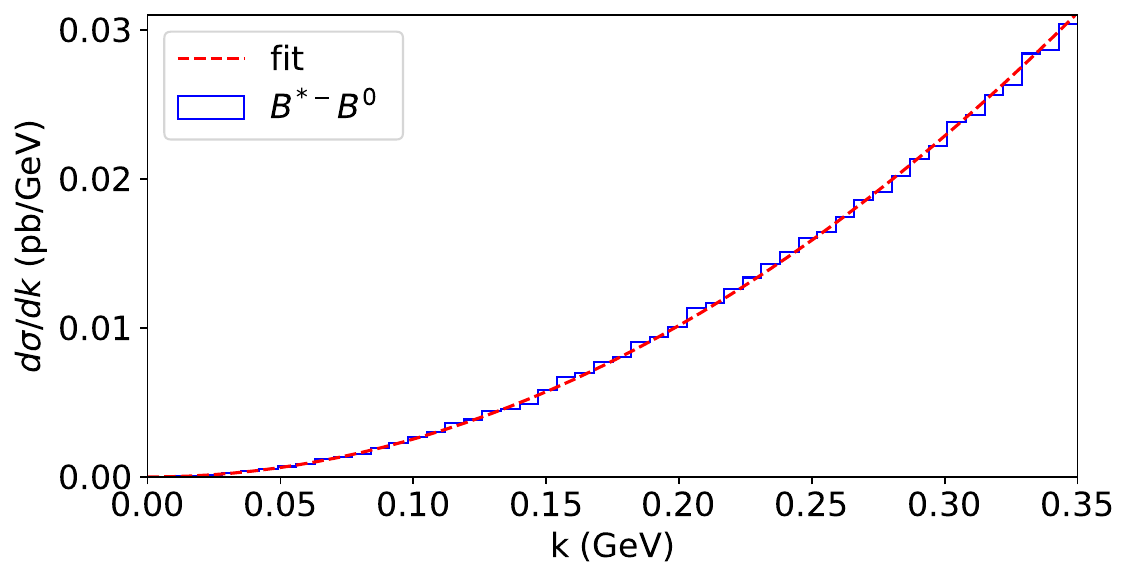}}
 \subfigure[]{\includegraphics[width=0.45\textwidth]{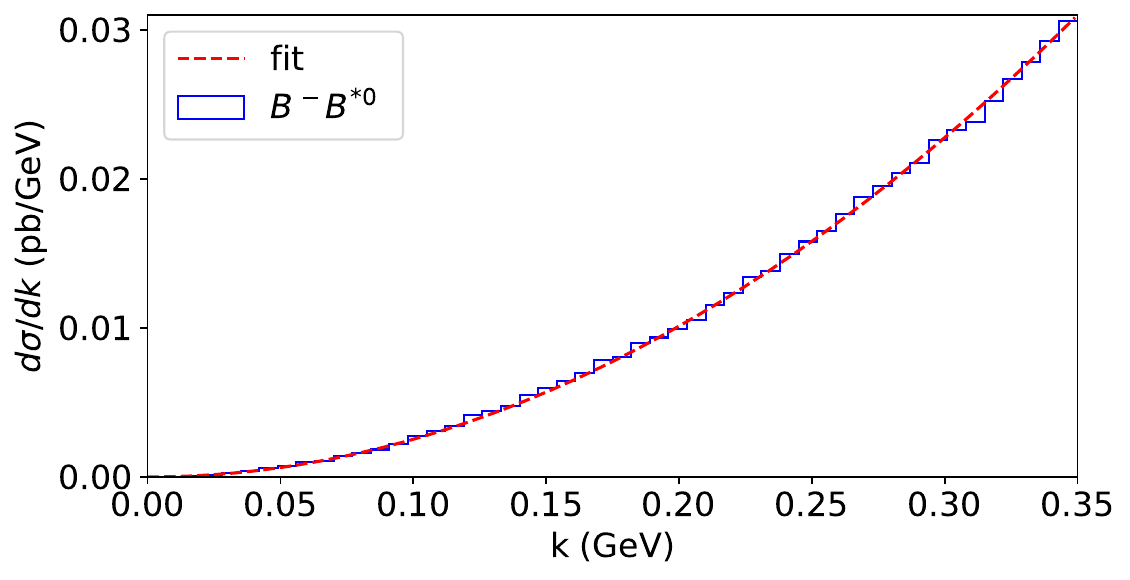}}
 \subfigure[]{\includegraphics[width=0.45\textwidth]{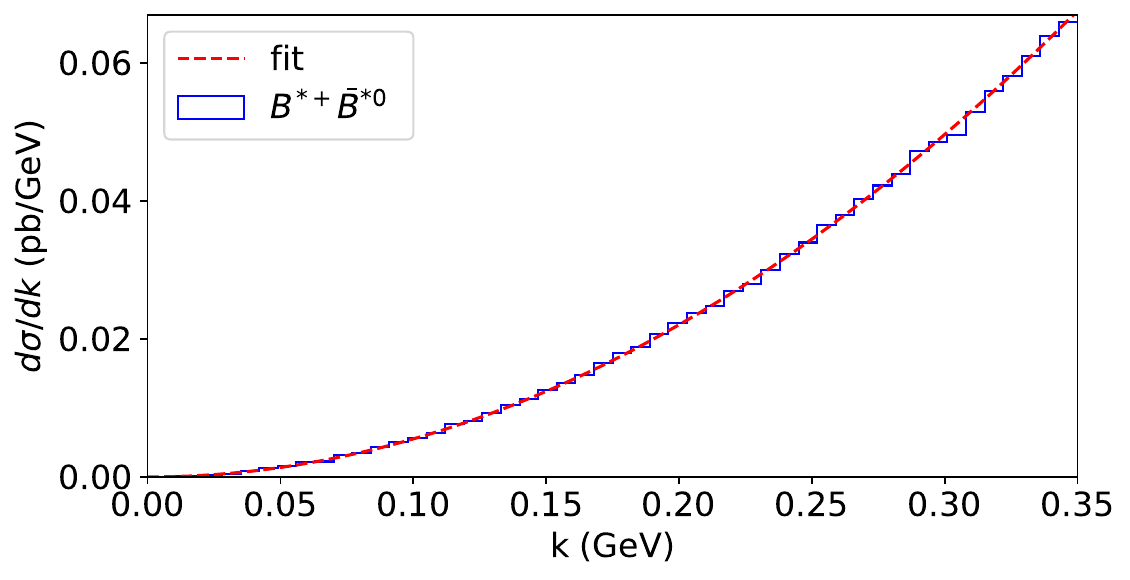}}
 \subfigure[]{\includegraphics[width=0.45\textwidth]{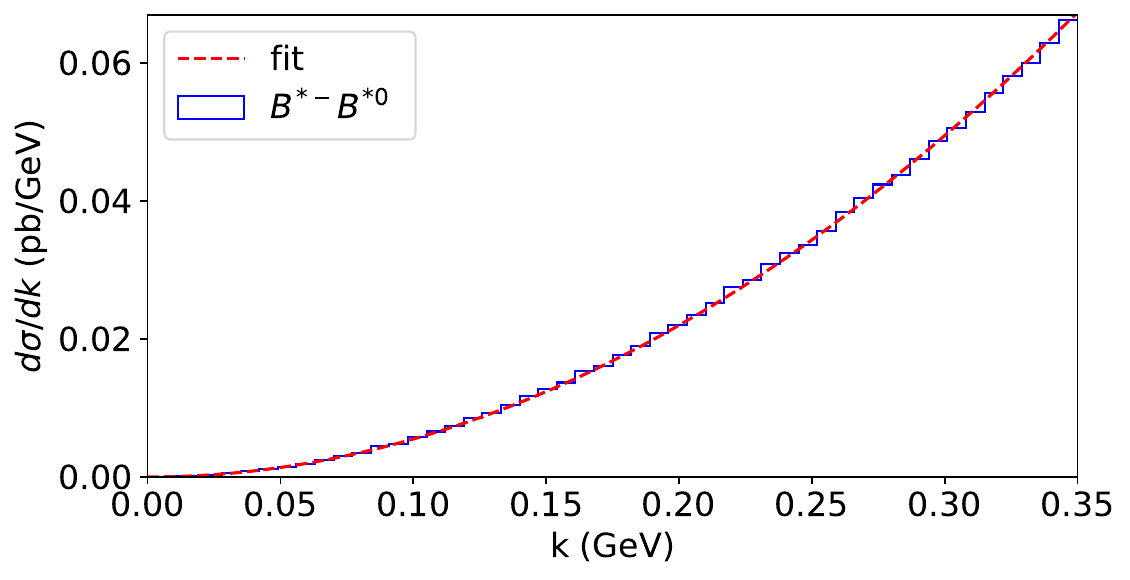}}
   \caption{Differential cross sections d$\sigma$/d$k$ (in units of $\rm{pb/GeV}$) for the processes $e^+e^- \to Z^0 \to B^{(*)+}\bar{B}^{(*)0}$ and $e^+e^- \to Z^0 \to B^{(*)-}B^{(*)0}$. The subfigures demonstrate the differential production cross sections for (a) $B^{\ast +}\bar{B}^0$; (b) $B^+\bar{B}^{\ast 0}$; (c) $B^{\ast -} B^0$; (d) $B^-B^{\ast 0}$; 
                            (e) $B^{\ast +}\bar{B}^{\ast 0}$; and (f) $B^{\ast -}B^{\ast 0}$.}
   \label{Fig:diffxsec_Zbpm}
\end{figure}

\begin{figure}[htbp]
 \subfigure[]{\includegraphics[width=0.45\textwidth]{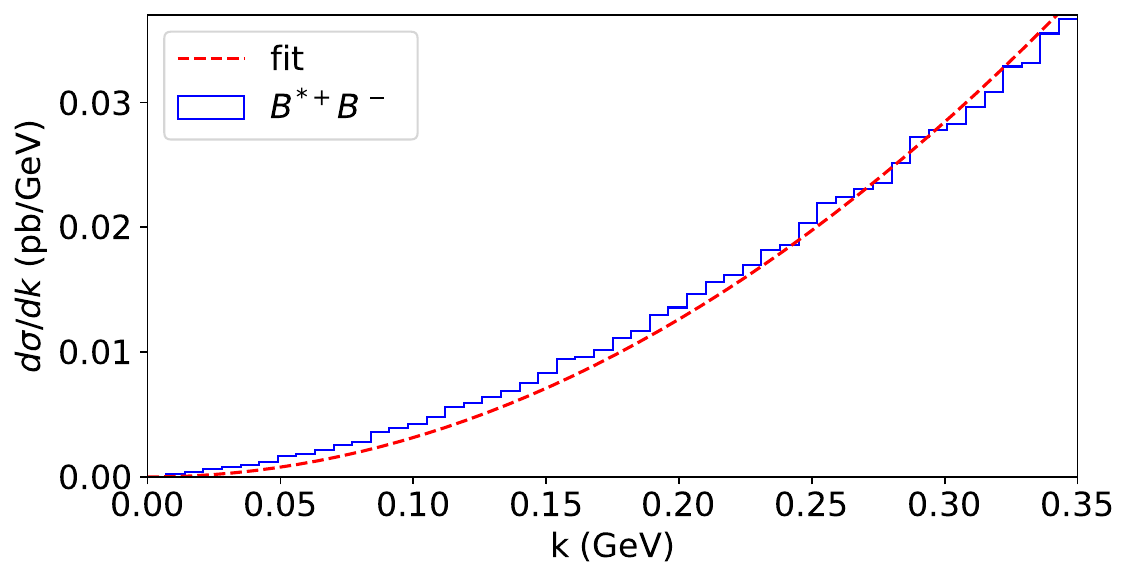}}
 \subfigure[]{\includegraphics[width=0.45\textwidth]{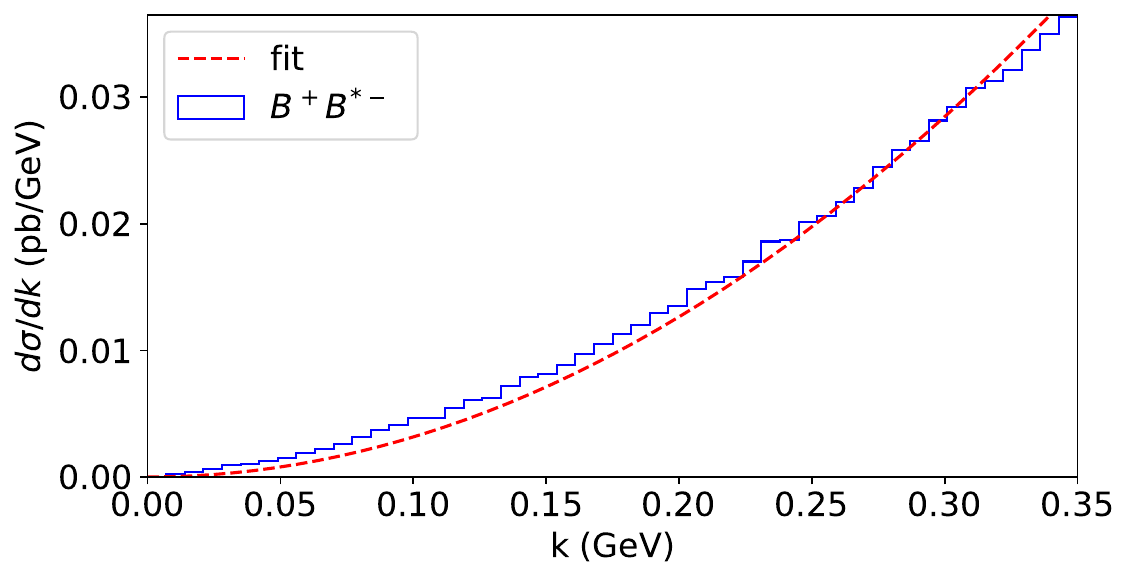}}
 \subfigure[]{\includegraphics[width=0.45\textwidth]{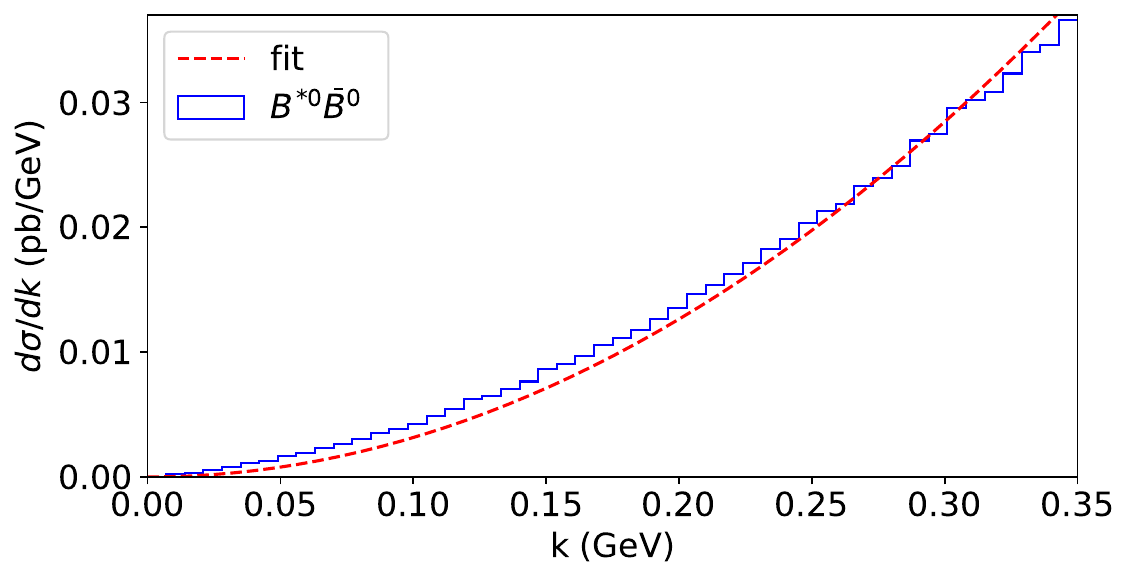}}
 \subfigure[]{\includegraphics[width=0.45\textwidth]{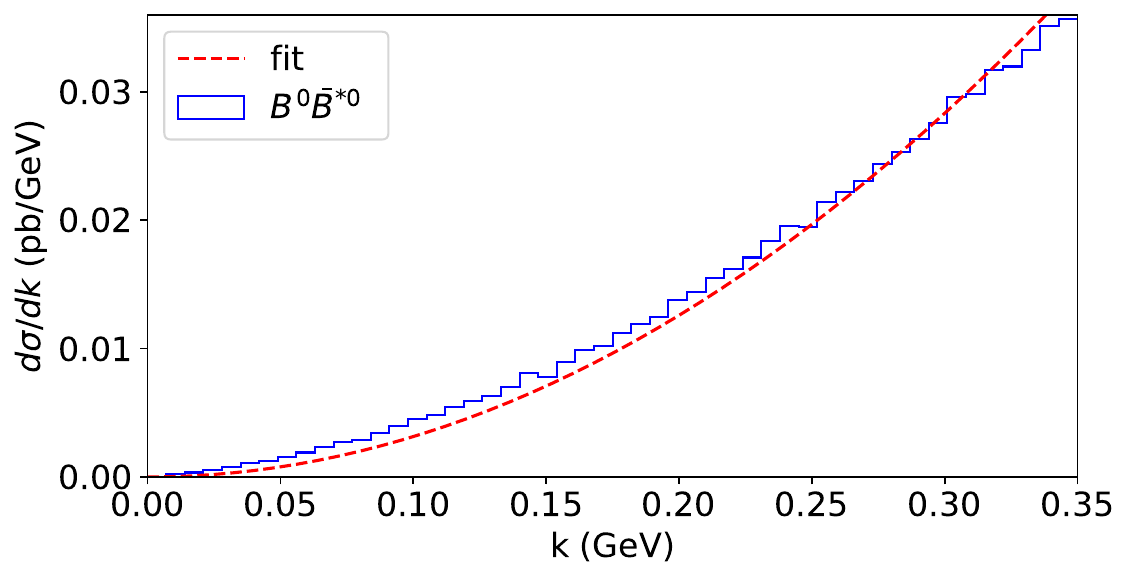}}
 \subfigure[]{\includegraphics[width=0.45\textwidth]{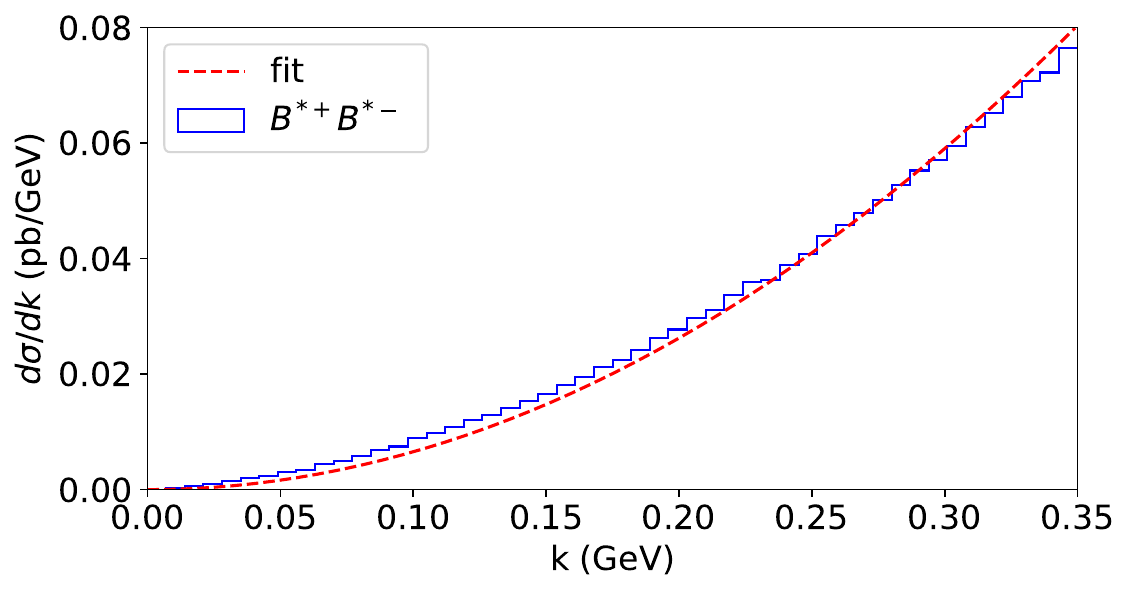}}
 \subfigure[]{\includegraphics[width=0.45\textwidth]{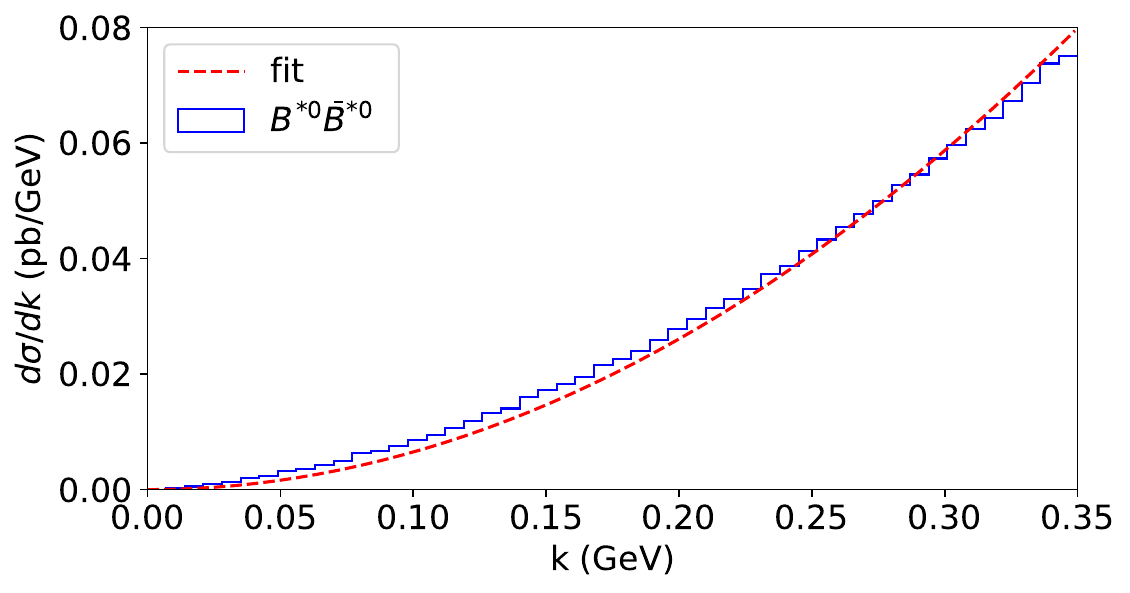}}
   \caption{Differential cross sections d$\sigma$/d$k$ (in units of $\rm{pb/GeV}$) for the process $e^+e^- \to Z^0 \to B^{(*)0}\bar{B}^{(*)0}$ and $e^+e^- \to Z^0 \to B^{(*)\pm}\bar{B}^{(*)\mp}$. The subfigures demonstrate the differential production cross sections for (a) $B^{\ast +}B^-$; (b) $B^+B^{\ast -}$; (c) $B^{\ast 0}\bar{B}^{0}$; (d) $B^0\bar{B}^{\ast 0}$; 
                            (e) $B^{\ast +}B^{\ast -}$; and (f) $B^{\ast 0}\bar{B}^{\ast 0}$.}
   \label{Fig:diffxsec_Zb0}
\end{figure}
\begin{figure}[htbp]
 \subfigure[]{\includegraphics[width=0.45\textwidth]{Fig_BstarmB0bar.pdf}}
 \subfigure[]{\includegraphics[width=0.45\textwidth]{Fig_Bstar0barBm.pdf}}
 \subfigure[]{\includegraphics[width=0.45\textwidth]{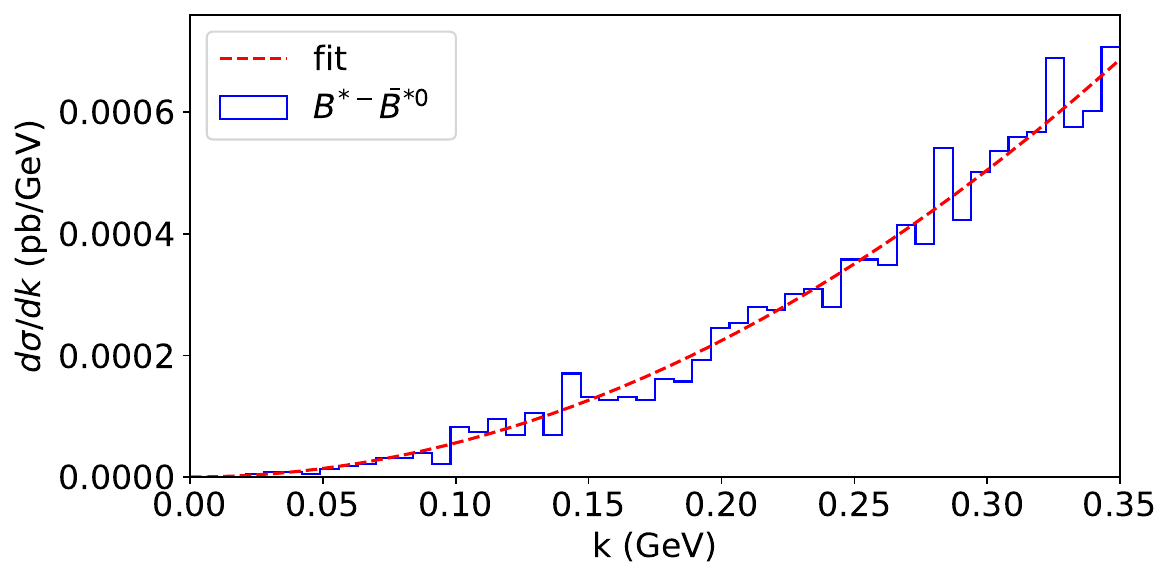}}
   \caption{Differential cross sections d$\sigma$/d$k$ (in units of $\rm{pb/GeV}$) for the process $e^+e^- \to Z^0 \to B^{(*)-}\bar{B}^{(*)0}$. The subfigures demonstrate the differential production cross sections for (a) $B^{\ast -} \bar{B}^0$; (b) $\bar{B}^{\ast 0}B^-$; and (c) $B^{\ast -}\bar{B}^{\ast 0}$.}
   \label{Fig:diffxsec_BB}
\end{figure}
\clearpage
\bibliography{ProdAtCEPC.bib}

\begin{thebibliography}{74}%
\makeatletter
\providecommand \@ifxundefined [1]{%
 \@ifx{#1\undefined}
}%
\providecommand \@ifnum [1]{%
 \ifnum #1\expandafter \@firstoftwo
 \else \expandafter \@secondoftwo
 \fi
}%
\providecommand \@ifx [1]{%
 \ifx #1\expandafter \@firstoftwo
 \else \expandafter \@secondoftwo
 \fi
}%
\providecommand \natexlab [1]{#1}%
\providecommand \enquote  [1]{``#1''}%
\providecommand \bibnamefont  [1]{#1}%
\providecommand \bibfnamefont [1]{#1}%
\providecommand \citenamefont [1]{#1}%
\providecommand \href@noop [0]{\@secondoftwo}%
\providecommand \href [0]{\begingroup \@sanitize@url \@href}%
\providecommand \@href[1]{\@@startlink{#1}\@@href}%
\providecommand \@@href[1]{\endgroup#1\@@endlink}%
\providecommand \@sanitize@url [0]{\catcode `\\12\catcode `\$12\catcode `\&12\catcode `\#12\catcode `\^12\catcode `\_12\catcode `\%12\relax}%
\providecommand \@@startlink[1]{}%
\providecommand \@@endlink[0]{}%
\providecommand \url  [0]{\begingroup\@sanitize@url \@url }%
\providecommand \@url [1]{\endgroup\@href {#1}{\urlprefix }}%
\providecommand \urlprefix  [0]{URL }%
\providecommand \Eprint [0]{\href }%
\providecommand \doibase [0]{http://dx.doi.org/}%
\providecommand \selectlanguage [0]{\@gobble}%
\providecommand \bibinfo  [0]{\@secondoftwo}%
\providecommand \bibfield  [0]{\@secondoftwo}%
\providecommand \translation [1]{[#1]}%
\providecommand \BibitemOpen [0]{}%
\providecommand \bibitemStop [0]{}%
\providecommand \bibitemNoStop [0]{.\EOS\space}%
\providecommand \EOS [0]{\spacefactor3000\relax}%
\providecommand \BibitemShut  [1]{\csname bibitem#1\endcsname}%
\let\auto@bib@innerbib\@empty
\bibitem [{\citenamefont {Choi}\ \emph {et~al.}(2003)\citenamefont {Choi} \emph {et~al.}}]{Belle:2003nnu}%
  \BibitemOpen
  \bibfield  {author} {\bibinfo {author} {\bibfnamefont {S.~K.}\ \bibnamefont {Choi}} \emph {et~al.} (\bibinfo {collaboration} {Belle}),\ }\href {\doibase 10.1103/PhysRevLett.91.262001} {\bibfield  {journal} {\bibinfo  {journal} {Phys. Rev. Lett.}\ }\textbf {\bibinfo {volume} {91}},\ \bibinfo {pages} {262001} (\bibinfo {year} {2003})},\ \Eprint {http://arxiv.org/abs/hep-ex/0309032} {arXiv:hep-ex/0309032} \BibitemShut {NoStop}%
\bibitem [{\citenamefont {Workman}\ \emph {et~al.}(2022)\citenamefont {Workman} \emph {et~al.}}]{ParticleDataGroup:2022pth}%
  \BibitemOpen
  \bibfield  {author} {\bibinfo {author} {\bibfnamefont {R.~L.}\ \bibnamefont {Workman}} \emph {et~al.} (\bibinfo {collaboration} {Particle Data Group}),\ }\href {\doibase 10.1093/ptep/ptac097} {\bibfield  {journal} {\bibinfo  {journal} {Prog. Theor. Exp. Phys.}\ }\textbf {\bibinfo {volume} {2022}},\ \bibinfo {pages} {083C01} (\bibinfo {year} {2022})}\BibitemShut {NoStop}%
\bibitem [{\citenamefont {Acosta}\ \emph {et~al.}(2004)\citenamefont {Acosta} \emph {et~al.}}]{CDF:2003cab}%
  \BibitemOpen
  \bibfield  {author} {\bibinfo {author} {\bibfnamefont {D.}~\bibnamefont {Acosta}} \emph {et~al.} (\bibinfo {collaboration} {CDF}),\ }\href {\doibase 10.1103/PhysRevLett.93.072001} {\bibfield  {journal} {\bibinfo  {journal} {Phys. Rev. Lett.}\ }\textbf {\bibinfo {volume} {93}},\ \bibinfo {pages} {072001} (\bibinfo {year} {2004})},\ \Eprint {http://arxiv.org/abs/hep-ex/0312021} {arXiv:hep-ex/0312021} \BibitemShut {NoStop}%
\bibitem [{\citenamefont {Abazov}\ \emph {et~al.}(2004)\citenamefont {Abazov} \emph {et~al.}}]{D0:2004zmu}%
  \BibitemOpen
  \bibfield  {author} {\bibinfo {author} {\bibfnamefont {V.~M.}\ \bibnamefont {Abazov}} \emph {et~al.} (\bibinfo {collaboration} {D0}),\ }\href {\doibase 10.1103/PhysRevLett.93.162002} {\bibfield  {journal} {\bibinfo  {journal} {Phys. Rev. Lett.}\ }\textbf {\bibinfo {volume} {93}},\ \bibinfo {pages} {162002} (\bibinfo {year} {2004})},\ \Eprint {http://arxiv.org/abs/hep-ex/0405004} {arXiv:hep-ex/0405004} \BibitemShut {NoStop}%
\bibitem [{\citenamefont {Aaij}\ \emph {et~al.}(2013)\citenamefont {Aaij} \emph {et~al.}}]{LHCb:2013kgk}%
  \BibitemOpen
  \bibfield  {author} {\bibinfo {author} {\bibfnamefont {R.}~\bibnamefont {Aaij}} \emph {et~al.} (\bibinfo {collaboration} {LHCb}),\ }\href {\doibase 10.1103/PhysRevLett.110.222001} {\bibfield  {journal} {\bibinfo  {journal} {Phys. Rev. Lett.}\ }\textbf {\bibinfo {volume} {110}},\ \bibinfo {pages} {222001} (\bibinfo {year} {2013})},\ \Eprint {http://arxiv.org/abs/1302.6269} {arXiv:1302.6269 [hep-ex]} \BibitemShut {NoStop}%
\bibitem [{\citenamefont {Chatrchyan}\ \emph {et~al.}(2013)\citenamefont {Chatrchyan} \emph {et~al.}}]{CMS:2013fpt}%
  \BibitemOpen
  \bibfield  {author} {\bibinfo {author} {\bibfnamefont {S.}~\bibnamefont {Chatrchyan}} \emph {et~al.} (\bibinfo {collaboration} {CMS}),\ }\href {\doibase 10.1007/JHEP04(2013)154} {\bibfield  {journal} {\bibinfo  {journal} {JHEP}\ }\textbf {\bibinfo {volume} {04}},\ \bibinfo {pages} {154} (\bibinfo {year} {2013})},\ \Eprint {http://arxiv.org/abs/1302.3968} {arXiv:1302.3968 [hep-ex]} \BibitemShut {NoStop}%
\bibitem [{\citenamefont {Ablikim}\ \emph {et~al.}(2014)\citenamefont {Ablikim} \emph {et~al.}}]{BESIII:2013fnz}%
  \BibitemOpen
  \bibfield  {author} {\bibinfo {author} {\bibfnamefont {M.}~\bibnamefont {Ablikim}} \emph {et~al.} (\bibinfo {collaboration} {BESIII}),\ }\href {\doibase 10.1103/PhysRevLett.112.092001} {\bibfield  {journal} {\bibinfo  {journal} {Phys. Rev. Lett.}\ }\textbf {\bibinfo {volume} {112}},\ \bibinfo {pages} {092001} (\bibinfo {year} {2014})},\ \Eprint {http://arxiv.org/abs/1310.4101} {arXiv:1310.4101 [hep-ex]} \BibitemShut {NoStop}%
\bibitem [{\citenamefont {Aaij}\ \emph {et~al.}(2014)\citenamefont {Aaij} \emph {et~al.}}]{LHCb:2014jvf}%
  \BibitemOpen
  \bibfield  {author} {\bibinfo {author} {\bibfnamefont {R.}~\bibnamefont {Aaij}} \emph {et~al.} (\bibinfo {collaboration} {LHCb}),\ }\href {\doibase 10.1016/j.nuclphysb.2014.06.011} {\bibfield  {journal} {\bibinfo  {journal} {Nucl. Phys. B}\ }\textbf {\bibinfo {volume} {886}},\ \bibinfo {pages} {665} (\bibinfo {year} {2014})},\ \Eprint {http://arxiv.org/abs/1404.0275} {arXiv:1404.0275 [hep-ex]} \BibitemShut {NoStop}%
\bibitem [{\citenamefont {Aaij}\ \emph {et~al.}(2012)\citenamefont {Aaij} \emph {et~al.}}]{LHCb:2011zzp}%
  \BibitemOpen
  \bibfield  {author} {\bibinfo {author} {\bibfnamefont {R.}~\bibnamefont {Aaij}} \emph {et~al.} (\bibinfo {collaboration} {LHCb}),\ }\href {\doibase 10.1140/epjc/s10052-012-1972-7} {\bibfield  {journal} {\bibinfo  {journal} {Eur. Phys. J. C}\ }\textbf {\bibinfo {volume} {72}},\ \bibinfo {pages} {1972} (\bibinfo {year} {2012})},\ \Eprint {http://arxiv.org/abs/1112.5310} {arXiv:1112.5310 [hep-ex]} \BibitemShut {NoStop}%
\bibitem [{\citenamefont {del Amo~Sanchez}\ \emph {et~al.}(2010)\citenamefont {del Amo~Sanchez} \emph {et~al.}}]{BaBar:2010wfc}%
  \BibitemOpen
  \bibfield  {author} {\bibinfo {author} {\bibfnamefont {P.}~\bibnamefont {del Amo~Sanchez}} \emph {et~al.} (\bibinfo {collaboration} {BaBar}),\ }\href {\doibase 10.1103/PhysRevD.82.011101} {\bibfield  {journal} {\bibinfo  {journal} {Phys. Rev. D}\ }\textbf {\bibinfo {volume} {82}},\ \bibinfo {pages} {011101} (\bibinfo {year} {2010})},\ \Eprint {http://arxiv.org/abs/1005.5190} {arXiv:1005.5190 [hep-ex]} \BibitemShut {NoStop}%
\bibitem [{\citenamefont {Aubert}\ \emph {et~al.}(2009)\citenamefont {Aubert} \emph {et~al.}}]{BaBar:2008flx}%
  \BibitemOpen
  \bibfield  {author} {\bibinfo {author} {\bibfnamefont {B.}~\bibnamefont {Aubert}} \emph {et~al.} (\bibinfo {collaboration} {BaBar}),\ }\href {\doibase 10.1103/PhysRevLett.102.132001} {\bibfield  {journal} {\bibinfo  {journal} {Phys. Rev. Lett.}\ }\textbf {\bibinfo {volume} {102}},\ \bibinfo {pages} {132001} (\bibinfo {year} {2009})},\ \Eprint {http://arxiv.org/abs/0809.0042} {arXiv:0809.0042 [hep-ex]} \BibitemShut {NoStop}%
\bibitem [{\citenamefont {Aubert}\ \emph {et~al.}(2008{\natexlab{a}})\citenamefont {Aubert} \emph {et~al.}}]{BaBar:2008qzi}%
  \BibitemOpen
  \bibfield  {author} {\bibinfo {author} {\bibfnamefont {B.}~\bibnamefont {Aubert}} \emph {et~al.} (\bibinfo {collaboration} {BaBar}),\ }\href {\doibase 10.1103/PhysRevD.77.111101} {\bibfield  {journal} {\bibinfo  {journal} {Phys. Rev. D}\ }\textbf {\bibinfo {volume} {77}},\ \bibinfo {pages} {111101} (\bibinfo {year} {2008}{\natexlab{a}})},\ \Eprint {http://arxiv.org/abs/0803.2838} {arXiv:0803.2838 [hep-ex]} \BibitemShut {NoStop}%
\bibitem [{\citenamefont {Aubert}\ \emph {et~al.}(2008{\natexlab{b}})\citenamefont {Aubert} \emph {et~al.}}]{BaBar:2007cmo}%
  \BibitemOpen
  \bibfield  {author} {\bibinfo {author} {\bibfnamefont {B.}~\bibnamefont {Aubert}} \emph {et~al.} (\bibinfo {collaboration} {BaBar}),\ }\href {\doibase 10.1103/PhysRevD.77.011102} {\bibfield  {journal} {\bibinfo  {journal} {Phys. Rev. D}\ }\textbf {\bibinfo {volume} {77}},\ \bibinfo {pages} {011102} (\bibinfo {year} {2008}{\natexlab{b}})},\ \Eprint {http://arxiv.org/abs/0708.1565} {arXiv:0708.1565 [hep-ex]} \BibitemShut {NoStop}%
\bibitem [{\citenamefont {Ablikim}\ \emph {et~al.}(2013)\citenamefont {Ablikim} \emph {et~al.}}]{BESIII:2013ris}%
  \BibitemOpen
  \bibfield  {author} {\bibinfo {author} {\bibfnamefont {M.}~\bibnamefont {Ablikim}} \emph {et~al.} (\bibinfo {collaboration} {BESIII}),\ }\href {\doibase 10.1103/PhysRevLett.110.252001} {\bibfield  {journal} {\bibinfo  {journal} {Phys. Rev. Lett.}\ }\textbf {\bibinfo {volume} {110}},\ \bibinfo {pages} {252001} (\bibinfo {year} {2013})},\ \Eprint {http://arxiv.org/abs/1303.5949} {arXiv:1303.5949 [hep-ex]} \BibitemShut {NoStop}%
\bibitem [{\citenamefont {Liu}\ \emph {et~al.}(2013)\citenamefont {Liu} \emph {et~al.}}]{Belle:2013yex}%
  \BibitemOpen
  \bibfield  {author} {\bibinfo {author} {\bibfnamefont {Z.~Q.}\ \bibnamefont {Liu}} \emph {et~al.} (\bibinfo {collaboration} {Belle}),\ }\href {\doibase 10.1103/PhysRevLett.110.252002} {\bibfield  {journal} {\bibinfo  {journal} {Phys. Rev. Lett.}\ }\textbf {\bibinfo {volume} {110}},\ \bibinfo {pages} {252002} (\bibinfo {year} {2013})},\ \bibinfo {note} {[Erratum: Phys.Rev.Lett. 111, 019901 (2013)]},\ \Eprint {http://arxiv.org/abs/1304.0121} {arXiv:1304.0121 [hep-ex]} \BibitemShut {NoStop}%
\bibitem [{\citenamefont {Ablikim}\ \emph {et~al.}(2015)\citenamefont {Ablikim} \emph {et~al.}}]{BESIII:2015cld}%
  \BibitemOpen
  \bibfield  {author} {\bibinfo {author} {\bibfnamefont {M.}~\bibnamefont {Ablikim}} \emph {et~al.} (\bibinfo {collaboration} {BESIII}),\ }\href {\doibase 10.1103/PhysRevLett.115.112003} {\bibfield  {journal} {\bibinfo  {journal} {Phys. Rev. Lett.}\ }\textbf {\bibinfo {volume} {115}},\ \bibinfo {pages} {112003} (\bibinfo {year} {2015})},\ \Eprint {http://arxiv.org/abs/1506.06018} {arXiv:1506.06018 [hep-ex]} \BibitemShut {NoStop}%
\bibitem [{\citenamefont {Aaij}\ \emph {et~al.}(2015)\citenamefont {Aaij} \emph {et~al.}}]{LHCb:2015yax}%
  \BibitemOpen
  \bibfield  {author} {\bibinfo {author} {\bibfnamefont {R.}~\bibnamefont {Aaij}} \emph {et~al.} (\bibinfo {collaboration} {LHCb}),\ }\href {\doibase 10.1103/PhysRevLett.115.072001} {\bibfield  {journal} {\bibinfo  {journal} {Phys. Rev. Lett.}\ }\textbf {\bibinfo {volume} {115}},\ \bibinfo {pages} {072001} (\bibinfo {year} {2015})},\ \Eprint {http://arxiv.org/abs/1507.03414} {arXiv:1507.03414 [hep-ex]} \BibitemShut {NoStop}%
\bibitem [{\citenamefont {Aaij}\ \emph {et~al.}(2019)\citenamefont {Aaij} \emph {et~al.}}]{LHCb:2019kea}%
  \BibitemOpen
  \bibfield  {author} {\bibinfo {author} {\bibfnamefont {R.}~\bibnamefont {Aaij}} \emph {et~al.} (\bibinfo {collaboration} {LHCb}),\ }\href {\doibase 10.1103/PhysRevLett.122.222001} {\bibfield  {journal} {\bibinfo  {journal} {Phys. Rev. Lett.}\ }\textbf {\bibinfo {volume} {122}},\ \bibinfo {pages} {222001} (\bibinfo {year} {2019})},\ \Eprint {http://arxiv.org/abs/1904.03947} {arXiv:1904.03947 [hep-ex]} \BibitemShut {NoStop}%
\bibitem [{\citenamefont {Aaij}\ \emph {et~al.}(2021)\citenamefont {Aaij} \emph {et~al.}}]{LHCb:2020jpq}%
  \BibitemOpen
  \bibfield  {author} {\bibinfo {author} {\bibfnamefont {R.}~\bibnamefont {Aaij}} \emph {et~al.} (\bibinfo {collaboration} {LHCb}),\ }\href {\doibase 10.1016/j.scib.2021.02.030} {\bibfield  {journal} {\bibinfo  {journal} {Sci. Bull.}\ }\textbf {\bibinfo {volume} {66}},\ \bibinfo {pages} {1278} (\bibinfo {year} {2021})},\ \Eprint {http://arxiv.org/abs/2012.10380} {arXiv:2012.10380 [hep-ex]} \BibitemShut {NoStop}%
\bibitem [{\citenamefont {Aaij}\ \emph {et~al.}(2023)\citenamefont {Aaij} \emph {et~al.}}]{LHCb:2022ogu}%
  \BibitemOpen
  \bibfield  {author} {\bibinfo {author} {\bibfnamefont {R.}~\bibnamefont {Aaij}} \emph {et~al.} (\bibinfo {collaboration} {LHCb}),\ }\href {\doibase 10.1103/PhysRevLett.131.031901} {\bibfield  {journal} {\bibinfo  {journal} {Phys. Rev. Lett.}\ }\textbf {\bibinfo {volume} {131}},\ \bibinfo {pages} {031901} (\bibinfo {year} {2023})},\ \Eprint {http://arxiv.org/abs/2210.10346} {arXiv:2210.10346 [hep-ex]} \BibitemShut {NoStop}%
\bibitem [{\citenamefont {Aaij}\ \emph {et~al.}(2022{\natexlab{a}})\citenamefont {Aaij} \emph {et~al.}}]{LHCb:2021vvq}%
  \BibitemOpen
  \bibfield  {author} {\bibinfo {author} {\bibfnamefont {R.}~\bibnamefont {Aaij}} \emph {et~al.} (\bibinfo {collaboration} {LHCb}),\ }\href {\doibase 10.1038/s41567-022-01614-y} {\bibfield  {journal} {\bibinfo  {journal} {Nature Phys.}\ }\textbf {\bibinfo {volume} {18}},\ \bibinfo {pages} {751} (\bibinfo {year} {2022}{\natexlab{a}})},\ \Eprint {http://arxiv.org/abs/2109.01038} {arXiv:2109.01038 [hep-ex]} \BibitemShut {NoStop}%
\bibitem [{\citenamefont {Chen}\ \emph {et~al.}(2016)\citenamefont {Chen}, \citenamefont {Chen}, \citenamefont {Liu},\ and\ \citenamefont {Zhu}}]{Chen:2016qju}%
  \BibitemOpen
  \bibfield  {author} {\bibinfo {author} {\bibfnamefont {H.-X.}\ \bibnamefont {Chen}}, \bibinfo {author} {\bibfnamefont {W.}~\bibnamefont {Chen}}, \bibinfo {author} {\bibfnamefont {X.}~\bibnamefont {Liu}}, \ and\ \bibinfo {author} {\bibfnamefont {S.-L.}\ \bibnamefont {Zhu}},\ }\href {\doibase 10.1016/j.physrep.2016.05.004} {\bibfield  {journal} {\bibinfo  {journal} {Phys. Rep.}\ }\textbf {\bibinfo {volume} {639}},\ \bibinfo {pages} {1} (\bibinfo {year} {2016})},\ \Eprint {http://arxiv.org/abs/1601.02092} {arXiv:1601.02092 [hep-ph]} \BibitemShut {NoStop}%
\bibitem [{\citenamefont {Hosaka}\ \emph {et~al.}(2016)\citenamefont {Hosaka}, \citenamefont {Iijima}, \citenamefont {Miyabayashi}, \citenamefont {Sakai},\ and\ \citenamefont {Yasui}}]{Hosaka:2016pey}%
  \BibitemOpen
  \bibfield  {author} {\bibinfo {author} {\bibfnamefont {A.}~\bibnamefont {Hosaka}}, \bibinfo {author} {\bibfnamefont {T.}~\bibnamefont {Iijima}}, \bibinfo {author} {\bibfnamefont {K.}~\bibnamefont {Miyabayashi}}, \bibinfo {author} {\bibfnamefont {Y.}~\bibnamefont {Sakai}}, \ and\ \bibinfo {author} {\bibfnamefont {S.}~\bibnamefont {Yasui}},\ }\href {\doibase 10.1093/ptep/ptw045} {\bibfield  {journal} {\bibinfo  {journal} {Prog. Theor. Exp. Phys.}\ }\textbf {\bibinfo {volume} {2016}},\ \bibinfo {pages} {062C01} (\bibinfo {year} {2016})},\ \Eprint {http://arxiv.org/abs/1603.09229} {arXiv:1603.09229 [hep-ph]} \BibitemShut {NoStop}%
\bibitem [{\citenamefont {Esposito}\ \emph {et~al.}(2017)\citenamefont {Esposito}, \citenamefont {Pilloni},\ and\ \citenamefont {Polosa}}]{Esposito:2016noz}%
  \BibitemOpen
  \bibfield  {author} {\bibinfo {author} {\bibfnamefont {A.}~\bibnamefont {Esposito}}, \bibinfo {author} {\bibfnamefont {A.}~\bibnamefont {Pilloni}}, \ and\ \bibinfo {author} {\bibfnamefont {A.~D.}\ \bibnamefont {Polosa}},\ }\href {\doibase 10.1016/j.physrep.2016.11.002} {\bibfield  {journal} {\bibinfo  {journal} {Phys. Rept.}\ }\textbf {\bibinfo {volume} {668}},\ \bibinfo {pages} {1} (\bibinfo {year} {2017})},\ \Eprint {http://arxiv.org/abs/1611.07920} {arXiv:1611.07920 [hep-ph]} \BibitemShut {NoStop}%
\bibitem [{\citenamefont {Lebed}\ \emph {et~al.}(2017)\citenamefont {Lebed}, \citenamefont {Mitchell},\ and\ \citenamefont {Swanson}}]{Lebed:2016hpi}%
  \BibitemOpen
  \bibfield  {author} {\bibinfo {author} {\bibfnamefont {R.~F.}\ \bibnamefont {Lebed}}, \bibinfo {author} {\bibfnamefont {R.~E.}\ \bibnamefont {Mitchell}}, \ and\ \bibinfo {author} {\bibfnamefont {E.~S.}\ \bibnamefont {Swanson}},\ }\href {\doibase 10.1016/j.ppnp.2016.11.003} {\bibfield  {journal} {\bibinfo  {journal} {Prog. Part. Nucl. Phys.}\ }\textbf {\bibinfo {volume} {93}},\ \bibinfo {pages} {143} (\bibinfo {year} {2017})},\ \Eprint {http://arxiv.org/abs/1610.04528} {arXiv:1610.04528 [hep-ph]} \BibitemShut {NoStop}%
\bibitem [{\citenamefont {Ali}\ \emph {et~al.}(2017)\citenamefont {Ali}, \citenamefont {Lange},\ and\ \citenamefont {Stone}}]{Ali:2017jda}%
  \BibitemOpen
  \bibfield  {author} {\bibinfo {author} {\bibfnamefont {A.}~\bibnamefont {Ali}}, \bibinfo {author} {\bibfnamefont {J.~S.}\ \bibnamefont {Lange}}, \ and\ \bibinfo {author} {\bibfnamefont {S.}~\bibnamefont {Stone}},\ }\href {\doibase 10.1016/j.ppnp.2017.08.003} {\bibfield  {journal} {\bibinfo  {journal} {Prog. Part. Nucl. Phys.}\ }\textbf {\bibinfo {volume} {97}},\ \bibinfo {pages} {123} (\bibinfo {year} {2017})},\ \Eprint {http://arxiv.org/abs/1706.00610} {arXiv:1706.00610 [hep-ph]} \BibitemShut {NoStop}%
\bibitem [{\citenamefont {Olsen}\ \emph {et~al.}(2018)\citenamefont {Olsen}, \citenamefont {Skwarnicki},\ and\ \citenamefont {Zieminska}}]{Olsen:2017bmm}%
  \BibitemOpen
  \bibfield  {author} {\bibinfo {author} {\bibfnamefont {S.~L.}\ \bibnamefont {Olsen}}, \bibinfo {author} {\bibfnamefont {T.}~\bibnamefont {Skwarnicki}}, \ and\ \bibinfo {author} {\bibfnamefont {D.}~\bibnamefont {Zieminska}},\ }\href {\doibase 10.1103/RevModPhys.90.015003} {\bibfield  {journal} {\bibinfo  {journal} {Rev. Mod. Phys.}\ }\textbf {\bibinfo {volume} {90}},\ \bibinfo {pages} {015003} (\bibinfo {year} {2018})},\ \Eprint {http://arxiv.org/abs/1708.04012} {arXiv:1708.04012 [hep-ph]} \BibitemShut {NoStop}%
\bibitem [{\citenamefont {Guo}\ \emph {et~al.}(2018)\citenamefont {Guo}, \citenamefont {Hanhart}, \citenamefont {Mei{\ss}ner}, \citenamefont {Wang}, \citenamefont {Zhao},\ and\ \citenamefont {Zou}}]{Guo:2017jvc}%
  \BibitemOpen
  \bibfield  {author} {\bibinfo {author} {\bibfnamefont {F.-K.}\ \bibnamefont {Guo}}, \bibinfo {author} {\bibfnamefont {C.}~\bibnamefont {Hanhart}}, \bibinfo {author} {\bibfnamefont {U.-G.}\ \bibnamefont {Mei{\ss}ner}}, \bibinfo {author} {\bibfnamefont {Q.}~\bibnamefont {Wang}}, \bibinfo {author} {\bibfnamefont {Q.}~\bibnamefont {Zhao}}, \ and\ \bibinfo {author} {\bibfnamefont {B.-S.}\ \bibnamefont {Zou}},\ }\href {\doibase 10.1103/RevModPhys.90.015004} {\bibfield  {journal} {\bibinfo  {journal} {Rev. Mod. Phys.}\ }\textbf {\bibinfo {volume} {90}},\ \bibinfo {pages} {015004} (\bibinfo {year} {2018})},\ \bibinfo {note} {[Erratum: Rev.Mod.Phys. 94, 029901 (2022)]},\ \Eprint {http://arxiv.org/abs/1705.00141} {arXiv:1705.00141 [hep-ph]} \BibitemShut {NoStop}%
\bibitem [{\citenamefont {Albuquerque}\ \emph {et~al.}(2019)\citenamefont {Albuquerque}, \citenamefont {Dias}, \citenamefont {Khemchandani}, \citenamefont {Mart\'\i{}nez~Torres}, \citenamefont {Navarra}, \citenamefont {Nielsen},\ and\ \citenamefont {Zanetti}}]{Albuquerque:2018jkn}%
  \BibitemOpen
  \bibfield  {author} {\bibinfo {author} {\bibfnamefont {R.~M.}\ \bibnamefont {Albuquerque}}, \bibinfo {author} {\bibfnamefont {J.~M.}\ \bibnamefont {Dias}}, \bibinfo {author} {\bibfnamefont {K.~P.}\ \bibnamefont {Khemchandani}}, \bibinfo {author} {\bibfnamefont {A.}~\bibnamefont {Mart\'\i{}nez~Torres}}, \bibinfo {author} {\bibfnamefont {F.~S.}\ \bibnamefont {Navarra}}, \bibinfo {author} {\bibfnamefont {M.}~\bibnamefont {Nielsen}}, \ and\ \bibinfo {author} {\bibfnamefont {C.~M.}\ \bibnamefont {Zanetti}},\ }\href {\doibase 10.1088/1361-6471/ab2678} {\bibfield  {journal} {\bibinfo  {journal} {J. Phys. G}\ }\textbf {\bibinfo {volume} {46}},\ \bibinfo {pages} {093002} (\bibinfo {year} {2019})},\ \Eprint {http://arxiv.org/abs/1812.08207} {arXiv:1812.08207 [hep-ph]} \BibitemShut {NoStop}%
\bibitem [{\citenamefont {Liu}\ \emph {et~al.}(2019)\citenamefont {Liu}, \citenamefont {Chen}, \citenamefont {Chen}, \citenamefont {Liu},\ and\ \citenamefont {Zhu}}]{Liu:2019zoy}%
  \BibitemOpen
  \bibfield  {author} {\bibinfo {author} {\bibfnamefont {Y.-R.}\ \bibnamefont {Liu}}, \bibinfo {author} {\bibfnamefont {H.-X.}\ \bibnamefont {Chen}}, \bibinfo {author} {\bibfnamefont {W.}~\bibnamefont {Chen}}, \bibinfo {author} {\bibfnamefont {X.}~\bibnamefont {Liu}}, \ and\ \bibinfo {author} {\bibfnamefont {S.-L.}\ \bibnamefont {Zhu}},\ }\href {\doibase 10.1016/j.ppnp.2019.04.003} {\bibfield  {journal} {\bibinfo  {journal} {Prog. Part. Nucl. Phys.}\ }\textbf {\bibinfo {volume} {107}},\ \bibinfo {pages} {237} (\bibinfo {year} {2019})},\ \Eprint {http://arxiv.org/abs/1903.11976} {arXiv:1903.11976 [hep-ph]} \BibitemShut {NoStop}%
\bibitem [{\citenamefont {Guo}\ \emph {et~al.}(2020)\citenamefont {Guo}, \citenamefont {Liu},\ and\ \citenamefont {Sakai}}]{Guo:2019twa}%
  \BibitemOpen
  \bibfield  {author} {\bibinfo {author} {\bibfnamefont {F.-K.}\ \bibnamefont {Guo}}, \bibinfo {author} {\bibfnamefont {X.-H.}\ \bibnamefont {Liu}}, \ and\ \bibinfo {author} {\bibfnamefont {S.}~\bibnamefont {Sakai}},\ }\href {\doibase 10.1016/j.ppnp.2020.103757} {\bibfield  {journal} {\bibinfo  {journal} {Prog. Part. Nucl. Phys.}\ }\textbf {\bibinfo {volume} {112}},\ \bibinfo {pages} {103757} (\bibinfo {year} {2020})},\ \Eprint {http://arxiv.org/abs/1912.07030} {arXiv:1912.07030 [hep-ph]} \BibitemShut {NoStop}%
\bibitem [{\citenamefont {Brambilla}\ \emph {et~al.}(2020)\citenamefont {Brambilla}, \citenamefont {Eidelman}, \citenamefont {Hanhart}, \citenamefont {Nefediev}, \citenamefont {Shen}, \citenamefont {Thomas}, \citenamefont {Vairo},\ and\ \citenamefont {Yuan}}]{Brambilla:2019esw}%
  \BibitemOpen
  \bibfield  {author} {\bibinfo {author} {\bibfnamefont {N.}~\bibnamefont {Brambilla}}, \bibinfo {author} {\bibfnamefont {S.}~\bibnamefont {Eidelman}}, \bibinfo {author} {\bibfnamefont {C.}~\bibnamefont {Hanhart}}, \bibinfo {author} {\bibfnamefont {A.}~\bibnamefont {Nefediev}}, \bibinfo {author} {\bibfnamefont {C.-P.}\ \bibnamefont {Shen}}, \bibinfo {author} {\bibfnamefont {C.~E.}\ \bibnamefont {Thomas}}, \bibinfo {author} {\bibfnamefont {A.}~\bibnamefont {Vairo}}, \ and\ \bibinfo {author} {\bibfnamefont {C.-Z.}\ \bibnamefont {Yuan}},\ }\href {\doibase 10.1016/j.physrep.2020.05.001} {\bibfield  {journal} {\bibinfo  {journal} {Phys. Rep.}\ }\textbf {\bibinfo {volume} {873}},\ \bibinfo {pages} {1} (\bibinfo {year} {2020})},\ \Eprint {http://arxiv.org/abs/1907.07583} {arXiv:1907.07583 [hep-ex]} \BibitemShut {NoStop}%
\bibitem [{\citenamefont {Chen}\ \emph {et~al.}(2023)\citenamefont {Chen}, \citenamefont {Chen}, \citenamefont {Liu}, \citenamefont {Liu},\ and\ \citenamefont {Zhu}}]{Chen:2022asf}%
  \BibitemOpen
  \bibfield  {author} {\bibinfo {author} {\bibfnamefont {H.-X.}\ \bibnamefont {Chen}}, \bibinfo {author} {\bibfnamefont {W.}~\bibnamefont {Chen}}, \bibinfo {author} {\bibfnamefont {X.}~\bibnamefont {Liu}}, \bibinfo {author} {\bibfnamefont {Y.-R.}\ \bibnamefont {Liu}}, \ and\ \bibinfo {author} {\bibfnamefont {S.-L.}\ \bibnamefont {Zhu}},\ }\href {\doibase 10.1088/1361-6633/aca3b6} {\bibfield  {journal} {\bibinfo  {journal} {Rept. Prog. Phys.}\ }\textbf {\bibinfo {volume} {86}},\ \bibinfo {pages} {026201} (\bibinfo {year} {2023})},\ \Eprint {http://arxiv.org/abs/2204.02649} {arXiv:2204.02649 [hep-ph]} \BibitemShut {NoStop}%
\bibitem [{\citenamefont {Ebert}\ \emph {et~al.}(2006)\citenamefont {Ebert}, \citenamefont {Faustov},\ and\ \citenamefont {Galkin}}]{Ebert:2005nc}%
  \BibitemOpen
  \bibfield  {author} {\bibinfo {author} {\bibfnamefont {D.}~\bibnamefont {Ebert}}, \bibinfo {author} {\bibfnamefont {R.~N.}\ \bibnamefont {Faustov}}, \ and\ \bibinfo {author} {\bibfnamefont {V.~O.}\ \bibnamefont {Galkin}},\ }\href {\doibase 10.1016/j.physletb.2006.01.026} {\bibfield  {journal} {\bibinfo  {journal} {Phys. Lett. B}\ }\textbf {\bibinfo {volume} {634}},\ \bibinfo {pages} {214} (\bibinfo {year} {2006})},\ \Eprint {http://arxiv.org/abs/hep-ph/0512230} {arXiv:hep-ph/0512230} \BibitemShut {NoStop}%
\bibitem [{\citenamefont {Hou}(2006)}]{Hou:2006it}%
  \BibitemOpen
  \bibfield  {author} {\bibinfo {author} {\bibfnamefont {W.-S.}\ \bibnamefont {Hou}},\ }\href {\doibase 10.1103/PhysRevD.74.017504} {\bibfield  {journal} {\bibinfo  {journal} {Phys. Rev. D}\ }\textbf {\bibinfo {volume} {74}},\ \bibinfo {pages} {017504} (\bibinfo {year} {2006})},\ \Eprint {http://arxiv.org/abs/hep-ph/0606016} {arXiv:hep-ph/0606016} \BibitemShut {NoStop}%
\bibitem [{\citenamefont {Bondar}\ \emph {et~al.}(2012)\citenamefont {Bondar} \emph {et~al.}}]{Belle:2011aa}%
  \BibitemOpen
  \bibfield  {author} {\bibinfo {author} {\bibfnamefont {A.}~\bibnamefont {Bondar}} \emph {et~al.} (\bibinfo {collaboration} {Belle}),\ }\href {\doibase 10.1103/PhysRevLett.108.122001} {\bibfield  {journal} {\bibinfo  {journal} {Phys. Rev. Lett.}\ }\textbf {\bibinfo {volume} {108}},\ \bibinfo {pages} {122001} (\bibinfo {year} {2012})},\ \Eprint {http://arxiv.org/abs/1110.2251} {arXiv:1110.2251 [hep-ex]} \BibitemShut {NoStop}%
\bibitem [{\citenamefont {Yang}\ \emph {et~al.}(2019)\citenamefont {Yang}, \citenamefont {Ping},\ and\ \citenamefont {Segovia}}]{Yang:2018oqd}%
  \BibitemOpen
  \bibfield  {author} {\bibinfo {author} {\bibfnamefont {G.}~\bibnamefont {Yang}}, \bibinfo {author} {\bibfnamefont {J.}~\bibnamefont {Ping}}, \ and\ \bibinfo {author} {\bibfnamefont {J.}~\bibnamefont {Segovia}},\ }\href {\doibase 10.1103/PhysRevD.99.014035} {\bibfield  {journal} {\bibinfo  {journal} {Phys. Rev. D}\ }\textbf {\bibinfo {volume} {99}},\ \bibinfo {pages} {014035} (\bibinfo {year} {2019})},\ \Eprint {http://arxiv.org/abs/1809.06193} {arXiv:1809.06193 [hep-ph]} \BibitemShut {NoStop}%
\bibitem [{\citenamefont {Sharma}\ and\ \citenamefont {Upadhyay}(2024)}]{Sharma:2024ern}%
  \BibitemOpen
  \bibfield  {author} {\bibinfo {author} {\bibfnamefont {A.}~\bibnamefont {Sharma}}\ and\ \bibinfo {author} {\bibfnamefont {A.}~\bibnamefont {Upadhyay}},\ }\href@noop {} {\  (\bibinfo {year} {2024})},\ \Eprint {http://arxiv.org/abs/2402.14885} {arXiv:2402.14885 [hep-ph]} \BibitemShut {NoStop}%
\bibitem [{\citenamefont {Ren}\ \emph {et~al.}(2022)\citenamefont {Ren}, \citenamefont {Wu},\ and\ \citenamefont {Zhu}}]{Ren:2021dsi}%
  \BibitemOpen
  \bibfield  {author} {\bibinfo {author} {\bibfnamefont {H.}~\bibnamefont {Ren}}, \bibinfo {author} {\bibfnamefont {F.}~\bibnamefont {Wu}}, \ and\ \bibinfo {author} {\bibfnamefont {R.}~\bibnamefont {Zhu}},\ }\href {\doibase 10.1155/2022/9103031} {\bibfield  {journal} {\bibinfo  {journal} {Adv. High Energy Phys.}\ }\textbf {\bibinfo {volume} {2022}},\ \bibinfo {pages} {9103031} (\bibinfo {year} {2022})},\ \Eprint {http://arxiv.org/abs/2109.02531} {arXiv:2109.02531 [hep-ph]} \BibitemShut {NoStop}%
\bibitem [{\citenamefont {Deng}\ and\ \citenamefont {Zhu}(2022)}]{Deng:2021gnb}%
  \BibitemOpen
  \bibfield  {author} {\bibinfo {author} {\bibfnamefont {C.}~\bibnamefont {Deng}}\ and\ \bibinfo {author} {\bibfnamefont {S.-L.}\ \bibnamefont {Zhu}},\ }\href {\doibase 10.1103/PhysRevD.105.054015} {\bibfield  {journal} {\bibinfo  {journal} {Phys. Rev. D}\ }\textbf {\bibinfo {volume} {105}},\ \bibinfo {pages} {054015} (\bibinfo {year} {2022})},\ \Eprint {http://arxiv.org/abs/2112.12472} {arXiv:2112.12472 [hep-ph]} \BibitemShut {NoStop}%
\bibitem [{\citenamefont {Gao}(2022)}]{Gao:2022lew}%
  \BibitemOpen
  \bibfield  {author} {\bibinfo {author} {\bibfnamefont {J.}~\bibnamefont {Gao}} (\bibinfo {collaboration} {CEPC Accelerator Study Group}),\ }\href@noop {} {\  (\bibinfo {year} {2022})},\ \Eprint {http://arxiv.org/abs/2203.09451} {arXiv:2203.09451 [physics.acc-ph]} \BibitemShut {NoStop}%
\bibitem [{\citenamefont {Sun}\ \emph {et~al.}(2023)\citenamefont {Sun}, \citenamefont {Wang}, \citenamefont {Liu}, \citenamefont {Zhu}, \citenamefont {Ruan},\ and\ \citenamefont {Wang}}]{Sun:2023lut}%
  \BibitemOpen
  \bibfield  {author} {\bibinfo {author} {\bibfnamefont {K.}~\bibnamefont {Sun}}, \bibinfo {author} {\bibfnamefont {Y.}~\bibnamefont {Wang}}, \bibinfo {author} {\bibfnamefont {J.}~\bibnamefont {Liu}}, \bibinfo {author} {\bibfnamefont {Y.}~\bibnamefont {Zhu}}, \bibinfo {author} {\bibfnamefont {M.}~\bibnamefont {Ruan}}, \ and\ \bibinfo {author} {\bibfnamefont {Y.}~\bibnamefont {Wang}},\ }\href@noop {} {\  (\bibinfo {year} {2023})},\ \Eprint {http://arxiv.org/abs/2306.11512} {arXiv:2306.11512 [hep-ex]} \BibitemShut {NoStop}%
\bibitem [{\citenamefont {Niu}\ \emph {et~al.}(2023)\citenamefont {Niu}, \citenamefont {Li}, \citenamefont {Bi},\ and\ \citenamefont {Ma}}]{Niu:2023ojf}%
  \BibitemOpen
  \bibfield  {author} {\bibinfo {author} {\bibfnamefont {J.-J.}\ \bibnamefont {Niu}}, \bibinfo {author} {\bibfnamefont {J.-B.}\ \bibnamefont {Li}}, \bibinfo {author} {\bibfnamefont {H.-Y.}\ \bibnamefont {Bi}}, \ and\ \bibinfo {author} {\bibfnamefont {H.-H.}\ \bibnamefont {Ma}},\ }\href {\doibase 10.1140/epjc/s10052-023-11958-1} {\bibfield  {journal} {\bibinfo  {journal} {Eur. Phys. J. C}\ }\textbf {\bibinfo {volume} {83}},\ \bibinfo {pages} {822} (\bibinfo {year} {2023})},\ \Eprint {http://arxiv.org/abs/2305.15362} {arXiv:2305.15362 [hep-ph]} \BibitemShut {NoStop}%
\bibitem [{\citenamefont {Ali}\ \emph {et~al.}(2018)\citenamefont {Ali}, \citenamefont {Parkhomenko}, \citenamefont {Qin},\ and\ \citenamefont {Wang}}]{Ali:2018ifm}%
  \BibitemOpen
  \bibfield  {author} {\bibinfo {author} {\bibfnamefont {A.}~\bibnamefont {Ali}}, \bibinfo {author} {\bibfnamefont {A.~Y.}\ \bibnamefont {Parkhomenko}}, \bibinfo {author} {\bibfnamefont {Q.}~\bibnamefont {Qin}}, \ and\ \bibinfo {author} {\bibfnamefont {W.}~\bibnamefont {Wang}},\ }\href {\doibase 10.1016/j.physletb.2018.05.055} {\bibfield  {journal} {\bibinfo  {journal} {Phys. Lett. B}\ }\textbf {\bibinfo {volume} {782}},\ \bibinfo {pages} {412} (\bibinfo {year} {2018})},\ \Eprint {http://arxiv.org/abs/1805.02535} {arXiv:1805.02535 [hep-ph]} \BibitemShut {NoStop}%
\bibitem [{\citenamefont {Ali}\ \emph {et~al.}(2019)\citenamefont {Ali}, \citenamefont {Parkhomenko}, \citenamefont {Qin},\ and\ \citenamefont {Wang}}]{Ali:2019tli}%
  \BibitemOpen
  \bibfield  {author} {\bibinfo {author} {\bibfnamefont {A.}~\bibnamefont {Ali}}, \bibinfo {author} {\bibfnamefont {A.~Y.}\ \bibnamefont {Parkhomenko}}, \bibinfo {author} {\bibfnamefont {Q.}~\bibnamefont {Qin}}, \ and\ \bibinfo {author} {\bibfnamefont {W.}~\bibnamefont {Wang}},\ }\href {\doibase 10.1134/S1547477119050029} {\bibfield  {journal} {\bibinfo  {journal} {Phys. Part. Nucl. Lett.}\ }\textbf {\bibinfo {volume} {16}},\ \bibinfo {pages} {481} (\bibinfo {year} {2019})}\BibitemShut {NoStop}%
\bibitem [{\citenamefont {Alwall}\ \emph {et~al.}(2014)\citenamefont {Alwall}, \citenamefont {Frederix}, \citenamefont {Frixione}, \citenamefont {Hirschi}, \citenamefont {Maltoni}, \citenamefont {Mattelaer}, \citenamefont {Shao}, \citenamefont {Stelzer}, \citenamefont {Torrielli},\ and\ \citenamefont {Zaro}}]{Alwall:2014hca}%
  \BibitemOpen
  \bibfield  {author} {\bibinfo {author} {\bibfnamefont {J.}~\bibnamefont {Alwall}}, \bibinfo {author} {\bibfnamefont {R.}~\bibnamefont {Frederix}}, \bibinfo {author} {\bibfnamefont {S.}~\bibnamefont {Frixione}}, \bibinfo {author} {\bibfnamefont {V.}~\bibnamefont {Hirschi}}, \bibinfo {author} {\bibfnamefont {F.}~\bibnamefont {Maltoni}}, \bibinfo {author} {\bibfnamefont {O.}~\bibnamefont {Mattelaer}}, \bibinfo {author} {\bibfnamefont {H.~S.}\ \bibnamefont {Shao}}, \bibinfo {author} {\bibfnamefont {T.}~\bibnamefont {Stelzer}}, \bibinfo {author} {\bibfnamefont {P.}~\bibnamefont {Torrielli}}, \ and\ \bibinfo {author} {\bibfnamefont {M.}~\bibnamefont {Zaro}},\ }\href {\doibase 10.1007/JHEP07(2014)079} {\bibfield  {journal} {\bibinfo  {journal} {JHEP}\ }\textbf {\bibinfo {volume} {07}},\ \bibinfo {pages} {079} (\bibinfo {year} {2014})},\ \Eprint {http://arxiv.org/abs/1405.0301} {arXiv:1405.0301 [hep-ph]} \BibitemShut {NoStop}%
\bibitem [{\citenamefont {Sjostrand}\ \emph {et~al.}(2006)\citenamefont {Sjostrand}, \citenamefont {Mrenna},\ and\ \citenamefont {Skands}}]{Sjostrand:2006za}%
  \BibitemOpen
  \bibfield  {author} {\bibinfo {author} {\bibfnamefont {T.}~\bibnamefont {Sjostrand}}, \bibinfo {author} {\bibfnamefont {S.}~\bibnamefont {Mrenna}}, \ and\ \bibinfo {author} {\bibfnamefont {P.~Z.}\ \bibnamefont {Skands}},\ }\href {\doibase 10.1088/1126-6708/2006/05/026} {\bibfield  {journal} {\bibinfo  {journal} {JHEP}\ }\textbf {\bibinfo {volume} {05}},\ \bibinfo {pages} {026} (\bibinfo {year} {2006})},\ \Eprint {http://arxiv.org/abs/hep-ph/0603175} {arXiv:hep-ph/0603175} \BibitemShut {NoStop}%
\bibitem [{\citenamefont {Artoisenet}\ and\ \citenamefont {Braaten}(2011)}]{Artoisenet:2010uu}%
  \BibitemOpen
  \bibfield  {author} {\bibinfo {author} {\bibfnamefont {P.}~\bibnamefont {Artoisenet}}\ and\ \bibinfo {author} {\bibfnamefont {E.}~\bibnamefont {Braaten}},\ }\href {\doibase 10.1103/PhysRevD.83.014019} {\bibfield  {journal} {\bibinfo  {journal} {Phys. Rev. D}\ }\textbf {\bibinfo {volume} {83}},\ \bibinfo {pages} {014019} (\bibinfo {year} {2011})},\ \Eprint {http://arxiv.org/abs/1007.2868} {arXiv:1007.2868 [hep-ph]} \BibitemShut {NoStop}%
\bibitem [{\citenamefont {Qin}\ \emph {et~al.}(2021)\citenamefont {Qin}, \citenamefont {Shen},\ and\ \citenamefont {Yu}}]{Qin:2020zlg}%
  \BibitemOpen
  \bibfield  {author} {\bibinfo {author} {\bibfnamefont {Q.}~\bibnamefont {Qin}}, \bibinfo {author} {\bibfnamefont {Y.-F.}\ \bibnamefont {Shen}}, \ and\ \bibinfo {author} {\bibfnamefont {F.-S.}\ \bibnamefont {Yu}},\ }\href {\doibase 10.1088/1674-1137/ac1b97} {\bibfield  {journal} {\bibinfo  {journal} {Chin. Phys. C}\ }\textbf {\bibinfo {volume} {45}},\ \bibinfo {pages} {103106} (\bibinfo {year} {2021})},\ \Eprint {http://arxiv.org/abs/2008.08026} {arXiv:2008.08026 [hep-ph]} \BibitemShut {NoStop}%
\bibitem [{\citenamefont {Yang}\ and\ \citenamefont {Guo}(2021)}]{Yang:2021jof}%
  \BibitemOpen
  \bibfield  {author} {\bibinfo {author} {\bibfnamefont {Z.}~\bibnamefont {Yang}}\ and\ \bibinfo {author} {\bibfnamefont {F.-K.}\ \bibnamefont {Guo}},\ }\href {\doibase 10.1088/1674-1137/ac2359} {\bibfield  {journal} {\bibinfo  {journal} {Chin. Phys. C}\ }\textbf {\bibinfo {volume} {45}},\ \bibinfo {pages} {123101} (\bibinfo {year} {2021})},\ \Eprint {http://arxiv.org/abs/2107.12247} {arXiv:2107.12247 [hep-ph]} \BibitemShut {NoStop}%
\bibitem [{\citenamefont {Shi}\ \emph {et~al.}(2022{\natexlab{a}})\citenamefont {Shi}, \citenamefont {Guo},\ and\ \citenamefont {Yang}}]{Shi:2022ipx}%
  \BibitemOpen
  \bibfield  {author} {\bibinfo {author} {\bibfnamefont {P.-P.}\ \bibnamefont {Shi}}, \bibinfo {author} {\bibfnamefont {F.-K.}\ \bibnamefont {Guo}}, \ and\ \bibinfo {author} {\bibfnamefont {Z.}~\bibnamefont {Yang}},\ }\href {\doibase 10.1103/PhysRevD.106.114026} {\bibfield  {journal} {\bibinfo  {journal} {Phys. Rev. D}\ }\textbf {\bibinfo {volume} {106}},\ \bibinfo {pages} {114026} (\bibinfo {year} {2022}{\natexlab{a}})},\ \Eprint {http://arxiv.org/abs/2208.02639} {arXiv:2208.02639 [hep-ph]} \BibitemShut {NoStop}%
\bibitem [{\citenamefont {Guo}\ \emph {et~al.}(2014{\natexlab{a}})\citenamefont {Guo}, \citenamefont {Mei\ss{}ner},\ and\ \citenamefont {Wang}}]{Guo:2013ufa}%
  \BibitemOpen
  \bibfield  {author} {\bibinfo {author} {\bibfnamefont {F.-K.}\ \bibnamefont {Guo}}, \bibinfo {author} {\bibfnamefont {U.-G.}\ \bibnamefont {Mei\ss{}ner}}, \ and\ \bibinfo {author} {\bibfnamefont {W.}~\bibnamefont {Wang}},\ }\href {\doibase 10.1088/0253-6102/61/3/14} {\bibfield  {journal} {\bibinfo  {journal} {Commun. Theor. Phys.}\ }\textbf {\bibinfo {volume} {61}},\ \bibinfo {pages} {354} (\bibinfo {year} {2014}{\natexlab{a}})},\ \Eprint {http://arxiv.org/abs/1308.0193} {arXiv:1308.0193 [hep-ph]} \BibitemShut {NoStop}%
\bibitem [{\citenamefont {Guo}\ \emph {et~al.}(2014{\natexlab{b}})\citenamefont {Guo}, \citenamefont {Mei\ss{}ner}, \citenamefont {Wang},\ and\ \citenamefont {Yang}}]{Guo:2014ppa}%
  \BibitemOpen
  \bibfield  {author} {\bibinfo {author} {\bibfnamefont {F.-K.}\ \bibnamefont {Guo}}, \bibinfo {author} {\bibfnamefont {U.-G.}\ \bibnamefont {Mei\ss{}ner}}, \bibinfo {author} {\bibfnamefont {W.}~\bibnamefont {Wang}}, \ and\ \bibinfo {author} {\bibfnamefont {Z.}~\bibnamefont {Yang}},\ }\href {\doibase 10.1007/JHEP05(2014)138} {\bibfield  {journal} {\bibinfo  {journal} {JHEP}\ }\textbf {\bibinfo {volume} {05}},\ \bibinfo {pages} {138} (\bibinfo {year} {2014}{\natexlab{b}})},\ \Eprint {http://arxiv.org/abs/1403.4032} {arXiv:1403.4032 [hep-ph]} \BibitemShut {NoStop}%
\bibitem [{\citenamefont {Guo}\ \emph {et~al.}(2014{\natexlab{c}})\citenamefont {Guo}, \citenamefont {Mei\ss{}ner}, \citenamefont {Wang},\ and\ \citenamefont {Yang}}]{Guo:2014sca}%
  \BibitemOpen
  \bibfield  {author} {\bibinfo {author} {\bibfnamefont {F.-K.}\ \bibnamefont {Guo}}, \bibinfo {author} {\bibfnamefont {U.-G.}\ \bibnamefont {Mei\ss{}ner}}, \bibinfo {author} {\bibfnamefont {W.}~\bibnamefont {Wang}}, \ and\ \bibinfo {author} {\bibfnamefont {Z.}~\bibnamefont {Yang}},\ }\href {\doibase 10.1140/epjc/s10052-014-3063-4} {\bibfield  {journal} {\bibinfo  {journal} {Eur. Phys. J. C}\ }\textbf {\bibinfo {volume} {74}},\ \bibinfo {pages} {3063} (\bibinfo {year} {2014}{\natexlab{c}})},\ \Eprint {http://arxiv.org/abs/1402.6236} {arXiv:1402.6236 [hep-ph]} \BibitemShut {NoStop}%
\bibitem [{\citenamefont {Albaladejo}\ \emph {et~al.}(2017)\citenamefont {Albaladejo}, \citenamefont {Guo}, \citenamefont {Hanhart}, \citenamefont {Mei\ss{}ner}, \citenamefont {Nieves}, \citenamefont {Nogga},\ and\ \citenamefont {Yang}}]{Albaladejo:2017blx}%
  \BibitemOpen
  \bibfield  {author} {\bibinfo {author} {\bibfnamefont {M.}~\bibnamefont {Albaladejo}}, \bibinfo {author} {\bibfnamefont {F.-K.}\ \bibnamefont {Guo}}, \bibinfo {author} {\bibfnamefont {C.}~\bibnamefont {Hanhart}}, \bibinfo {author} {\bibfnamefont {U.-G.}\ \bibnamefont {Mei\ss{}ner}}, \bibinfo {author} {\bibfnamefont {J.}~\bibnamefont {Nieves}}, \bibinfo {author} {\bibfnamefont {A.}~\bibnamefont {Nogga}}, \ and\ \bibinfo {author} {\bibfnamefont {Z.}~\bibnamefont {Yang}},\ }\href {\doibase 10.1088/1674-1137/41/12/121001} {\bibfield  {journal} {\bibinfo  {journal} {Chin. Phys. C}\ }\textbf {\bibinfo {volume} {41}},\ \bibinfo {pages} {121001} (\bibinfo {year} {2017})},\ \Eprint {http://arxiv.org/abs/1709.09101} {arXiv:1709.09101 [hep-ph]} \BibitemShut {NoStop}%
\bibitem [{\citenamefont {Ling}\ \emph {et~al.}(2021)\citenamefont {Ling}, \citenamefont {Dai}, \citenamefont {Du},\ and\ \citenamefont {Wang}}]{Ling:2021sld}%
  \BibitemOpen
  \bibfield  {author} {\bibinfo {author} {\bibfnamefont {P.}~\bibnamefont {Ling}}, \bibinfo {author} {\bibfnamefont {X.-H.}\ \bibnamefont {Dai}}, \bibinfo {author} {\bibfnamefont {M.-L.}\ \bibnamefont {Du}}, \ and\ \bibinfo {author} {\bibfnamefont {Q.}~\bibnamefont {Wang}},\ }\href {\doibase 10.1140/epjc/s10052-021-09613-8} {\bibfield  {journal} {\bibinfo  {journal} {Eur. Phys. J. C}\ }\textbf {\bibinfo {volume} {81}},\ \bibinfo {pages} {819} (\bibinfo {year} {2021})},\ \Eprint {http://arxiv.org/abs/2104.11133} {arXiv:2104.11133 [hep-ph]} \BibitemShut {NoStop}%
\bibitem [{\citenamefont {Shi}\ \emph {et~al.}(2022{\natexlab{b}})\citenamefont {Shi}, \citenamefont {Zhang}, \citenamefont {Guo},\ and\ \citenamefont {Yang}}]{Shi:2021hzm}%
  \BibitemOpen
  \bibfield  {author} {\bibinfo {author} {\bibfnamefont {P.-P.}\ \bibnamefont {Shi}}, \bibinfo {author} {\bibfnamefont {Z.-H.}\ \bibnamefont {Zhang}}, \bibinfo {author} {\bibfnamefont {F.-K.}\ \bibnamefont {Guo}}, \ and\ \bibinfo {author} {\bibfnamefont {Z.}~\bibnamefont {Yang}},\ }\href {\doibase 10.1103/PhysRevD.105.034024} {\bibfield  {journal} {\bibinfo  {journal} {Phys. Rev. D}\ }\textbf {\bibinfo {volume} {105}},\ \bibinfo {pages} {034024} (\bibinfo {year} {2022}{\natexlab{b}})},\ \Eprint {http://arxiv.org/abs/2111.13496} {arXiv:2111.13496 [hep-ph]} \BibitemShut {NoStop}%
\bibitem [{\citenamefont {Jin}\ \emph {et~al.}(2021)\citenamefont {Jin}, \citenamefont {Li}, \citenamefont {Liu}, \citenamefont {Qin}, \citenamefont {Si},\ and\ \citenamefont {Yu}}]{Jin:2021cxj}%
  \BibitemOpen
  \bibfield  {author} {\bibinfo {author} {\bibfnamefont {Y.}~\bibnamefont {Jin}}, \bibinfo {author} {\bibfnamefont {S.-Y.}\ \bibnamefont {Li}}, \bibinfo {author} {\bibfnamefont {Y.-R.}\ \bibnamefont {Liu}}, \bibinfo {author} {\bibfnamefont {Q.}~\bibnamefont {Qin}}, \bibinfo {author} {\bibfnamefont {Z.-G.}\ \bibnamefont {Si}}, \ and\ \bibinfo {author} {\bibfnamefont {F.-S.}\ \bibnamefont {Yu}},\ }\href {\doibase 10.1103/PhysRevD.104.114009} {\bibfield  {journal} {\bibinfo  {journal} {Phys. Rev. D}\ }\textbf {\bibinfo {volume} {104}},\ \bibinfo {pages} {114009} (\bibinfo {year} {2021})},\ \Eprint {http://arxiv.org/abs/2109.05678} {arXiv:2109.05678 [hep-ph]} \BibitemShut {NoStop}%
\bibitem [{\citenamefont {Hua}\ \emph {et~al.}(2023)\citenamefont {Hua}, \citenamefont {Li}, \citenamefont {Wang}, \citenamefont {Yang}, \citenamefont {Zhao},\ and\ \citenamefont {Zou}}]{Hua:2023zpa}%
  \BibitemOpen
  \bibfield  {author} {\bibinfo {author} {\bibfnamefont {X.-L.}\ \bibnamefont {Hua}}, \bibinfo {author} {\bibfnamefont {Y.-Y.}\ \bibnamefont {Li}}, \bibinfo {author} {\bibfnamefont {Q.}~\bibnamefont {Wang}}, \bibinfo {author} {\bibfnamefont {S.}~\bibnamefont {Yang}}, \bibinfo {author} {\bibfnamefont {Q.}~\bibnamefont {Zhao}}, \ and\ \bibinfo {author} {\bibfnamefont {B.-S.}\ \bibnamefont {Zou}},\ }\href@noop {} {\  (\bibinfo {year} {2023})},\ \Eprint {http://arxiv.org/abs/2310.04258} {arXiv:2310.04258 [hep-ph]} \BibitemShut {NoStop}%
\bibitem [{\citenamefont {Bignamini}\ \emph {et~al.}(2009)\citenamefont {Bignamini}, \citenamefont {Grinstein}, \citenamefont {Piccinini}, \citenamefont {Polosa},\ and\ \citenamefont {Sabelli}}]{Bignamini:2009sk}%
  \BibitemOpen
  \bibfield  {author} {\bibinfo {author} {\bibfnamefont {C.}~\bibnamefont {Bignamini}}, \bibinfo {author} {\bibfnamefont {B.}~\bibnamefont {Grinstein}}, \bibinfo {author} {\bibfnamefont {F.}~\bibnamefont {Piccinini}}, \bibinfo {author} {\bibfnamefont {A.~D.}\ \bibnamefont {Polosa}}, \ and\ \bibinfo {author} {\bibfnamefont {C.}~\bibnamefont {Sabelli}},\ }\href {\doibase 10.1103/PhysRevLett.103.162001} {\bibfield  {journal} {\bibinfo  {journal} {Phys. Rev. Lett.}\ }\textbf {\bibinfo {volume} {103}},\ \bibinfo {pages} {162001} (\bibinfo {year} {2009})},\ \Eprint {http://arxiv.org/abs/0906.0882} {arXiv:0906.0882 [hep-ph]} \BibitemShut {NoStop}%
\bibitem [{\citenamefont {Bierlich}\ \emph {et~al.}(2022)\citenamefont {Bierlich} \emph {et~al.}}]{Bierlich:2022pfr}%
  \BibitemOpen
  \bibfield  {author} {\bibinfo {author} {\bibfnamefont {C.}~\bibnamefont {Bierlich}} \emph {et~al.},\ }\href {\doibase 10.21468/SciPostPhysCodeb.8} {\bibfield  {journal} {\bibinfo  {journal} {SciPost Phys. Codeb.}\ }\textbf {\bibinfo {volume} {2022}},\ \bibinfo {pages} {8} (\bibinfo {year} {2022})},\ \Eprint {http://arxiv.org/abs/2203.11601} {arXiv:2203.11601 [hep-ph]} \BibitemShut {NoStop}%
\bibitem [{\citenamefont {Dong}\ \emph {et~al.}(2021{\natexlab{a}})\citenamefont {Dong}, \citenamefont {Guo},\ and\ \citenamefont {Zou}}]{Dong:2021bvy}%
  \BibitemOpen
  \bibfield  {author} {\bibinfo {author} {\bibfnamefont {X.-K.}\ \bibnamefont {Dong}}, \bibinfo {author} {\bibfnamefont {F.-K.}\ \bibnamefont {Guo}}, \ and\ \bibinfo {author} {\bibfnamefont {B.-S.}\ \bibnamefont {Zou}},\ }\href {\doibase 10.1088/1572-9494/ac27a2} {\bibfield  {journal} {\bibinfo  {journal} {Commun. Theor. Phys.}\ }\textbf {\bibinfo {volume} {73}},\ \bibinfo {pages} {125201} (\bibinfo {year} {2021}{\natexlab{a}})},\ \Eprint {http://arxiv.org/abs/2108.02673} {arXiv:2108.02673 [hep-ph]} \BibitemShut {NoStop}%
\bibitem [{\citenamefont {Dong}\ \emph {et~al.}(2021{\natexlab{b}})\citenamefont {Dong}, \citenamefont {Guo},\ and\ \citenamefont {Zou}}]{Dong:2021juy}%
  \BibitemOpen
  \bibfield  {author} {\bibinfo {author} {\bibfnamefont {X.-K.}\ \bibnamefont {Dong}}, \bibinfo {author} {\bibfnamefont {F.-K.}\ \bibnamefont {Guo}}, \ and\ \bibinfo {author} {\bibfnamefont {B.-S.}\ \bibnamefont {Zou}},\ }\href {\doibase 10.13725/j.cnki.pip.2021.02.001} {\bibfield  {journal} {\bibinfo  {journal} {Progr. Phys.}\ }\textbf {\bibinfo {volume} {41}},\ \bibinfo {pages} {65} (\bibinfo {year} {2021}{\natexlab{b}})},\ \Eprint {http://arxiv.org/abs/2101.01021} {arXiv:2101.01021 [hep-ph]} \BibitemShut {NoStop}%
\bibitem [{\citenamefont {Braaten}\ and\ \citenamefont {Kusunoki}(2005)}]{Braaten:2005jj}%
  \BibitemOpen
  \bibfield  {author} {\bibinfo {author} {\bibfnamefont {E.}~\bibnamefont {Braaten}}\ and\ \bibinfo {author} {\bibfnamefont {M.}~\bibnamefont {Kusunoki}},\ }\href {\doibase 10.1103/PhysRevD.72.014012} {\bibfield  {journal} {\bibinfo  {journal} {Phys. Rev. D}\ }\textbf {\bibinfo {volume} {72}},\ \bibinfo {pages} {014012} (\bibinfo {year} {2005})},\ \Eprint {http://arxiv.org/abs/hep-ph/0506087} {arXiv:hep-ph/0506087} \BibitemShut {NoStop}%
\bibitem [{\citenamefont {Artoisenet}\ and\ \citenamefont {Braaten}(2010)}]{Artoisenet:2009wk}%
  \BibitemOpen
  \bibfield  {author} {\bibinfo {author} {\bibfnamefont {P.}~\bibnamefont {Artoisenet}}\ and\ \bibinfo {author} {\bibfnamefont {E.}~\bibnamefont {Braaten}},\ }\href {\doibase 10.1103/PhysRevD.81.114018} {\bibfield  {journal} {\bibinfo  {journal} {Phys. Rev. D}\ }\textbf {\bibinfo {volume} {81}},\ \bibinfo {pages} {114018} (\bibinfo {year} {2010})},\ \Eprint {http://arxiv.org/abs/0911.2016} {arXiv:0911.2016 [hep-ph]} \BibitemShut {NoStop}%
\bibitem [{\citenamefont {Nieves}\ and\ \citenamefont {Valderrama}(2012)}]{Nieves:2012tt}%
  \BibitemOpen
  \bibfield  {author} {\bibinfo {author} {\bibfnamefont {J.}~\bibnamefont {Nieves}}\ and\ \bibinfo {author} {\bibfnamefont {M.~P.}\ \bibnamefont {Valderrama}},\ }\href {\doibase 10.1103/PhysRevD.86.056004} {\bibfield  {journal} {\bibinfo  {journal} {Phys. Rev. D}\ }\textbf {\bibinfo {volume} {86}},\ \bibinfo {pages} {056004} (\bibinfo {year} {2012})},\ \Eprint {http://arxiv.org/abs/1204.2790} {arXiv:1204.2790 [hep-ph]} \BibitemShut {NoStop}%
\bibitem [{\citenamefont {Guo}\ \emph {et~al.}(2013)\citenamefont {Guo}, \citenamefont {Hidalgo-Duque}, \citenamefont {Nieves},\ and\ \citenamefont {Valderrama}}]{Guo:2013sya}%
  \BibitemOpen
  \bibfield  {author} {\bibinfo {author} {\bibfnamefont {F.-K.}\ \bibnamefont {Guo}}, \bibinfo {author} {\bibfnamefont {C.}~\bibnamefont {Hidalgo-Duque}}, \bibinfo {author} {\bibfnamefont {J.}~\bibnamefont {Nieves}}, \ and\ \bibinfo {author} {\bibfnamefont {M.~P.}\ \bibnamefont {Valderrama}},\ }\href {\doibase 10.1103/PhysRevD.88.054007} {\bibfield  {journal} {\bibinfo  {journal} {Phys. Rev. D}\ }\textbf {\bibinfo {volume} {88}},\ \bibinfo {pages} {054007} (\bibinfo {year} {2013})},\ \Eprint {http://arxiv.org/abs/1303.6608} {arXiv:1303.6608 [hep-ph]} \BibitemShut {NoStop}%
\bibitem [{\citenamefont {Aaij}\ \emph {et~al.}(2022{\natexlab{b}})\citenamefont {Aaij} \emph {et~al.}}]{LHCb:2021auc}%
  \BibitemOpen
  \bibfield  {author} {\bibinfo {author} {\bibfnamefont {R.}~\bibnamefont {Aaij}} \emph {et~al.} (\bibinfo {collaboration} {LHCb}),\ }\href {\doibase 10.1038/s41467-022-30206-w} {\bibfield  {journal} {\bibinfo  {journal} {Nature Commun.}\ }\textbf {\bibinfo {volume} {13}},\ \bibinfo {pages} {3351} (\bibinfo {year} {2022}{\natexlab{b}})},\ \Eprint {http://arxiv.org/abs/2109.01056} {arXiv:2109.01056 [hep-ex]} \BibitemShut {NoStop}%
\bibitem [{\citenamefont {Du}\ \emph {et~al.}(2022)\citenamefont {Du}, \citenamefont {Baru}, \citenamefont {Dong}, \citenamefont {Filin}, \citenamefont {Guo}, \citenamefont {Hanhart}, \citenamefont {Nefediev}, \citenamefont {Nieves},\ and\ \citenamefont {Wang}}]{Du:2021zzh}%
  \BibitemOpen
  \bibfield  {author} {\bibinfo {author} {\bibfnamefont {M.-L.}\ \bibnamefont {Du}}, \bibinfo {author} {\bibfnamefont {V.}~\bibnamefont {Baru}}, \bibinfo {author} {\bibfnamefont {X.-K.}\ \bibnamefont {Dong}}, \bibinfo {author} {\bibfnamefont {A.}~\bibnamefont {Filin}}, \bibinfo {author} {\bibfnamefont {F.-K.}\ \bibnamefont {Guo}}, \bibinfo {author} {\bibfnamefont {C.}~\bibnamefont {Hanhart}}, \bibinfo {author} {\bibfnamefont {A.}~\bibnamefont {Nefediev}}, \bibinfo {author} {\bibfnamefont {J.}~\bibnamefont {Nieves}}, \ and\ \bibinfo {author} {\bibfnamefont {Q.}~\bibnamefont {Wang}},\ }\href {\doibase 10.1103/PhysRevD.105.014024} {\bibfield  {journal} {\bibinfo  {journal} {Phys. Rev. D}\ }\textbf {\bibinfo {volume} {105}},\ \bibinfo {pages} {014024} (\bibinfo {year} {2022})},\ \Eprint {http://arxiv.org/abs/2110.13765} {arXiv:2110.13765 [hep-ph]} \BibitemShut {NoStop}%
\bibitem [{\citenamefont {Dai}\ \emph {et~al.}(2023)\citenamefont {Dai}, \citenamefont {Fleming}, \citenamefont {Hodges},\ and\ \citenamefont {Mehen}}]{Dai:2023mxm}%
  \BibitemOpen
  \bibfield  {author} {\bibinfo {author} {\bibfnamefont {L.}~\bibnamefont {Dai}}, \bibinfo {author} {\bibfnamefont {S.}~\bibnamefont {Fleming}}, \bibinfo {author} {\bibfnamefont {R.}~\bibnamefont {Hodges}}, \ and\ \bibinfo {author} {\bibfnamefont {T.}~\bibnamefont {Mehen}},\ }\href {\doibase 10.1103/PhysRevD.107.076001} {\bibfield  {journal} {\bibinfo  {journal} {Phys. Rev. D}\ }\textbf {\bibinfo {volume} {107}},\ \bibinfo {pages} {076001} (\bibinfo {year} {2023})},\ \Eprint {http://arxiv.org/abs/2301.11950} {arXiv:2301.11950 [hep-ph]} \BibitemShut {NoStop}%
\bibitem [{\citenamefont {Wu}\ \emph {et~al.}(2019)\citenamefont {Wu}, \citenamefont {Chen},\ and\ \citenamefont {Guo}}]{Wu:2018xaa}%
  \BibitemOpen
  \bibfield  {author} {\bibinfo {author} {\bibfnamefont {Q.}~\bibnamefont {Wu}}, \bibinfo {author} {\bibfnamefont {D.-Y.}\ \bibnamefont {Chen}}, \ and\ \bibinfo {author} {\bibfnamefont {F.-K.}\ \bibnamefont {Guo}},\ }\href {\doibase 10.1103/PhysRevD.99.034022} {\bibfield  {journal} {\bibinfo  {journal} {Phys. Rev. D}\ }\textbf {\bibinfo {volume} {99}},\ \bibinfo {pages} {034022} (\bibinfo {year} {2019})},\ \Eprint {http://arxiv.org/abs/1810.09696} {arXiv:1810.09696 [hep-ph]} \BibitemShut {NoStop}%
\bibitem [{\citenamefont {Garmash}\ \emph {et~al.}(2015)\citenamefont {Garmash} \emph {et~al.}}]{Belle:2014vzn}%
  \BibitemOpen
  \bibfield  {author} {\bibinfo {author} {\bibfnamefont {A.}~\bibnamefont {Garmash}} \emph {et~al.} (\bibinfo {collaboration} {Belle}),\ }\href {\doibase 10.1103/PhysRevD.91.072003} {\bibfield  {journal} {\bibinfo  {journal} {Phys. Rev. D}\ }\textbf {\bibinfo {volume} {91}},\ \bibinfo {pages} {072003} (\bibinfo {year} {2015})},\ \Eprint {http://arxiv.org/abs/1403.0992} {arXiv:1403.0992 [hep-ex]} \BibitemShut {NoStop}%
\bibitem [{\citenamefont {Aoki}\ \emph {et~al.}(2023)\citenamefont {Aoki}, \citenamefont {Aoki},\ and\ \citenamefont {Inoue}}]{Aoki:2023nzp}%
  \BibitemOpen
  \bibfield  {author} {\bibinfo {author} {\bibfnamefont {T.}~\bibnamefont {Aoki}}, \bibinfo {author} {\bibfnamefont {S.}~\bibnamefont {Aoki}}, \ and\ \bibinfo {author} {\bibfnamefont {T.}~\bibnamefont {Inoue}},\ }\href {\doibase 10.1103/PhysRevD.108.054502} {\bibfield  {journal} {\bibinfo  {journal} {Phys. Rev. D}\ }\textbf {\bibinfo {volume} {108}},\ \bibinfo {pages} {054502} (\bibinfo {year} {2023})},\ \Eprint {http://arxiv.org/abs/2306.03565} {arXiv:2306.03565 [hep-lat]} \BibitemShut {NoStop}%
\bibitem [{\citenamefont {Alexandrou}\ \emph {et~al.}(2024)\citenamefont {Alexandrou}, \citenamefont {Finkenrath}, \citenamefont {Leontiou}, \citenamefont {Meinel}, \citenamefont {Pflaumer},\ and\ \citenamefont {Wagner}}]{Alexandrou:2024iwi}%
  \BibitemOpen
  \bibfield  {author} {\bibinfo {author} {\bibfnamefont {C.}~\bibnamefont {Alexandrou}}, \bibinfo {author} {\bibfnamefont {J.}~\bibnamefont {Finkenrath}}, \bibinfo {author} {\bibfnamefont {T.}~\bibnamefont {Leontiou}}, \bibinfo {author} {\bibfnamefont {S.}~\bibnamefont {Meinel}}, \bibinfo {author} {\bibfnamefont {M.}~\bibnamefont {Pflaumer}}, \ and\ \bibinfo {author} {\bibfnamefont {M.}~\bibnamefont {Wagner}},\ }\href@noop {} {\  (\bibinfo {year} {2024})},\ \Eprint {http://arxiv.org/abs/2404.03588} {arXiv:2404.03588 [hep-lat]} \BibitemShut {NoStop}%
\end{thebibliography}%

\end{document}